\documentclass[aps,prd,longbibliography,preprintnumbers,nofootinbib,
superscriptaddress,amsmath,floatfix,10pt,twocolumn]{revtex4}

\usepackage{amssymb}  
\usepackage{amsmath}
\usepackage{hyperref}
\usepackage{graphicx
}
\usepackage{stmaryrd}
\usepackage{exscale}
\usepackage{bbold}
\usepackage{textcomp}
\usepackage{enumerate}
\usepackage [english]{babel}
\usepackage[utf8]{inputenc}
\usepackage[T1]{fontenc}
\usepackage [autostyle, english = american]{csquotes}
\MakeOuterQuote{"}

\usepackage{enumitem}
\setlist[itemize]{leftmargin=*}

\usepackage{color}

\usepackage[normalem]{ulem}  

\newcommand{\non}{\nonumber\\}

\newcommand{\MKK}{M_{\rm KK}}
\newcommand{\uKK}{u_{\rm KK}}

\newcommand{\der}{\partial}

\newcommand{\be}{\begin{equation}}
\newcommand{\ee}{\end{equation}}
\newcommand{\bea}{\begin{eqnarray}}
\newcommand{\eea}{\end{eqnarray}}
\newcommand{\ba}[1]{\begin{array}{#1}}
\newcommand{\ea}{\end{array}}

\newcommand{\Tr}{{\rm Tr}}
\newcommand{\STr}{{\rm STr}}

\newcommand{\Aa}{\mathcal{A}}

\newcommand{\of}[1]{\left(#1\right)}
\newcommand{\off}[1]{\left[#1\right]}

\begin{document}

\title{ 
Advances in the holographic description of hot and dense QCD matter
 }

\author{Nicolas Kovensky}
\email{nicolas.kovensky@iflp.unlp.edu.ar}
\affiliation{Instituto de F\'isica de La Plata - CONICET
Diagonal 113 e/ 63 y 64, 1900 - La Plata, Argentina}

\author{Andreas Schmitt}
\email{a.schmitt@soton.ac.uk}
\affiliation{Mathematical Sciences and STAG Research Centre, University of Southampton, Southampton SO17 1BJ, United Kingdom}

\date{27 August 2026}


\begin{abstract} 
We review and extend recent progress on QCD-like phases at nonzero temperature and chemical potentials within the gauge-gravity duality. We focus on the Witten-Sakai-Sugimoto model, a top-down approach based on string theory which allows for a unified treatment of mesons, baryons and quarks. Phases such as pion condensation, baryonic and quarkyonic matter are located in the phase diagram, and we discuss  predictions as well as weaknesses and limitations of this holographic approach. One of the novelties included in this review is the first holographic construction of a pionic phase between BEC and BCS regimes, which however is found to be metastable. We also discuss future directions in holographic QCD, within the model used here and in a wider context. 
\end{abstract}

\maketitle

\tableofcontents

\section{Introduction}

\subsection{Context -- QCD}

Quantum Chromodynamics (QCD) describes the interactions of quarks and gluons \cite{Gross:1973id,Politzer:1973fx}. It constitutes the microscopic theory underlying a variety of physical phenomena, ranging from hadron structure, quark-gluon plasma dynamics in heavy-ion collisions to the cosmological evolution of the early universe and the physics of neutron stars and their mergers \cite{Brambilla:2014jmp}. For many of these applications, a considerable amount of experimental observations and data from numerical simulations has been collected over the last few decades. 
Nevertheless, our theoretical understanding of QCD is patchy in the large and multi-dimensional parameter space spanned by thermodynamic quantities such as baryon chemical potential \cite{Alford:2007xm,Fukushima:2025ujk,Karkkainen:2025nkz,Schmitt:2025cqi}, isospin chemical potential \cite{Brandt:2017oyy,Brandt:2022hwy,Abbott:2023coj}, temperature \cite{HotQCD:2018pds,Bresciani:2025mcu}, and magnetic field \cite{Bali:2011qj,Kharzeev:2013jha,Bruckmann:2013oba}. Similarly, important fundamental questions  remain unanswered for the phase structure in the space spanned by the parameters of the theory itself, such as quark masses \cite{Bazavov:2017xul,Klinger:2025xxb}, number of flavors $N_f$  \cite{Cuteri:2021ikv,Fejos:2024bgl}, and number of colors $N_c$ \cite{McLerran:2007qj,Cotter:2012mb}. The reason is the enormous difficulty of first-principle calculations away from the perturbative regime, manifest for instance in the infamous sign problem of lattice QCD in the presence of a baryon chemical potential \cite{deForcrand:2009zkb,Nagata:2021ugx}.

In practice, it is therefore often necessary to resort to effective field theories such as chiral perturbation theory ($\chi$PT) 
\cite{GasserLeutwyler:1983CPT,Ecker:1994gg,Son:2000xc}, chiral effective theories for nucleons 
\cite{Epelbaum:2008ga,Tews:2018kmu,Alp:2025wjn}, or to phenomenological models, for instance the linear sigma model 
\cite{Gell-Mann:1960mvl,Lee:1968da,Pisarski:1983ms} or the Nambu-Jona--Lasinio  (NJL) model  
\cite{Nambu:1961tp,Nambu:1961fr,Buballa:2003qv}. Non-perturbative effects, without solving full QCD, can be included for instance with the help of Dyson-Schwinger equations \cite{Nickel:2006vf,Fischer:2018sdj} or the functional renormalization group \cite{Gies:2002hq,Braun:2018svj}. The results obtained from these approaches -- as well as those obtained from perturbative QCD \cite{Freedman:1976ub,Kraemmer:2003gd} -- are either trustworthy only in a particular region of the phase diagram, for example at very low or asymptotically high baryon densities, or may predict  qualitative results that may or may not be realized in full QCD -- to be confirmed or ruled out ultimately by first-principle calculations or experiment. 

This ongoing effort has led to very valuable insights into and conjectures about hot and dense QCD matter, albeit far from a complete picture of the QCD phase diagram. For instance, a particularly intriguing regime is that of cold and dense matter, where several exotic phases have been proposed. Examples are, at large baryon chemical potential, color-flavor-locking \cite{Alford:1998mk}, quarkyonic matter 
\cite{McLerran:2007qj,McLerran:2008ua} and, at large 
isospin chemical potential, a crossover from Bose-Einstein condensation of pions (BEC) to a phase of 
weakly coupled quark-antiquark pairs as in the Bardeen-Cooper-Schrieffer (BCS) theory of superconductivity 
\cite{Son:2000xc,Brandt:2019hel}. Dense QCD is also expected to exhibit phases that spontaneously break rotational and/or translational invariance 
\cite{Anglani:2013gfu,Buballa:2014tba,Brauner:2016pko,Evans:2023hms}.

\subsection{Context -- holography}

The Anti-de Sitter/Conformal field theory (AdS/CFT) correspondence (or ``gauge-gravity duality'' or ``holography'') \cite{Maldacena:1997re,Gubser:1998bc,Witten:1998qj} provides a complementary approach. In general terms, holography can be understood as a framework for studying the strongly-coupled regime of gauge theories (at large $N_c$) via classical gravity and/or string theory in a higher-dimensional curved space (the ``bulk''). Its application to QCD would be straightforward if the precise gravity dual of QCD was known. This is not the case, and thus a multitude of models within the framework of holography are being used -- different from, but ideally as close as possible to real-world QCD.  Holography is well suited to tackle non-equilibrium phenomena of QCD-like theories \cite{Kovtun:2004de,Chesler:2008hg,Casalderrey-Solana:2013aba,Hoyos:2021njg,Bea:2024bls,Bigazzi:2026ppu}. Here, however, we are only concerned with phases in thermodynamic equilibrium. Different approaches within the holographic framework can, broadly speaking, be divided into two classes. 

Firstly, ``top-down'' models that are based on string theory and whose dual field theory is known -- at least in the simplest cases. The prime example employs a stack of $N_c$ D3-branes, which generates a spacetime solution of the form ${\rm AdS}_5 \times S^5$, with a string theory on this background that is dual to ${\cal N}=4$ supersymmetric Yang-Mills (YM) theory. Adding fermions in the fundamental representation of the gauge group is achieved by $N_f$ D7-branes on top of the gluonic  background. This setup has been employed extensively to understand strongly-coupled hot and dense matter \cite{Kobayashi:2006sb,Mateos:2007vn,Mateos:2007vc,Evans:2010iy,Evans:2011mu,Evans:2011eu,Hoyos:2016zke}. 

Secondly, ``bottom-up'' models, which are of phenomenological nature. In this case, a higher-dimensional action is ``guessed'' without reference to an underlying string theory and without rigorous knowledge of the dual field theory. Examples are hard-wall \cite{Polchinski:2001tt,Erlich:2005qh,Bartolini:2022rkl} and soft-wall \cite{Karch:2006pv,Gherghetta:2009ac} models, ``Einstein-Maxwell-Dilaton'' theory   \cite{Gubser:2008ny,DeWolfe:2010he,Zhang:2022uin,Rougemont:2023gfz,Deng:2026aht} and holographic QCD in the Veneziano limit (``V-QCD'') \cite{Jarvinen:2011qe,Alho:2012mh,Gursoy:2017wzz,Tootle:2022pvd,CruzRojas:2023ugm,Demircik:2024aig,Bartolini:2025sag}.  The V-QCD approach is remarkable in the sense that the backreaction of the flavor sector on the gluon background is taken into account, and significant progress has been made on topics similar to the ones discussed here, see Refs.\ \cite{Jarvinen:2021jbd,Jarvinen:2022doa} for reviews. 

In this review we focus on recent advances in the Witten-Sakai-Sugimoto (WSS) model \cite{Witten:1998qj,Sakai:2004cn,Sakai:2005yt} regarding the structure of QCD at nonzero temperatures, baryon and isospin chemical potentials, mainly based on Refs.\   \cite{Li:2015uea,Kovensky:2019bih,Kovensky:2020xif,Kovensky:2021ddl,Kovensky:2021kzl,Kovensky:2021wzu,Kovensky:2023mye,Kovensky:2024oqx,Ecker:2025sjb}. The WSS model is a non-supersymmetric top-down construction. It is closer to QCD than most of the other holographic approaches, e.g., there is a limit, albeit inaccessible because string corrections become important, where it is dual to large-$N_c$ QCD. Nevertheless, the results have to be taken with a lot of care, due to the natural limitations of the gauge-gravity duality and due to additional approximations and simplifications made within the given framework that we shall explain. Therefore, our results apply to a ``distorted'' version of QCD, where some of the distortions can be identified and some remain obscured. For instance, asymptotic freedom at very large energy densities  is not built-in: The coupling of the field theory remains large throughout our analysis.

Remarkably, and rarely achieved in more traditional models, the WSS model allows us to treat quarks, mesons, and baryons within a single theoretical framework. Therefore, we are able to include the physics of pion and rho meson condensates, dense isospin-asymmetric baryonic matter, quark contributions, and temperature effects on an equal footing.  A considerable number of relevant phases can be constructed, including some  exotic ones where different types of QCD matter coexist. In all cases, we compute and compare the corresponding free energies fully dynamically.

\subsection{QCD matter from holography}
\label{sec:QCDmatter}

The holographic dictionary \cite{Maldacena:1997re,Witten:1998qj} provides an understanding of fields, symmetries, sources, and observables of the gauge theory in terms of string-theoretic objects in its dual description. We start with a brief reminder of the dictionary, tailored to our purposes within the WSS model. 

\begin{itemize}

\item Pure SU($N_c$) glue physics is captured by the AdS-type geometry in the bulk. Upon going from low to high temperature, the gravitational background undergoes a Hawking-Page phase transition \cite{Hawking:1982dh} from a thermal AdS-type geometry to an AdS black hole. This is the holographic realization of the confinement-deconfinement transition \cite{Witten:1998zw}. Therefore, we shall often refer to the different backgrounds as ``confined geometry'' and ``deconfined geometry''. 
    
    \item Flavor degrees of freedom are associated with  the dynamics of a set of $N_f$ D-branes and $N_f$ anti-D-branes, embedded in this curved background. They are separated asymptotically by a fixed distance and can connect in the bulk, thus providing a geometric realization of the chiral symmetry breaking pattern ${\rm SU}(N_f)_L\times {\rm SU}(N_f)_R\to {\rm  SU}(N_f)_{L+R}$ \cite{Sakai:2004cn}. We work in the so-called quenched limit $N_f \ll N_c$. This means, in particular, that we ignore the backreaction of the flavor branes on the geometric background. 
    
    \item The relevant low-energy fluctuations of these branes are captured by a set of gauge fields $\mathcal{A}_\mu^a$ with $\mu=0,1,2,3,U$, where $U$ is the holographic coordinate, and $a = 1,\dots, N_f$. We will always work in the $\mathcal{A}_U^a=0$ gauge. Sources such as the chemical potentials, as well as the pion condensate, are encoded in the asymptotic boundary conditions of the temporal components $\Aa_0^a$. Including vector meson condensates induces nonzero spatial components $\Aa_i^a$. 

    \item Fundamental quarks correspond to strings with one endpoint connected to the flavor branes. The remaining endpoint must end on a different (color) brane, or behind the black hole horizon, when there is one. 
    How much these strings are stretched in a given configuration is related to the constituent quark mass \cite{Kovensky:2019bih}.  

    \item Baryons are non-perturbative objects, namely additional D-branes wrapped around a compact sphere in the string-theoretic extra dimensions, with $N_c$ strings attached to  each of them \cite{Witten:1998xy,Gross:1998gk,Sakai:2004cn}. In a simple approximation, these are pointlike objects from the point of view of the effective five-dimensional geometry. A more precise description is given in terms of topologically non-trivial  configurations of the gauge fields $\Aa_\mu^a$ on the flavor branes \cite{Douglas:1995bn,Hata:2007mb,Hashimoto:2008zw}. This provides a holographic realization of  ``skyrmions from instantons'' \cite{Atiyah:1989dq,Sutcliffe:2010et}.

    \item For applications to neutron stars it is necessary   to include leptons. Here we simply combine the holographic model with a non-interacting  gas of electrons and muons. Leptons only affect our strongly interacting matter through the conditions of equilibrium with respect to electroweak processes and electric charge neutrality.
\end{itemize}

In its original version \cite{Sakai:2004cn,Sakai:2005yt}, the WSS model has two free parameters: the 't Hooft coupling $\lambda$ of the field theory, which is related to the curvature of the gravitational background, and the Kaluza-Klein mass $\MKK$, which sets the scale for the deconfinement phase transition. At low energies, the effective action matches that of $\chi$PT. This was used in fitting the parameters $\lambda\simeq 16.6$ and $\MKK \simeq 949\, {\rm  MeV}$  \cite{Sakai:2005yt} to  match the experimental values for the pion decay constant and the rho meson mass.  In this fit, the asymptotic separation of the flavor branes $L$ is antipodal, $L=\pi/M_{\rm KK}$. It is also possible to choose $L$ non-antipodal \cite{Aharony:2006da,Parnachev:2006dn,Horigome:2006xu,Evans:2007jr}. 

In the results presented here, we shall work with two variants of the model. Firstly, the confined geometry with antipodal separation, used for zero temperature.  Secondly, the ``decompactified limit'', where $L$ is non-antipodal and which has a nontrivial temperature dependence. For the latter, we shall employ the deconfined geometry down to zero temperature. This is motivated by the observation that for $L\ll \pi/M_{\rm KK}$ the scales for deconfinement and chiral symmetry breaking are widely separated. This decompactified limit can be viewed as a holographic version of an NJL-like model \cite{Antonyan:2006vw,Davis:2007ka,Preis:2012fh}. Although some of the top-down control is lost, this version of the model might well be closer to $N_c=3$ QCD because the gluon dynamics, which dominate at large $N_c$, are suppressed. In this setup, the two relevant combinations of the parameters can be chosen as $L$ and the dimensionless ratio $\lambda_5/L$, where $\lambda_5$ is the five-dimensional 't Hooft coupling. In both versions of the model, the pion mass is included through an effective action \cite{Aharony:2008an,Kovensky:2019bih} such that the vacuum mass of the pion $m_\pi$ is an additional parameter.

\begin{figure*} [t]
\begin{center}
{\includegraphics[width=0.8\textwidth]{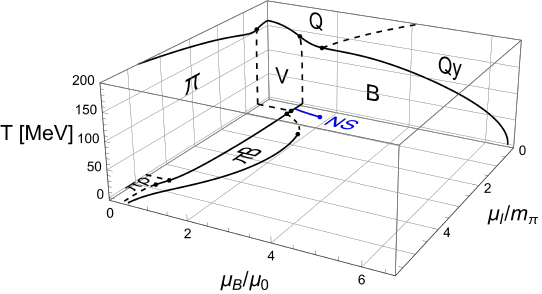}}
\caption{QCD-like phase diagram obtained from the Witten-Sakai-Sugimoto model in the space of temperature $T$, baryon chemical potential $\mu_B$, and isospin chemical potential $\mu_I$, at physical pion mass $m_\pi=140\, {\rm MeV}$. Solid (dashed) curves are phase transitions of first (second) order. Neutron stars are located on the blue curve. See text for the definition of the various phases, and Fig.\ \ref{fig:embeddings} for their geometries in the holographic bulk. The phase transition curves are taken from Refs.\ \cite{Kovensky:2023mye,Kovensky:2024oqx} and (with improvements in this review)  Ref.\ \cite{Kovensky:2020xif}. Inconsistencies between $T=0$ and $T>0$ (mismatching phase transitions, no chiral restoration in the $T=0$ plane) arise from using different versions of the model. The $\mu_B$ axis is given in units of the chemical potential for the baryon onset $\mu_0$. Details of all phases are discussed in this review, using this phase diagram as a guide.} 
\label{fig:phases3D}
\end{center}
\end{figure*}

\begin{figure*} [t]
\begin{center}
\includegraphics[width=0.75\textwidth]{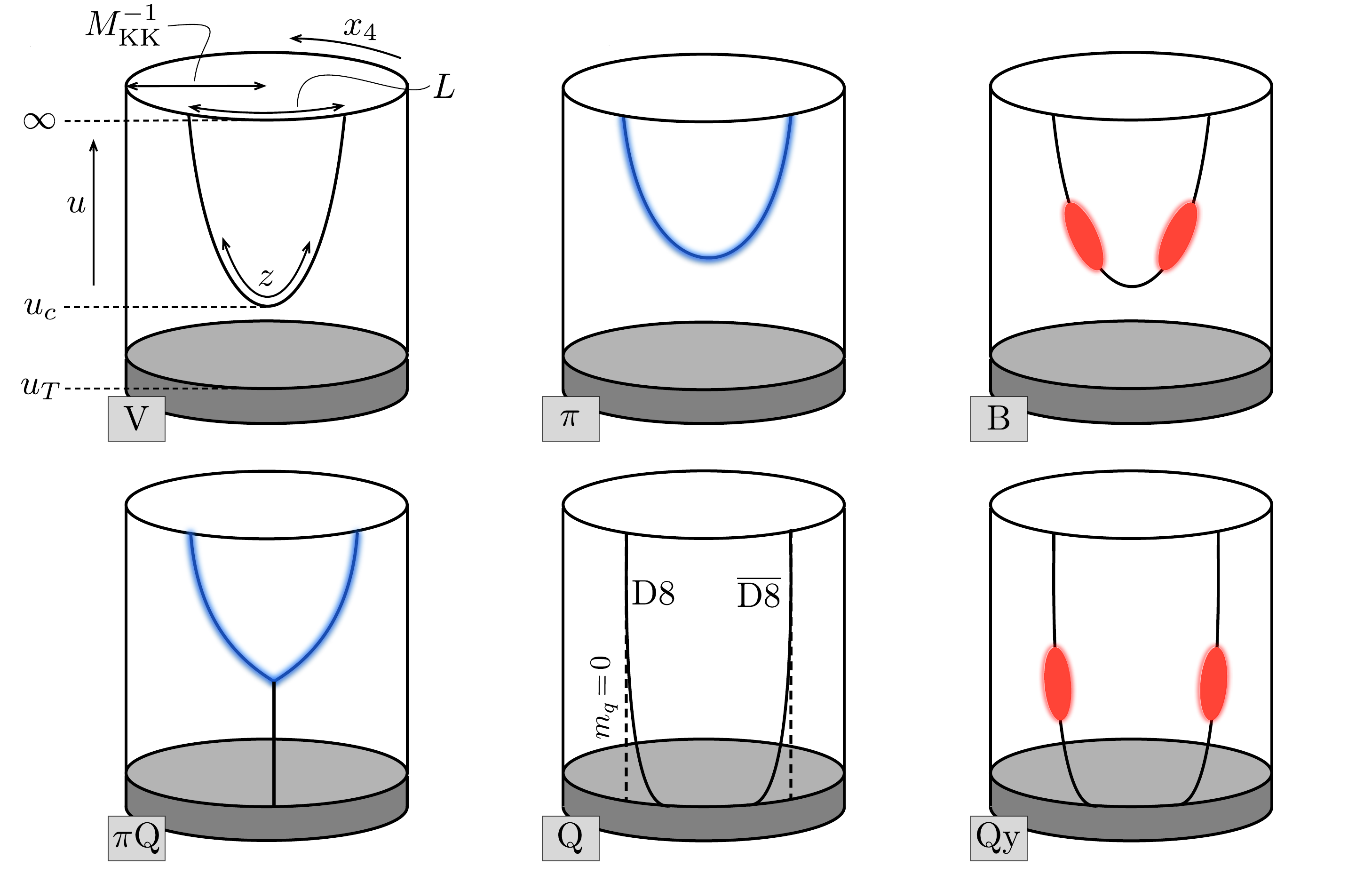}
\caption{Schematic view of the configurations needed for the phase diagram in Fig.\ \ref{fig:phases3D}. The space spanned by the holographic coordinate $u$ (or $z$ along the flavor branes) and the compactified direction $x_4$ (with radius $M_{\rm KK}^{-1}$) is shown for the deconfined geometry. This geometry is used for the $T>0$ phases in Fig.\ \ref{fig:phases3D}, while analogous configurations in the confined geometry, with antipodal separation $L=\pi/M_{\rm KK}$, are used for $T=0$. Connected flavor branes correspond to phases with spontaneously broken chiral symmetry. Pion condensation (blue) and baryonic matter (red) are described by particular gauge field configurations, while quark degrees of freedom are introduced by adding string sources or when the flavor branes reach the horizon at $u=u_T$ (straight branes for vanishing quark masses $m_q$). The different sources can be combined, for instance to obtain pion condensation coexisting with baryonic matter, resulting in phase $\pi$B in Fig.\ \ref{fig:phases3D}.} 
\label{fig:embeddings}
\end{center}
\end{figure*}

\subsection{Phase diagram}

The guiding principle of this review is the phase diagram in Fig.\ \ref{fig:phases3D}, spanned by temperature $T$, baryon chemical potential $\mu_B$, and isospin chemical potential $\mu_I$. It summarizes the main results obtained in recent years in the WSS model regarding hot and dense QCD matter. The theme of the review is to go through the features of the phase diagram step by step. The geometry of the phases in the bulk is illustrated in Fig.\ \ref{fig:embeddings} for the deconfined geometry, and we now give a short description for each of them. 

\begin{itemize}
    \item \textbf{Vacuum} (V). 
    This is the simplest configuration, which occurs at sufficiently low temperatures and chemical potentials. String and baryon sources are absent, and the gauge fields either vanish or are constant. In the non-antipodal case, the embedding of the flavor branes remains non-trivial and is temperature-dependent. It is $\cup$-shaped, signaling that chiral symmetry is broken, see upper left panel of Fig.\ \ref{fig:embeddings}.

    \item \textbf{Pion condensation} ($\pi$).  
    For $\mu_I > m_\pi/2$ and sufficiently small $T$ and $\mu_B$, pion condensation occurs. The temporal components of the non-abelian gauge fields acquire non-trivial profiles, indicated by blue colored branes in Fig.\ \ref{fig:embeddings}. The pion condensate is computed dynamically, and the brane embedding adjusts accordingly.

    \item \textbf{Rho meson condensation} ($\rho$ and $\pi\rho$).
    If pion condensation is ignored, a phase with rho meson condensation ($\rho$) occurs for $\mu_I> m_\rho/2$, realized by a non-trivial profile for the spatial components of the non-abelian gauge fields. In the presence of pion condensation, the $\rho$ phase is metastable. However, rho and pion condensates can coexist. The resulting phase is denoted by $\pi\rho$ and it sets in at the medium-dependent rho meson mass. (In Fig.\ \ref{fig:phases3D}, the second-order onset of the $\pi\rho$ phase does not continue to nonzero temperatures due to the different versions of the model. This is explained in detail in Sec.\ \ref{sec: rho meson deconfined}.) 
        
    \item \textbf{Quark matter} (Q).
    When a black hole is present in the geometry, the possibility of strings sources stretched between the horizon and the flavor branes must be taken into account. At low temperatures, one obtains a configuration with a $\Ydown$-shaped embedding \cite{Bergman:2007wp}, as in the lower left panel of Fig.\ \ref{fig:embeddings}. This is the low-temperature quark phase (LTQ) of Ref.\ \cite{Kovensky:2019bih}, which is metastable unless baryons are artificially ignored, and thus does not occur in Fig.\ \ref{fig:phases3D}. As the temperature is increased, the strings pull the flavor branes towards the horizon, and one eventually reaches the $\sqcup$-shaped configuration of the high-temperature quark phase (HTQ in Ref.\ \cite{Kovensky:2019bih}), see lower middle panel of Fig.\ \ref{fig:embeddings}. In the chiral limit, the branes and anti-branes become straight and separated, signaling that chiral symmetry is restored. The HTQ phase -- simply denoted by Q in this review -- is the holographic representation of the quark-gluon plasma.  
    
    \item \textbf{Baryonic matter} (B). As mentioned above, a single baryon is described by a topologically non-trivial solution for the gauge fields on the flavor branes. Using multi-instanton solutions in the bulk to discuss many-baryon physics quickly becomes very  complicated \cite{Li:2015uea,BitaghsirFadafan:2018uzs,Preis:2016fsp,Bartolini:2026dbr}. Instead, we may treat the baryons as pointlike objects \cite{Bergman:2007wp} or consider a distribution which is homogeneous in the spatial directions, first discussed in Ref.\ \cite{Rozali:2007rx} and expected to work reasonably well at high densities. Importantly, as the density increases it becomes favorable for the baryons to move towards the ultraviolet (UV), creating a number of baryonic ``layers'' at different energy scales \cite{Kaplunovsky:2012gb,Kovensky:2020xif,Ecker:2025sjb}, see upper right panel of Fig.\ \ref{fig:embeddings}.

    \item \textbf{Pion condensate with quarks} ($\pi$Q). In this configuration a pion condensate as well as string sources, providing an isospin density from  quarks (and antiquarks), coexist, see lower left panel of Fig.\ \ref{fig:embeddings}. Geometrically, this configuration is continuously connected to the (chirally broken) pure pion-condensed phase and to the (chirally restored) pure quark phase. Hence, it is a candidate configuration for realizing a BEC-BCS crossover in holography. However, we find that this phase is never energetically favored in our model, and hence does not appear in Fig.\ \ref{fig:phases3D}. 
    
    \item \textbf{Meson condensates in baryonic matter} ($\pi$B, $\rho$B, $\pi\rho$B). We can also study the dynamics of a pion condensate in the presence of baryonic matter ($\pi$B). At zero temperature, this phase plays an important role in the region where $\mu_B$ and $\mu_I$ are both nonzero and not too large. At large values of $\mu_I$, rho meson condensation sets in and the coexistence phases $\rho$B and $\pi\rho$B form. Since the approximation used here for isospin-asymmetric baryons breaks rotational symmetry, this additional vector meson condensation is smoothly connected to the pure baryonic phase \cite{Kovensky:2023mye}, and thus we have not marked the $\rho$B and $\pi\rho$B phases explicitly in Fig.\ \ref{fig:phases3D}. 
    
    \item \textbf{Quarkyonic matter} (Qy). String sources can coexist with topologically nontrivial gauge field configurations. This is shown in the lower right panel of Fig.\ \ref{fig:embeddings} for the HTQ configuration, relevant for the phase diagram, but can also be constructed with the LTQ configuration. This is a holographic realization of quarkyonic matter \cite{McLerran:2007qj}, where quark and baryon degrees of freedom coexist. A layered structure in the energy scale given by the holographic direction emerges dynamically, similar to the layered Fermi spheres conjectured in  weak-coupling models of quarkyonic matter \cite{McLerran:2018hbz,Cao:2020byn,Margueron:2021dtx}. 
    
\end{itemize}

The phase diagram also shows the location of neutron stars. Neutron stars, unless they are in the very early stages of their lives or undergo a merger process, essentially have zero temperature on the scale of the QCD phase diagram. Using the isospin-asymmetric baryonic phase in the WSS model, adding leptons, and imposing the constraints of electroweak equilibrium and electric charge neutrality, realistic neutron stars can be constructed entirely from holography, including the crust \cite{Kovensky:2021kzl}. The blue curve in Fig.\ \ref{fig:phases3D} results from this calculation and has an endpoint that corresponds to the center of the most massive stable star. 

What can we actually learn about real-world QCD from the holographic phase diagram in Fig.~\ref{fig:phases3D} (and more generally from the predictions based on the WSS model)? The results reviewed here should not be taken as a precise, quantitative prediction for hot and dense QCD.  Nevertheless, the analytic tools that the WSS model provides for studying QCD-like matter at strong coupling, including several of the exotic candidate phases proposed in the non-holographic literature, create a theoretical framework complementary to the more traditional effective theories and phenomenological models. As we shall discuss in detail, the holographic toolkit can shed light on  general qualitative questions, such as  the role of the quarkyonic phase in the cold and dense regime, or  whether meson condensates and quark matter are connected continuously at large $\mu_I$, and also on more concrete ones, such as whether pions condense in neutron stars.

\subsection{Overview}

The remainder of this review is organized as follows. We start with a discussion of the setup of the model in Sec.\ \ref{sec: setup}. This discussion is adapted to the purpose of this review: it describes the general framework needed for all the phases of Fig.\ \ref{fig:phases3D}, but omits many details, for which we refer the reader to the original works of the WSS model \cite{Sakai:2004cn,Sakai:2005yt},  other reviews \cite{Peeters:2007ab,Gubser:2009md,Sugimoto:2012vpa,Kim:2012ey,Rebhan:2014rxa,Li:2023iuf}, and the specific works quoted in the text. We then address the three planes shown in Fig.\ \ref{fig:phases3D} (the three-dimensional phase structure in the space spanned by these planes has not been calculated yet). More explicitly, in Sec.\ \ref{sec: muB=0 plane}, we discuss the $\mu_B=0$ plane. This is, in large parts, a discussion of pion condensation and its variants and a comparison to lattice QCD, which is free of the sign problem in this regime. The $\mu_I=0$ plane is discussed in Sec.\ \ref{sec:baryons}, which centers around baryonic matter and its various approximations. Isospin and baryon number are combined in Sec.\ \ref{sec: baryons with isospin}, which addresses the $T=0$ plane and the application to neutron stars. 

Mostly we review previously existing results. We have, however, added various novel calculations: the $\pi\rho$ phase in the decompactified limit in Sec.\ \ref{sec: rho meson deconfined}, the construction of a phase with coexisting string sources and pion condensation in Sec.\ \ref{sec:BECBCS}, the generalization of the energetically preferred ``block-like'' baryonic configuration to the deconfined geometry in Sec.\ \ref{sec: muB-T baryonic phase}, and the generalization of quarkyonic matter from the pointlike approximation for the baryons to a smeared charge distribution in Sec.\ \ref{sec: muB-T quarkyonic phase}. 

There is a significant amount of literature using holography for matter in strong magnetic fields \cite{Gursoy:2021efc}, including many studies within the WSS model \cite{Bergman:2008sg,Bergman:2008qv,Thompson:2008qw,Rebhan:2008ur,Lifschytz:2009sz,Rebhan:2009vc,Preis:2010cq,Callebaut:2011ab,Preis:2011sp,Preis:2012fh,Callebaut:2013ria,Callebaut:2013wba,Li:2016gtz,Ballon-Bayona:2017dvv,Amano:2025iwi}. The magnetic field provides a natural extension into a fourth dimension of the phase diagram in Fig.\ \ref{fig:phases3D}. We have decided not to include these results here to keep the review as focused as possible. Also, many of the studies with a magnetic field can and should be improved in the future with the help of the progress presented here. Hence, we avoid, for instance, to discuss results of magnetized baryonic matter in an approximation that would be somewhat outdated in view of the phase diagram in Fig.\ \ref{fig:phases3D}. Another possible axis to be added in the phase diagram is given by an externally imposed rotation, which has been studied on the lattice \cite{Yamamoto:2013zwa,Braguta:2020biu} and within holography \cite{Golubtsova:2022ldm,Zhao:2022uxc,Chen:2024jet}. Including rotation in the WSS model has only begun very recently \cite{Amano:2026faw} and thus we leave this discussion to the future.

\section{Setup of the model}
\label{sec: setup}

\subsection{Geometries}

\begin{figure}
    \centering
    \includegraphics[width=\columnwidth]{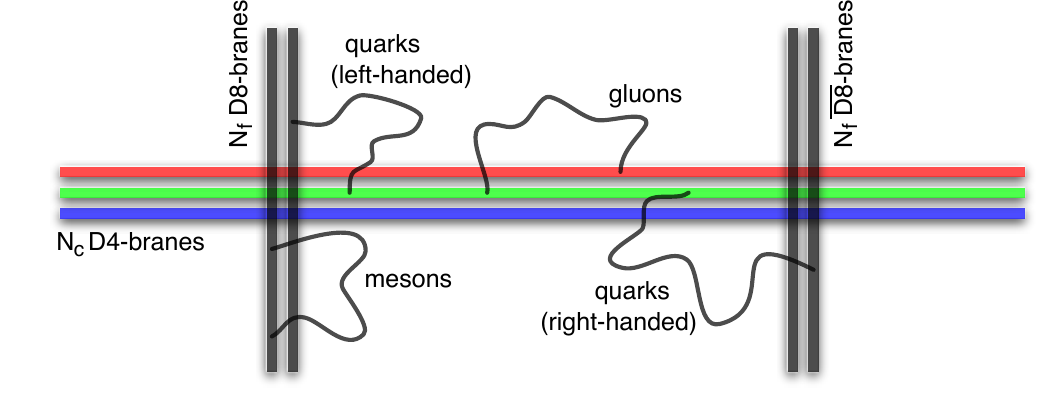}
    \caption{Schematic picture of the brane setup of the WSS model with $N_c$ D4-branes and pairs of $N_f$ D8-$\overline{\rm D8}$-branes separated along a compact $X_4$ direction with asymptotic distance $L$. Quarks, gluons, and mesons correspond to open strings with endpoints on the D-branes.  }
    \label{fig:setup}
\end{figure}

The setup of the WSS model constitutes a solution of type-IIA string theory sourced by $N_c$  D4-branes and $N_f$ pairs of D8-$\overline{\rm D8}$-branes, the latter being separated along the compact $X_4$ direction. We refer to them as color and flavor branes, respectively. The infrared (IR) excitations are described in terms of open strings, whose endpoints must lie on one of these sets of branes. D4-D4-strings transform in the adjoint representation of SU($N_c$) and correspond to gluon degrees of freedom. D4-D8 and D4-$\overline{\rm D8}$-strings transform in the fundamental representations of SU($N_c$) and of either U($N_f$)$_L$ or U($N_f$)$_R$, thus representing left- and right-handed quarks. Finally, the low-energy dynamics of D8-$\overline{\rm D8}$-strings capture the physics of mesons. This is illustrated in Fig.\ \ref{fig:setup}. The radius of the $X_4$ direction is $\MKK^{-1}$, as already shown in Fig.\ \ref{fig:embeddings}, and imposing the appropriate boundary conditions for the fermions breaks supersymmetry completely \cite{Witten:1998qj}. 

The topology of the preferred bulk configuration depends on the temperature by means of a Hawking-Page transition at $T_c = M_{\rm KK}/(2\pi)$ between a thermal background and a black hole (confined and deconfined geometries). Both geometries can be described by the following form of the metric,
\bea\label{ds2}
    ds^2 &=& \of{\frac{U}{R}}^{3/2} \off{f_T(U)dX_0^2 + d\vec{X}^2 + f(U)dX_4^2}\non[2ex]
    &&+\of{\frac{R}{U}}^{3/2} \off{\frac{dU^2}{f(U)} + U^2 d\Omega_4^2} \, , 
\eea
where $X_0$ is the (Euclidean) time direction, $\vec{X}$ are the flat directions of three-dimensional position space, $U$ is the holographic radial coordinate and $d\Omega_4$ describes an internal $S^4$. The curvature radius $R$ is defined in terms of the string length $\ell_s$ and the 't Hooft coupling $\lambda$,
\be
R^3 = \frac{\lambda \ell_s^2}{2 \MKK} \,, \qquad \lambda =2\pi g_s \ell_s N_c  \MKK \, , 
\ee 
where $g_s$ is the string coupling. The difference between the confined and deconfined geometries is captured by the functions $f_T(U)$ and $f(U)$,  
\begin{subequations}
\bea
    {\rm conf.:}&&   f_T(U)=1 \, , \quad  f(U) = 1 - \frac{U_{\rm KK}^3}{U^3} 
    \, , \\[2ex] 
    {\rm deconf.:}&& f_T(U)=1 - \frac{U_{T}^3}{U^3}  \, , \quad f(U) = 1  \, .
\eea
\end{subequations}
In the confined geometry,  $X_4$ caps off at $U_{\rm KK}$, while in the deconfined geometry $U_T$ is the location of the black hole horizon, with
\be
U_{\rm KK}=\frac{2 \lambda \MKK \ell_s^2}{9} \, , \qquad 
U_T = \left(\frac{4 \pi}{3}\right)^2 T^2R^3  \, .
\ee

\subsection{Action -- general form}

We work in the limit $N_c \gg N_f$, where the physics of the flavor branes and other sources can be studied as probes in the background geometries given by Eq.\ (\ref{ds2}). (For attempts going beyond that limit in the WSS model see Refs. \cite{Burrington:2007qd,Bigazzi:2014qsa,Li:2016gtz}.) Throughout this review, we will work with 2 flavors, $N_f=2$. We write the action of the flavor branes as 
\begin{equation}
\label{SHQCD}
    S = S_{\rm DBI} + S_{\rm CS} + S_q + S_{m_\pi}   \, .
\end{equation}
The first term is the Dirac-Born-Infeld (DBI) action
\begin{equation}
    S_{\mathrm{DBI}} = 2 T_8 V_4 \int d^4X \int_{U_{\rm c}}^{\infty}
    dU e^{-\phi}\, \STr\sqrt{\det(g + 2\pi\alpha' {\cal{F}})}\, ,
    \label{DBI00}
\end{equation}
where $T_8 = 1/\left[(2\pi)^8 \ell_s^9\right]$ is the D8-brane tension, $V_4=8\pi^2/3$ is the volume of the unit 4-sphere, and the dilaton is given by $e^{\phi} = g_s (U/R)^{3/4}$. We have denoted the endpoint of the branes by $U_c$. In the case of the confined geometry, where we always work in the antipodal scenario, this has to be replaced by $U_{\rm KK}$. In the deconfined geometry, it indicates the tip of the connected flavor branes, see Fig.\ \ref{fig:embeddings}, or, if the branes reach the horizon, has to be replaced by $U_T$. The induced metric is denoted by $g$ and contains the embedding function $X_4(U)$. Moreover, $\alpha'=\ell_s^2$, and ${\cal F}_{\mu\nu} = \partial_\mu {\cal A}_\nu - \partial_\nu{\cal A}_\mu+i[{\cal A}_\mu,{\cal A}_\nu]$ is the world-volume field strength tensor. The gauge symmetry is non-abelian and thus a prescription for evaluating the DBI action is required. We choose to work with the  symmetrized-trace prescription \cite{Tseytlin:1999dj}, denoted by STr. At lowest non-trivial order, this reduces to the YM action. 

We decompose the gauge field into its U(1) and SU(2) parts, \be \label{AhatA}
{\cal A}_\mu = \hat{A}_\mu+A_\mu^a\tau_a \, , 
\ee
where $\tau_a$ are the Pauli matrices and $a=1,2,3$. This decomposition is used to write the Chern-Simons (CS) contribution as 
\cite{Hata:2007mb,Rebhan:2008ur}
\bea \label{SCS}
&&S_{\rm CS}=-i \frac{N_c}{12\pi^2} \int d^4X\int_{U_c}^\infty dU \non[2ex]
&&\times \left\{\frac{3}{2}\hat{A}_\mu \left(F_{\nu\rho}^aF_{\sigma\lambda}^a+\frac{1}{3}
\hat{F}_{\nu\rho}\hat{F}_{\sigma\lambda}\right) \right.\non[2ex]
&&\left.+2\partial_\mu\left[\hat{A}_\nu\left(F_{\rho\sigma}^aA_\lambda^a+\frac{1}{4}\epsilon_{abc}A_\rho^aA_\sigma^bA_\lambda^c\right)\right]
\right\}\epsilon^{\mu\nu\rho\sigma\lambda}  \, . 
\eea
If the flavor branes are connected, it is often useful to work with the coordinate $Z$ defined by 
\be
U^3 = U_c^3 + U_c Z^2 \, ,
\ee
such that $Z$ is positive (negative) on the right (left) half of the connected branes, i.e., $Z\to\pm \infty$ corresponds to the D8 and $\overline{{\rm D8}}$ boundaries in the UV.  From the bulk perspective, we have a five-dimensional theory with a {\it local} U($N_f$) symmetry on the flavor branes. In order to describe mesons, we must restrict ourselves to configurations for which ${\cal F}_{\mu\nu}$ is trivial at $Z\to \pm \infty$. This can be achieved using a gauge for which ${\cal{A}}_\mu(X,Z\to \pm \infty) = 0$. There are, however, residual transformations generated by $g(X,Z) \in$ U($N_f$) for which $\partial_\mu g$ vanishes in the UV. The corresponding boundary values $g_{L/R}= g(X,Z\to \pm \infty)$ implement the \textit{global} U($N_f$)$_L \times $U($N_f$)$_R$ transformation in the boundary field theory, identified with chiral symmetry transformations of QCD. The corresponding chiral field, parameterizing the three pion degrees of freedom, is  
\begin{equation}
    \Sigma(X) = {\cal{P}} \exp \left[  i \int_{-\infty}^{\infty} dZ \, {\cal{A}}_Z(X,Z) \right] \, , 
    \label{Udef}
\end{equation}
which transforms as $\Sigma \to g_L \Sigma g_R^{-1}$. The effective four-dimensional action obtained from $S_{\rm DBI} $ then reproduces, for low energies, the Lagrangian of $\chi$PT -- the pion decay constant $f_\pi$ can thus be  written in terms of the parameters of the model; in the confined geometry \cite{Sakai:2004cn},
\be \label{fpiconf}
f_\pi^2 = \frac{2N_c M_{\rm KK}^2\lambda_0}{27\pi^3} \, ,
\ee
where we have introduced the convenient abbreviation
\be
\lambda_0 \equiv \frac{\lambda}{4\pi} \, .
\ee
The CS term is associated with the chiral anomaly and reduces to the Wess-Zumino-Witten term \cite{WessZumino:1971WZW,Witten:1983WZW}. In the  gauge where $\Aa_Z = 0$, the chiral field $\Sigma$ is trivial and the pion fields instead appear in the boundary conditions of the other gauge field components.

Baryons are constructed as topologically non-trivial configurations of the gauge field. The baryon number is obtained from the instanton number 
\be \label{NB1}
N_B = \frac{1}{8\pi^2} \int d^3X\, dZ \, \Tr {\cal{F}}^2 \, ,
\ee
which couples to the abelian potential through the first term in Eq.~\eqref{SCS}. 
At very large values of the 't Hooft coupling $\lambda$, one can think of a single instanton as a pointlike object
located at $Z=0$, and the baryon mass $m_B$ is approximated by that of a probe D4 brane wrapping the internal $S^4$, which in the confined geometry  yields
\be\label{mB}
m_B = \frac{\MKK \lambda N_c}{27 \pi} \, . 
\ee
Static properties of single baryons are studied by using the single-instanton Belavin-Polyakov-Schwarz-Tyupkin (BPST) solution \cite{1975PhLB...59...85B}, centered at $Z=0$, where the induced geometry is nearly flat. Assuming the single instanton also to be centered at $\vec{X}=0$, this solution
has non-abelian gauge field components 
\be
\label{Amu BPST}
    A_\mu (X,Z) = - i \phi \psi   \partial_\mu \psi^{-1}  \, , 
    \ee
where
\be
    \phi \equiv \frac{\xi^2}{\xi^2 + \rho^2} \, , \qquad
    \psi \equiv \frac{Z- i \vec{X}\cdot \vec{\tau}}{\xi^2}\,, 
\ee
with $\xi^2 = X^2 + \rho^2$. The resulting instanton width scales as  $\rho \sim 1/\sqrt{\lambda}$, while the abelian potential is 
\begin{equation}
    \hat{A}_0(X,Z) = \frac{24\pi}{\xi^2} \off{1-\frac{\rho^4}{(\rho^2+\xi^2)^2}} \, .
\end{equation}
The holographic baryon spectrum and the corresponding form factors \cite{Hashimoto:2008zw} can be derived by  quantizing a slowly moving instanton by means of the moduli space approximation \cite{Manton:1981mp,Hata:2007mb}. Many-baryon systems then correspond to multi-instanton solutions. They  can also be  obtained from different approximations. This will be discussed in detail in Sec.\ \ref{sec:baryons}.

\begin{table*}[t]
    \centering
    \begin{tabular}{|c|c|c|c|c|}
    \hline
    $\;\;x_i^{-1},x_4^{-1},t\;\;$ & $u,z$ & $x_0^{-1},\bar{m}_\pi,\bar{\mu}_I, N_c\bar{\mu}_B$  & $\bar{n}_I, \bar{n}_B$, $\bar{n}_q$ & $\bar{\Omega}$ \\[1ex]
    \hline 
    $\MKK$ & $\;\;\MKK^{2} R^{3}\;\;$ & $\;\;\lambda_0 \MKK\;\;$ & $\;\;N_f\lambda_0^{2} \MKK^{3}/(6\pi^2)\;\;$ & $\;\;N_f {\cal N} \;\;$\\[1ex]
    \hline 
    \end{tabular}
    \caption{Definitions of dimensionless quantities. In each case, multiplying the dimensionless quantity in the first row by the expression in the second row gives the dimensionful counterpart, e.g., $x_i=M_{\rm KK}X_i$, $T=M_{\rm KK} t$. In these units, $u_{\rm KK}=4/9$. Gauge fields scale like the inverse of the corresponding coordinate, e.g., $A_i^a = M_{\rm KK}a_i^a$, $a_z^a = M_{\rm KK}^2R^3 A_Z^a$. The factor $N_c$ in front of $\bar{\mu}_B$ ensures that $\mu_B$ is the baryon chemical potential, while the dimensionless version $\bar{\mu}_B$, somewhat confusingly but notationally convenient, should be thought of as a quark chemical potential. This factor does not appear in front of $\bar{\mu}_I$, except for Sec.\ \ref{sec:neutron}, see comments below Eq.\ (\ref{neutral}).}
    \label{table: dimensionless defs}
\end{table*}

Let us now get back to the action \eqref{SHQCD} and its remaining contributions. The term $S_q$ accounts for string sources, which we assume to be uniformly distributed along the flat directions. The strings are stretched vertically between the flavor branes and the horizon, hence $S_q$ can only be non-trivial in the presence of a black hole, see lower left panel of Fig.\ \ref{fig:embeddings}. The string sources represent quarks and thus can give a contribution to baryon and/or isospin number. The Nambu-Goto (NG) string action can be written as \cite{Bergman:2007wp} \begin{equation}
    S_{q} = N_cN_q \int dX_0\int_{U_c}^{\infty} dU \delta(U-U_q)
    \left(\frac{U - U_T}{2\pi\alpha'} - \hat{A}_0\right) \, ,
    \label{Sstring}
    \end{equation}
where $U_q$ is the point at which the strings attach to the flavor branes, and $N_cN_q$ is the total number of string sources. The factor $N_c$ is pulled out such that $N_q$ counts baryon number, which is convenient later. The subscript $q$ indicates that this baryon number is generated by quarks. In Eq.\ (\ref{Sstring}) we have assumed that the strings act as a pointlike source only for the abelian gauge field $\hat{A}_0$, assuming baryon, but no isospin, charge. We will adjust Eq.~\eqref{Sstring} for the case of nonzero isospin number in Sec.\ \ref{sec:BECBCS}.

Finally, the factor $S_{m_\pi}$ in Eq.\ \eqref{SHQCD} accounts for the pion mass. Including this effect is more difficult in the WSS model compared to other holographic approaches. A mass for fundamental quarks is usually included by the separation of the color branes from the flavor branes. However, in the D4-D8-$\overline{\rm D8}$ setup of the WSS model there are no directions left to do so. We will thus follow the effective approach of Refs.\ \cite{Kovensky:2019bih,Kovensky:2020xif,Kovensky:2023mye}, based on the original idea proposed and evaluated in the vacuum in Refs.\ \cite{Aharony:2008an,Hashimoto:2008sr,McNees:2008km,Argyres:2008sw,Hashimoto:2009hj,Seki:2012tt}. (A different approach based on a tachyonic scalar field was explored in  \cite{Bergman:2007pm,Dhar:2007bz,Dhar:2008um}.) To this end, we add to the action a bilinear quark term which breaks chiral symmetry explicitly; since left- and right-handed quarks are separated, this operator includes an open Wilson line along the $X_4$ direction. This generates a mass term for the chiral field \eqref{Udef}: At strong coupling, one can approximate the expectation value of this operator by  the exponential of the NG action for the string connecting both sets of branes, plus the boundary term that gives rise to the chiral field. The resulting contribution to the action is only determined up to a proportionality constant, which can be fixed from the Gell--Mann-Oakes-Renner relation \cite{Gell-Mann:1968hlm}. One obtains \cite{Kovensky:2019bih}
\begin{equation}
    S_{m_\pi} =  - \frac{m_\pi^2 f_\pi^2}{4}  e^{- (S_{\mathrm{NG}}-S_{\rm NG}^0)} \int d^4X \, \Tr [
    \Sigma + \Sigma^\dagger]  \label{Sm} \, , 
\end{equation}
 matching the mass term in lowest-order $\chi$PT. Here, the NG action is
\be
S_{\mathrm{NG}} = \frac{1}{2\pi \alpha'} \int_{U_c}^\infty dU \int_0^{X_4(U)} dX_4 \, \sqrt{g} \, . 
\label{SNG1}
\ee
It depends implicitly on the medium through the embedding function $X_4(U)$; the notation $S_{\rm NG}^0$ indicates that the NG action is evaluated with the vacuum embedding.

For notational convenience we introduce dimensionless quantities, which we will work with throughout the review. In most cases, they are denoted by lower case letters, for instance $x_4 = \MKK X_4$ or $\lambda_0 \MKK \hat{a}_0 = \hat{A}_0$.
In some cases, where this is inconvenient, dimensionless quantities are indicated by a bar, for instance $\mu_B = N_c \lambda_0 \MKK \bar{\mu}_B$. All relevant definitions are collected in Table \ref{table: dimensionless defs}. In addition, we rescale the gauge potentials $\hat{A}_0$, $A_0^a$ by a factor of $i$ because we work in Euclidean signature.

\subsection{Action -- simplified and specific}\label{sec:action2}

In all our main results we assume the dynamical fields, i.e., the gauge fields, the embedding function, and the pion condensate, to depend only on the holographic coordinate $u$ (and not on $x_i$). This is obviously a very convenient simplification because the equations of motion become ordinary differential equations. It is justified for spatially homogeneous systems, which means that our phase diagram in Fig.\ \ref{fig:phases3D} does not include the possibility of inhomogeneous phases. For baryonic matter, this means that the results in Fig.\ \ref{fig:phases3D} rely on a homogeneous approximation in which the single instantons  are assumed to strongly overlap. To explain approximations of baryonic matter based on multi-instanton systems, we will temporarily reinstate spatial dependence in Sec.\ \ref{sec:multi}.

Moreover, for all phases discussed here we will work in the gauge 
\be \label{az0}
\hat{a}_{u} = a_{u}^a  = 0 \, , 
\ee
set the off-diagonal components of the non-abelian gauge field to zero, and denote (no summation over $i$)
\begin{equation}
    a_i^{a}(u) = - \frac{\lambda_0}{2}h_i(u) \delta^a_i \, .
    \label{diagonal ansatz}
\end{equation}
 In the most complicated situation we consider -- baryonic matter in the presence of an isospin chemical potential and away from the chiral limit -- ignoring the off-diagonal components is, strictly speaking, not allowed. This can be seen from the equations of motion \cite{Kovensky:2023mye}, which we do not write down in full generality here. However, when any of these conditions is relaxed, that is, when at least one of $\mu_I$, $\mu_B$, $m_\pi$ vanish, the diagonal ansatz is consistent. This means that our treatment is accurate in the $\mu_B=0$ plane, and also for nonzero $\mu_B$, if $\mu_I$ is either small, where the isotropic limit $h_1=h_2=h_3$ is a good approximation, or large, where $m_\pi$ is negligible. In the regime of intermediate values of $\mu_I$, our ansatz is not accurate but, as argued in Ref.\ \cite{Kovensky:2023mye}, the effects from the off-diagonal components are not expected to change the phase diagram qualitatively.

With these simplifications we can write the CS term, which is independent of the background metric, as 
\bea
\label{SCS homogeneous}
    &&S_{\rm CS} = -{\cal N} N_f \frac{3V}{2T}\int_{u_c}^\infty du\, \Bigg[ \frac{\lambda_0}{2} \hat{a}_0 (h_1h_2h_3)' 
     \non[2ex]
    &&
   - \hat{a}_1(a_0^1 h_2 h_3)' - \hat{a}_2(a_0^2 h_1 h_3)'
    - \hat{a}_3(a_0^3 h_1 h_2)'\Bigg]  \, , 
    \eea
where $V$ is the three-volume, the prime denotes the derivative with respect to $u$, and 
    \bea   
    {\cal N}  \equiv  \frac{N_cM_{\rm KK}^4\lambda_0^3}{6\pi^2} \, . 
\eea
We have omitted the boundary term in Eq.\ (\ref{SCS}); for a discussion of this contribution see Ref.\ \cite{Bartolini:2023eam}.
The baryon number density is given by 
\begin{equation} \label{nBh123}
\bar{n}_B = -\frac{3\lambda_0}{8}
 \int_{-\infty}^\infty dz\,\partial_z(h_1h_2h_3)     \, .
\end{equation}
This shows one of the caveats of the homogeneous ansatz: Since $h_i(z=\pm \infty)=0$, the total derivative in Eq.\ \eqref{nBh123} can only give a non-vanishing integral if the functions $h_i(z)$ are allowed to be discontinuous in the bulk \cite{Rozali:2007rx}.

To write down the explicit forms of the other contributions to the action we need to distinguish the two versions of the model explained at the end of Sec.\ \ref{sec:QCDmatter}. 
In the original version, the embedding function is trivial and independent of the medium, $x_4'(u) = 0$.
 Moreover, the background geometry is temperature-independent from $T=0$ up to $T=T_c$, the regime where the confined geometry is energetically favored. In this version of the model we will employ the YM approximation for the DBI action. Then, in our dimensionless quantities and with the above ansatz, we obtain the one-dimensional effective DBI action\footnote{Here and in the following, the label ``conf'' stands for the confined geometry and implicitly implies antipodal separation of the branes, i.e., trivial embedding. Non-antipodal separation and thus non-trivial embedding in the confined geometry, although possible, is not considered in this review.} 
\bea
    \label{Lag DBI conf}
&&S_{\rm DBI}^{\rm conf} \simeq  S_{\rm YM}^{\rm conf} = {\cal N} N_f \frac{V}{T} \int_{u_{\rm KK}}^\infty du \, \frac{u^{5/2}}{2\sqrt{f}}\non[2ex]
&&\times \left(-f\hat{a}_0'^2-f {a_0^a}' {a_0^a}'+\frac{f\hat{a}_i'\hat{a}_i'}{\lambda_0^2}+g_1+g_2-g_3\right)  \, , \;\;
\eea
where we have abbreviated 
\begin{subequations} \label{g123}
\bea
g_1 &\equiv& \frac{f}{4}(h_1'^2+h_2'^2+h_3'^2) \, , \\[2ex]
g_2&\equiv&\frac{\lambda_0^2}{4u^3}(h_1^2h_2^2+h_1^2h_3^2+h_2^2h_3^2) \, , \\[2ex]
g_3&\equiv& \frac{\lambda_0^2}{u^3}\left[ (a_0^1)^2(h_2^2+h_3^2)+(a_0^2)^2(h_1^2+h_3^2)\right.\non[2ex]
&&\left.+(a_0^3)^2(h_1^2+h_2^2)\right] \, .
\eea
\end{subequations}
The mass term (\ref{Sm}) is made explicit by using the parameterization of the chiral field 
\begin{equation}
    \Sigma = \cos \theta \, \mathbb{1} + i \sin \theta ( \tau_1 \cos \varphi   + \tau_2 \sin \varphi ) \, .  
    \label{U0cond}
\end{equation}
Here we have set the contribution from the neutral pion condensate to zero, while $\sin\theta$ with $\theta\in [0,\pi]$ is the modulus of the charged pion condensate, and $\varphi\in[0,2\pi]$ corresponds to a residual U$(1)_{L+R}$ symmetry  that is intact even in the presence of a pion condensate. Therefore, our results must be independent of $\varphi$. Next, we notice that due to the trivial embedding the NG action in Eq.\ (\ref{Sm}) does not give any nontrivial factor, and we obtain 
\begin{equation}
\label{S_mpi conf}
    S_{m_\pi}^{\rm conf} = - {\cal N} N_f \frac{V}{T} \frac{2\bar{m}_\pi^2}{9\pi} \cos\theta \, , 
\end{equation}
where we have used the pion decay constant from Eq.\ (\ref{fpiconf}). 
Finally, in this version of the model the quark contribution  is zero, $S_q^{\rm conf}=0$, because there is no black hole in the background, hence the string sources have nowhere to end. 

For the second version of the model, the decompactified limit, we use the prescription for the DBI action motivated in Ref.\ \cite{Kovensky:2021ddl} [generalized to the case where the functions $h_1(u)$, $h_2(u)$ and $h_3(u)$ are not necessarily the same]\footnote{The label ``deconf'' implicitly implies the decompactified limit $LM_{\rm KK} \ll \pi$, the only context in which we employ the deconfined geometry. Smallness of $LM_{\rm KK}$ is, however, never explicitly used in our derivations.}, 
\bea
    \label{Lag DBI deconf}
&&\hspace{-0.5cm}S_{\rm DBI}^{\rm deconf} = {\cal N} N_f \frac{V}{T} \int_{u_{c}}^\infty du \, u^{5/2} \non[2ex]
&&\hspace{-0.5cm}\times \sqrt{(1+u^3 f_T x_4'^2 + g_1 -\hat{a}_0'^2- {a_0^a}' {a_0^a}')(1+g_2-g_3)} \, , \qquad
\eea 
where the definitions for $g_1$, $g_2$ and $g_3$ are analogous to those of Eq.~\eqref{g123}, up to small differences due to the blackening factor of the background metric, 
\begin{subequations} \label{g123deconf}
\bea
g_1 &\equiv& \frac{f_T}{4}(h_1'^2+h_2'^2+h_3'^2) \, , \\[2ex]
g_2&\equiv&\frac{\lambda_0^2}{4u^3}(h_1^2h_2^2+h_1^2h_3^2+h_2^2h_3^2) \, , \\[2ex]
g_3&\equiv& \frac{\lambda_0^2}{u^3f_T}\left[ (a_0^1)^2(h_2^2+h_3^2)+(a_0^2)^2(h_1^2+h_3^2)\right.\non[2ex]
&&\left.+(a_0^3)^2(h_1^2+h_2^2)\right] \, .
\eea
\end{subequations}
The DBI action (\ref{Lag DBI deconf}) agrees with the YM result and the symmetrized trace prescription at order ${\cal{F}}^2$. (See appendix B of Ref.\ \cite{Kovensky:2021ddl} for the result of the symmetrized trace prescription to all orders, which is much too complicated to be applied for our purposes.) Moreover, it arranges the higher order terms in a way that reproduces the isospin symmetric $N_f=1$ limit of Ref.\ \cite{Li:2015uea}, while keeping the argument of the square root factorized. This allows for an immediate first integration of the equations of motion, which is extremely useful for the subsequent algebraic manipulations. 

The mass term is more complicated compared to Eq.~\eqref{S_mpi conf} because now the embedding function $x_4(u)$ is nontrivial. We obtain the NG action 
\begin{equation}
    \label{SNG}
S_{\rm NG} = -2\lambda_0\left[\phi_T(u_c)x_4(u_c)+\int_{u_c}^\infty du\, \phi_T(u) x_4'(u)\right]\, , 
\end{equation}
where
\bea\label{phiT}
    &&\phi_T(u) \equiv \int \frac{du}{\sqrt{f_T(u)}} = \frac{u}{\sqrt{f_T(u)}} \non[2ex]
    &&\times \left\{
    1-\frac{3u_T^3}{4 u^3 f_T^{1/6}(u)} {}_2F_1 \left[
    \frac{1}{6},\frac{2}{3},\frac{5}{3}, -\frac{u_T^3}{u^3 f_T(u)}
    \right]\right\} \, , \;\;
\eea
with the hypergeometric function ${}_2F_1$.
We then abbreviate 
\be\label{Aalpha}
A \equiv \frac{2 \alpha}{\lambda_0^2} 
    e^{-S_{\rm NG}} \, \, , \qquad \alpha \equiv \frac{3\pi^2f_\pi^2m_\pi^2}{N_cM_{\rm KK}^4e^{-S_{\rm NG}^{0}}} \, ,
\ee
where $A$ depends on the medium and needs to be determined self-consistently later, while $\alpha$ serves as a convenient dimensionless ``mass parameter''.  The mass term (\ref{Sm}) can now be written compactly as  
\begin{equation}
    \label{S_mpi deconf}
    S_{m_\pi}^{\rm deconf} = - {\cal{N}} N_f \frac{V}{T} \frac{A}{2\lambda_0} \cos \theta \, ,
\end{equation}
where, again, we have set the neutral pion condensate to zero. 

Finally, the deconfined geometry allows for 
string sources. In our dimensionless quantities, the string contribution to the action (\ref{Sstring}) reads   
  \begin{equation}
    S_{q}^{\rm deconf} = {\cal{N}}N_f \frac{V}{T} \int_{u_c}^{\infty} du\, \bar{n}_q \left[
    u - u_T - \hat{a}_0(u)\right]\delta(u-u_q) \, ,
    \label{Sstringdeconf}
    \end{equation}
where the dimensionless baryon number from quarks $\bar{n}_q$ is related to the dimensionful $N_q/V$ by the factor given in Table \ref{table: dimensionless defs}.  

In both versions of the model we define the dimensionless free energy density $\bar{\Omega}$ as 
\be\label{Sonshell}
\frac{T}{V} S|_{\mathrm{on-shell}} = \Omega = N_f{\cal N}\bar{\Omega} \, .
\ee
This (infinite) free energy density needs to be renormalized by subtracting the free energy density of the vacuum. 

We also collect the UV boundary conditions that are universal for our purposes. The boundary value of the embedding function encodes the asymptotic separation of the branes, 
\begin{equation}
\label{x4 bc}
x_4(u\to \infty) = \frac{\ell}{2} \, ,
\end{equation} 
where
\be
\ell \equiv LM_{\rm KK} \, .
\ee
In practice, this condition is conveniently implemented by imposing
\be \label{ell2}
\int_0^\infty du\, x_4' = \frac{\ell}{2}\, ,
\ee
because our formalism always allows us to find a semi-analytic form of $x_4'$. 
In fact, we can always rescale all quantities by suitable powers of $\ell$ such that $\ell$ is eliminated from this (and all other) equations, and is only reinstated later to translate our results into physical quantities. 

The temporal components of the gauge fields contain the information of the field-theoretic chemical potentials. The matrix for the (dimensionless) chemical potentials of the up and down quarks can be written as ${\rm diag}(\bar{\mu}_u,\bar{\mu}_d)=\bar{\mu}_B\mathbb{1} +\bar{\mu}_I \tau_3$, hence we identify\footnote{In our convention for the isospin chemical potential, the onset of pion condensation is at $\mu_I = m_\pi/2$. In this review we use this convention throughout, thereby adjusting the convention of Ref.\ \cite{Kovensky:2023mye}.} 
\begin{equation}
\label{bc potentials UV without pion}
    \hat{a}_0(z\to \pm \infty) = \bar{\mu}_B \, , \qquad a_0^3(z\to \pm \infty) = \bar{\mu}_{I} \, . 
\end{equation}
If pion condensation is taken into account, the boundary value of $a_0^3$ is changed and nonzero boundary values for $a_0^1$ and $a_0^2$ are induced, as we shall explain in the next section. All other gauge field components vanish at infinity, except for $\hat{a}_3$, which is nonzero in the isospin-asymmetric baryonic phase and whose boundary value is dynamically adjusted to make the net baryon current vanish.


\section{Phases without baryons}
\label{sec: muB=0 plane}

In this section we focus on the $\mu_B=0$ plane of the phase diagram in Fig.\ \ref{fig:phases3D}, where  baryons can be ignored. Isospin and temperature effects are described by the decompactified version of the WSS model, as in an early version of the $T$-$\mu_I$ plane without quark mass effects  \cite{Parnachev:2007bc}. We first review the results of Ref.\ \cite{Kovensky:2024oqx}, computing properties and comparing free energies of vacuum, pion-condensed and deconfined phases. Since the pion-condensed phase is also relevant in the confined geometry for the $T=0$ plane, we shall also quote the relevant -- much simpler -- equations for this case. We then extend the analysis to include rho meson condensation, and present a holographic construction that potentially realizes a BEC-BCS crossover, as conjectured for QCD at nonzero $\mu_I$. 


\subsection{Pion condensation}
\label{sec: pion condensation}

The chiral field $\Sigma$ is given in terms of the gauge field ${\cal A}_Z$ (\ref{Udef}) and, at the same time, has the form (\ref{U0cond}) with a nonzero $\theta$ (and an arbitrary $\varphi$). In view of our gauge choice (\ref{az0}) this seems contradictory. The solution is to apply a global chiral rotation such that the rotated chiral field is trivial, $\Sigma' = g_L \Sigma g_R^\dagger = \mathbb{1}$ \cite{Aharony:2007uu,Rebhan:2008ur}. The choice of $g_L$ and $g_R$ is not unique. In the presence of a pion mass,  it is useful to set \cite{Kovensky:2024oqx}
\begin{equation}
   g\equiv g_R = g_L^\dagger = \cos\frac{\theta}{2} \, \mathbb{1} +i\sin\frac{\theta}{2}(\tau_1\cos\varphi+\tau_2\sin\varphi) \, ,
\label{gLgR2}
\end{equation}
such that $g^2 = \Sigma$. The physics is unchanged in the rotated frame, but the rotation modifies the boundary values of the non-abelian gauge potentials. Instead of Eq.\ (\ref{bc potentials UV without pion}) we have 
\begin{subequations} \label{aaaboundary}
\bea
    a_0^1 (z\to \pm \infty) &=& \mp \bar{\mu}_I \sin \varphi \sin \theta \, , \\[2ex]
    a_0^2 (z\to \pm \infty) &=& \pm \bar{\mu}_I \cos \varphi \sin \theta \, , \\[2ex]
    a_0^3 (z\to \pm \infty) &=& \bar{\mu}_I \cos \theta \, .
\eea
\end{subequations}
In the confined version of the model, the only nontrivial equations of motion for the pion-condensed phase are  
\begin{equation}
    \partial_u(u^{5/2}\sqrt{f} {a_0^a} ') = 0 \, , \qquad a = 1,2,3 \, , 
\end{equation}
with analytic solutions
\begin{subequations}\label{piconfined}
\bea
    a_0^1 (z) &=& - \frac{2\bar{\mu}_I}{\pi} \sin \varphi \sin \theta \arctan \frac{z}{\uKK} \, , \\[2ex] 
    a_0^2 (z) &=& \frac{2\bar{\mu}_I}{\pi} \cos \varphi \sin \theta \arctan \frac{z}{\uKK} \, , \\[2ex]
    a_0^3 (z) &=& \bar{\mu}_I \cos \theta  \, .
\eea
\end{subequations}
Minimizing the free energy density with respect to the condensate gives $\cos\theta=\bar{m}_\pi^2/(4\bar{\mu}_I^2)$, and evaluating free energy  and isospin densities at the minimum yields
\bea \label{thOmnI}
  \bar{\Omega}_\pi &=& - \frac{4\bar{\mu}_I^2}{9\pi}\left(1-\frac{\bar{m}_\pi^2}{4\bar{\mu}_I^2}\right)^2 \,, \non[2ex]
   \bar{n}_I &=& -\frac{\der \bar{\Omega}_{\pi} }{\der \bar{\mu}_I} = \frac{8\bar{\mu}_I}{9\pi}\left(1-\frac{\bar{m}_\pi^4}{16\bar{\mu}_I^4}\right) \, , 
\eea
exactly reproducing $\chi$PT \cite{Son:2000xc}. 

The decompactified version of the model goes beyond $\chi$PT as the non-trivial embedding function $x_4(u)$ couples to the gauge potentials, hence capturing the (flavor brane) backreaction due to the presence of the pion condensate. The integrated equations of motion are
\begin{equation}
    {a_0^a}' = \frac{\kappa_a}{u^{5/2}}\zeta \,, \qquad  
x_4' = \frac{k+A\phi_T\cos\theta}{u^{11/2}f_T}\zeta \, ,
\end{equation}
where $k$ and $\kappa_a$ are integration constants, and 
\be
\zeta = \left[1-\frac{(k+A\phi_T\cos\theta)^2}{u^8f_T}+\frac{\kappa_a \kappa_a}{u^5}\right]^{-1/2} 
 \, .
\ee
By determining the stationary points of the free energy one finds that $\kappa_3=0$, and thus $a_0^3(u) = \bar{\mu}_I\cos\theta$ is constant. 
Moreover, due to the invariance with respect to rotations by the angle $\varphi$, only $\kappa^2=\kappa_1^2+\kappa_2^2$, not $\kappa_1$ and $\kappa_2$ separately, is relevant. The dimensionless free energy can then be written as 
\bea \label{Omega0}
\bar{\Omega}_\pi &=& 
\int_{u_c}^\infty du\left(\frac{u^{5/2}}{\zeta}+A\phi_Tx_4'\cos\theta\right)-\frac{A\cos\theta}{2\lambda_0} \non[2ex]
&& +k \, \frac{\ell}{2} -\bar{\mu}_I\bar{n}_I \, , 
\eea
with the isospin density
\be \label{nIpi}
\bar{n}_I= \left(\int_{u_c}^\infty du\, \frac{\zeta}{u^{5/2}}\right)^{-1}\bar{\mu}_I \sin^2\theta \, . 
\ee
By comparing this to chiral perturbation theory we can identify the inverse integral as the  medium-dependent pion decay constant squared (up to a constant factor). If $\zeta$ is evaluated in the vacuum and at zero pion mass [note the difference to the confined-antipodal result (\ref{fpiconf})],
\be \label{fpideconf}
f_\pi^2(m_\pi=0) = \frac{128 N_c \tilde{\lambda}_0 }{3\pi L^2} \left(\frac{\Gamma[9/16]}{\Gamma[1/16]}\right)^3\frac{\Gamma[11/16]}{\Gamma[3/16]} \, , 
\ee
 where we have abbreviated
\be
\tilde{\lambda}_0 \equiv \frac{\lambda_0}{\ell} = \frac{\lambda_5}{8\pi^2 L} \, ,
\ee
with the 5-dimensional 't Hooft coupling $\lambda_5 = 2\pi \lambda/M_{\rm KK}$.

It remains to compute the free parameters $k$, $u_c$, $\kappa$, $\theta$, $A$ from the following coupled equations:  minimization of the free energy with respect to $\theta$, the condition that the embedding function $x_4(u)$ be consistent with the definition of $A$ according to Eqs.~\eqref{SNG}-\eqref{Aalpha}, the boundary condition for $x_4(u)$ in Eq.~\eqref{x4 bc}, and the smoothness of the embedding at $u=u_c$ (which is equivalent to stationarity of the free energy with respect to $u_c$). This is obviously much more complicated than in the confined case but can be done numerically. More technical details  can be found in Ref.\ \cite{Kovensky:2024oqx}.

\subsection{Vacuum and quark matter} 

The isospin vacuum, see upper left panel of Fig.\ \ref{fig:embeddings}, can be obtained from the above expressions by setting $\theta=\kappa_a=0$. In this case all  gauge potentials are constant and thus the physics is independent of $\bar{\mu}_I$. In particular, the isospin density is zero for all temperatures in this configuration. This is different from QCD and explains why there cannot be a smooth crossover from the vacuum to quark matter in our holographic setup, not even in the decompactified limit and for nonzero $m_\pi$. In the vacuum configuration, stationarity with respect to $u_c$ yields $k = u_c^4\sqrt{f_T(u_c)}-A\phi_T(u_c)$. The rest of the calculation needs to be done numerically (only the zero-temperature, $m_\pi=0$ limit allows for an analytic result, see for instance appendix B of Ref.\ \cite{Li:2015uea}): $A$ and $u_c$ need to be determined from the self-consistency condition for $A$ and the boundary condition for $x_4(u)$. 

In the quark matter phase, see the lower middle panel of Fig.\ \ref{fig:embeddings}, the flavor branes become $\sqcup$-shaped and reach the black hole horizon, $u_c = u_T$. The nonzero isospin density comes from string sources which are, so to speak, hidden behind this horizon. (The second version of the quark phase, where isospin density is generated by strings stretching from the horizon to the connected flavor branes, does not play a role in the phase diagram \cite{Kovensky:2024oqx}.) By definition, the pion condensate vanishes in this phase, $\theta=0$, such that the only non-trivial function besides the embedding $x_4(u)$ is $a_0^3(u)$. In the massless limit, the embedding is straight, $x_4'(u)=0$, and $x_4(u_T)=\ell/2$. With nonzero pion mass, however, $x_4(u_T)$ must be determined dynamically (and numerically).  In this phase we find $k = - A \phi_T(u_T)$ with 
\be
\phi_T(u_T)=\frac{3\sqrt{\pi}\,\Gamma[5/3]}{2\Gamma[1/6]}\, u_T \, ,
\ee
and the free energy can be written as 
\bea
\bar{\Omega}_{\rm Q} &=& \int_{u_T}^\infty du\left(\frac{u^{5/2}}{\zeta}+A\phi_Tx_4'\right)-\frac{A}{2\lambda_0}-\bar{\mu}_I\bar{n}_I\non[2ex]
&&+k\left[\frac{\ell}{2}-x_4(u_T)\right] \, .
\label{OM HTQ}
\eea

\subsection{Model parameters and phase structure} 

The three relevant parameters for the decompactified limit are $L$, $\tilde{\lambda}_0= \lambda_0/\ell$, and $\tilde{\alpha}=\ell^4\alpha$, where $\alpha$ is the mass parameter introduced in Eq.~\eqref{S_mpi deconf}. As usual in holographic applications, we set $N_c=3$, extrapolating down from large $N_c$ where our approximation is valid, just like choosing a finite value of $\lambda$ is an extrapolation from the controlled large-coupling limit. We fix the parameters to reproduce the vacuum mass of the pion $m_\pi=140\, {\rm MeV}$ and the pion decay constant in the vacuum $f_\pi = 92\, {\rm MeV}$ (at physical pion mass, which requires a -- numerical -- generalization of the massless limit (\ref{fpideconf}), as explained in Ref.\ \cite{Kovensky:2024oqx}). This ensures that the zero-temperature pion onset is as in QCD. As a third physical quantity we choose the critical temperature $T_c$ for the chiral phase transition at $\mu_I=0$. In QCD this is a crossover, while our model predicts a first-order phase transition. We thus require our critical temperature to (roughly) reproduce the pseudocritical temperature of QCD, $T_c=160\, {\rm MeV}$. This  yields  the model parameters 
\be
\label{Lfit}
\tilde{\lambda}_0 \simeq 0.6856 \, , \quad \tilde{\alpha} \simeq 1.005\times 10^{-3} \, , \quad 
L \simeq  
0.1950\, {\rm fm} \, .
\ee
In the following subsection, we shall also discuss a different fit, where the vacuum mass of the rho meson is used instead of $T_c$.

\begin{figure} [t]
\begin{center}
\includegraphics[width=\columnwidth]{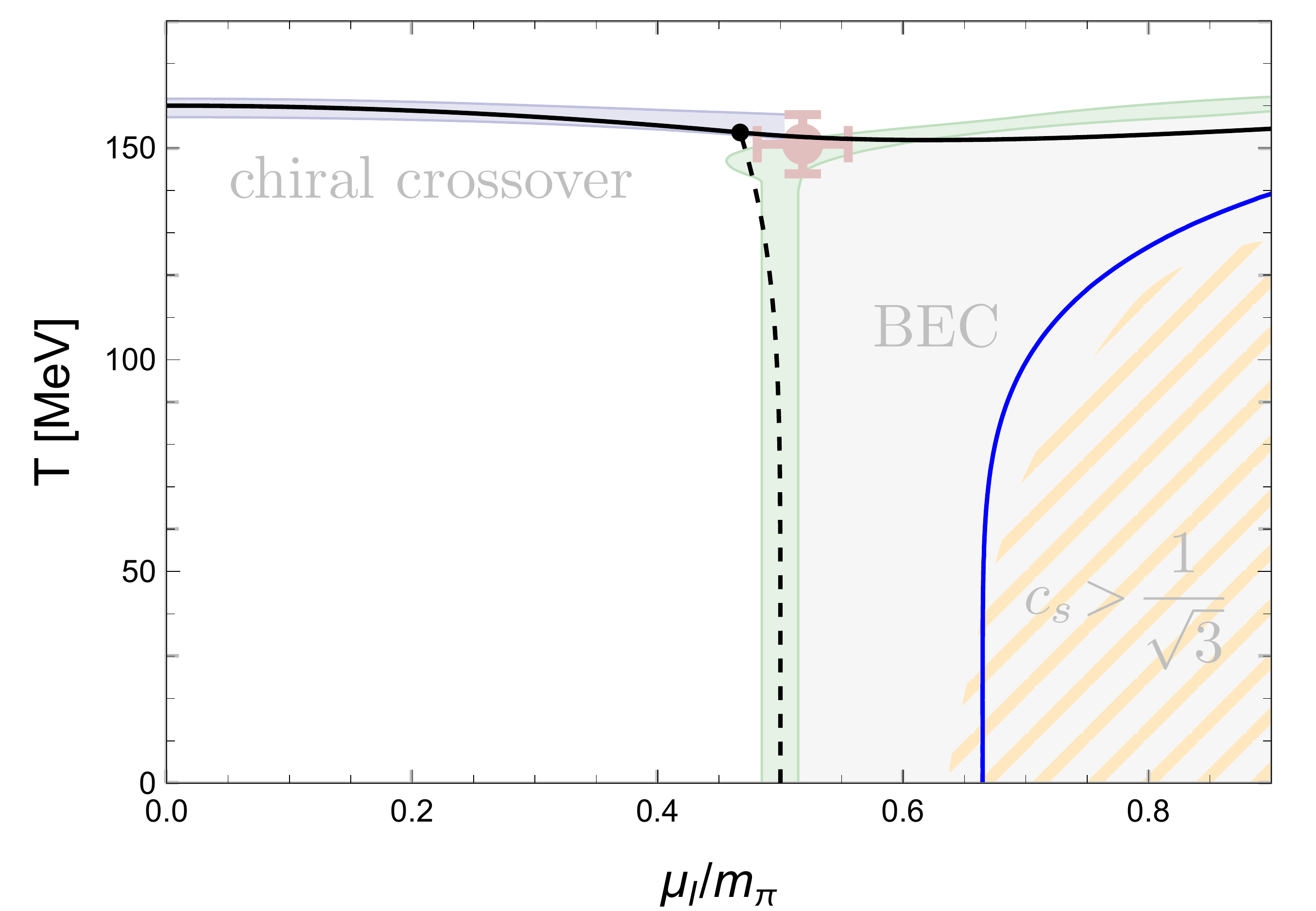}
\caption{
Phase diagram for $\mu_B=0$ in the decompactified limit of the WSS model. Black solid (dashed) lines are first-(second-)order phase transitions and are compared to the lattice QCD results from Ref.\ \cite{Brandt:2022hwy} (background). The blue line is defined by $c_s^2=1/3$. Figure taken from Ref.\ \cite{Kovensky:2024oqx}. 
}
\label{fig:PionT}
\end{center}
\end{figure}

\begin{figure} [t]
\begin{center}
\includegraphics[width=\columnwidth]{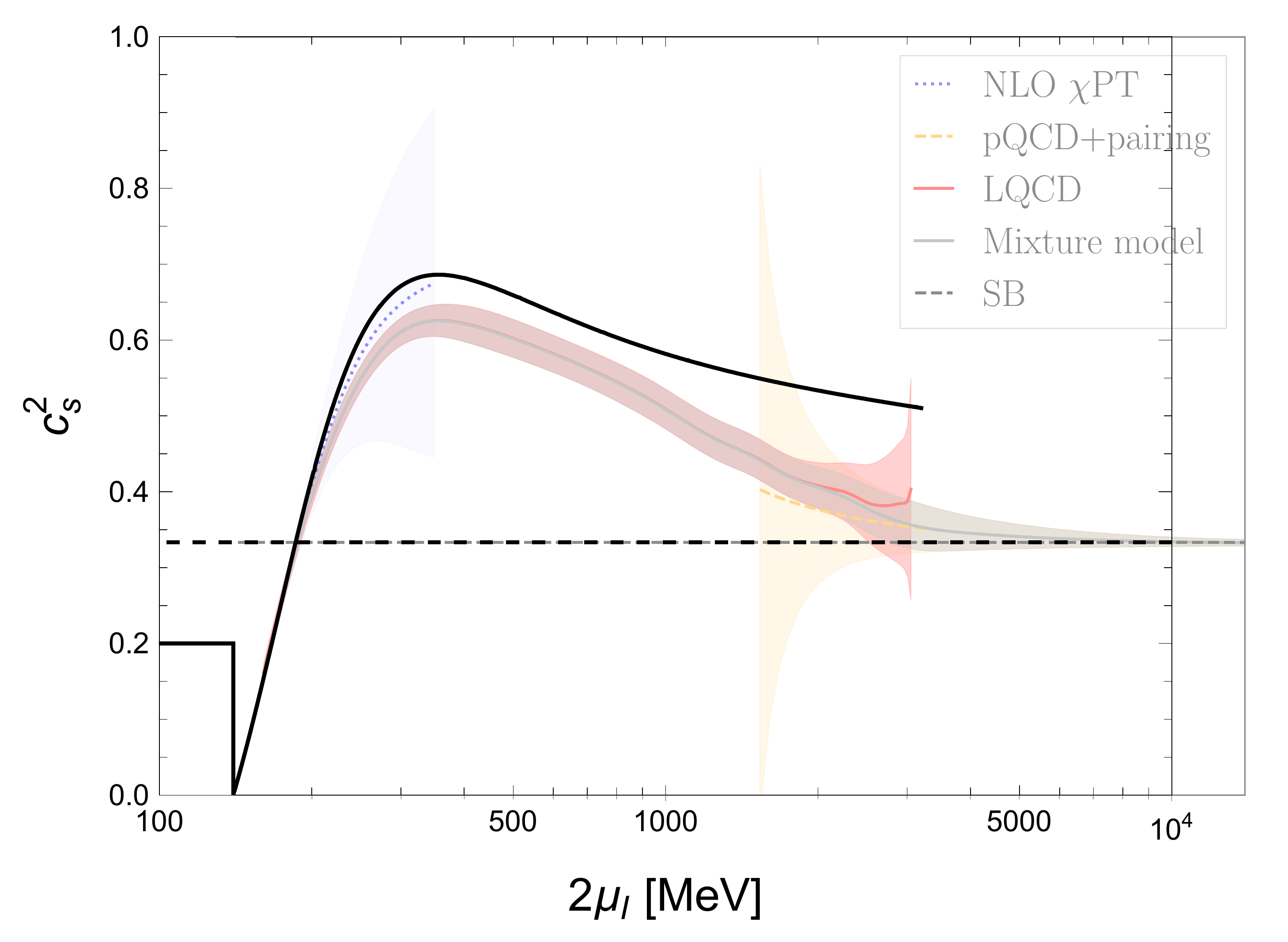}
\caption{
 Speed of sound for $T=0$ compared to the lattice QCD results (red band) from Ref.\ \cite{Abbott:2024vhj} (to match the  different convention in that paper, the horizontal axis is scaled by a factor 2). Figure taken from Ref.\ \cite{Kovensky:2024oqx}. 
}
\label{fig:PionTcs}
\end{center}
\end{figure}

Detailed results for various thermodynamic properties of the pion-condensed phase can be found in Ref.\ \cite{Kovensky:2024oqx}. Here we focus on the main points which are shown in Figs.\ \ref{fig:PionT} and \ref{fig:PionTcs}. The phase diagram in Fig.\ \ref{fig:PionT} is used for the $\mu_I$-$T$ plane of Fig.\ \ref{fig:phases3D}. The most important observations are, firstly, the (unphysical) first-order transition between the vacuum V and the chirally restored phase Q, which nevertheless follows the QCD crossover curve remarkably well. Secondly, the second-order pion onset curve, which is almost vertical, again reproducing the lattice QCD results to very good accuracy. This behavior cannot be obtained from $\chi$PT, even beyond lowest order \cite{Adhikari:2020kdn}, but has also been seen in a quark-meson model with Polyakov loop \cite{Adhikari:2018cea,Chiba:2024cny}. Thirdly, within the pion-condensed phase $\pi$ the speed of sound exceeds the conformal value in a region comparable to the one found on the lattice. This is shown in more detail in Fig.\ \ref{fig:PionTcs}: the zero-temperature speed of sound follows the lattice QCD results (and next-to-leading-order $\chi$PT) for small $\mu_I$, exceeds the conformal value and exhibits a maximum. Since our holographic model does not show asymptotic freedom, our results deviate significantly from the lattice at very large $\mu_I$, in particular in the region where perturbative QCD starts to take over. In fact, our holographic speed of sound has an asymptotic value $c_s^2 = 2/5$, in contrast to the asymptotic QCD result $c_s^2 = 1/3$.

The fact that these results match with the phase structure from lattice QCD extremely well, is reassuring and gives us some confidence that the predictions of the model can be taken seriously also in the regimes where lattice calculations are not available, i.e., away from the $\mu_B=0$ plane, see Secs.\ \ref{sec:baryons} and \ref{sec: baryons with isospin}.

\subsection{Rho meson condensation}
\label{sec: rho meson deconfined}

As the isospin chemical potential is increased beyond the scale shown in Fig.\ \ref{fig:PionT} we expect rho meson condensation to play a role (in the larger scale of Fig.\ \ref{fig:PionTcs}, this effect was ignored in the holographic calculation). As a first rough estimate based on the vacuum mass of the rho meson $m_\rho \simeq 776\, {\rm MeV}$, we might expect condensation at $\mu_I\sim m_\rho/2 \simeq 2.75 \, m_\pi$. However, the relevant quantity is the in-medium rho meson mass, i.e., one must take into account the effect of the pion condensate. This was discussed in the confined version of the model for the first time in Ref.\ \cite{Aharony:2007uu} and later generalized in Ref.\ \cite{Kovensky:2023mye}. This calculation will be reviewed in Sec.\ \ref{sec: baryons with isospin}. It results in the $T=0$ second-order phase transition line between $\pi$ and $\pi\rho$ phases in Fig.\ \ref{fig:phases3D}. Here we generalize this calculation to nonzero temperatures, including the backreaction on the embedding of the flavor branes, and will explain why the transition line in Fig.\ \ref{fig:phases3D} does not have a counterpart at nonzero temperature  in the depicted range. 

The bulk mode associated with the rho meson of the boundary theory resides in the spatial components of the non-abelian  gauge fields. Due to the particular rotation (\ref{gLgR2}) we have performed to describe pion condensation, the rho meson sits in the third isospin component. Hence, in order to construct the configuration where pion and rho meson condensates coexist we need to repeat the above analysis with a nontrivial function $h_3(u)$, introduced in Eq.~(\ref{diagonal ansatz}). The boundary conditions for the gauge potentials are unchanged, we require $h_3(\infty)=0$, and we use the residual ${\rm U}(1)_{L+R}$ symmetry to set $\varphi=0$ from the beginning. The equation of motion for $a_0^3(u)$ is the same as before since it does not couple to $h_3(u)$, see Eq.~\eqref{g123}, hence the solution is still a constant, namely $a_0^3(u) = \bar{\mu}_I \cos \theta$. It remains to solve the equations of motion for $x_4$, $a_0^2$ and $h_3$. The corresponding system can be expressed as  
\bea \label{EOMpirho}
&&x_4' = \frac{k+A\phi_T\cos\theta}{u^{11/2}f_T}\zeta \, , \qquad
{a_0^2}' = \frac{\kappa+\xi(u)}{u^{5/2}}\zeta \, ,  \non[2ex]
&& \left(\frac{u^{5/2}f_Th_3'}{\zeta}\right)' = -\frac{4\lambda_0^2({a_0^2})^2  h_3\zeta }{u^{1/2}f_T}\, ,
\eea
with an integration constant $\kappa$, with 
\begin{equation}
\zeta = \frac{\sqrt{1+g_1+u^3f_Tx_4'^2-{a_0^a}'{a_0^a}'}}{\sqrt{1-g_3}}   
\end{equation}
[$g_1$ and $g_3$ from Eq.\ (\ref{g123deconf})], and with  the auxiliary function 
\be
\xi(u)\equiv \lambda_0^2\int_{u_c}^u du\frac{{a_0^2} h_3^2\zeta}{u^{1/2} f_T} \, .
\ee
After some algebra we can eliminate $x_4'$ and ${a_0^2}'$ from $\zeta$, which can be rewritten as 
\be \label{zetapirho}
\zeta = \frac{\sqrt{1+g_1}}{\sqrt{1-g_3-\frac{(k+A\phi_T\cos\theta)^2}{u^8f_T}+\frac{(\kappa+\xi)^2}{u^5}}} \, .
\ee
To compute the isospin density and to set up the  stationarity conditions for the free energy we need the derivatives of $\bar{\Omega}$ with respect to $x=\bar{\mu}_I,h_{3c},\theta$. Here we have denoted the IR boundary value $h_{3c}\equiv h_3(u_c)$. With the help of the equations of motion we find 
\be\label{dOmdx}
\frac{\partial \bar{\Omega}}{\partial x} = \pi_{{a_0^2}}\left.\frac{\partial a_0^2}{\partial x}\right|_{u_c}^\infty +  \pi_{h_3}\left.\frac{\partial h_3}{\partial x}\right|_{u_c}^\infty+k\left.\frac{\partial x_4}{\partial x}\right|_{u_c}^\infty - \frac{A}{2\lambda_0}\frac{\partial \cos\theta}{\partial x} \, ,
\ee
with the conjugate momenta 
\be
\pi_{a_0^2} = -\frac{u^{5/2}{a_0^2}'}{\zeta} \, , \qquad \pi_{h_3} = \frac{u^{5/2}f_Th_3'}{4\zeta} \, .
\ee
For $x=\bar{\mu}_I$ only the first term in Eq.\ (\ref{dOmdx}) contributes, due to $a_0^2(\infty)= \bar{\mu}_I\sin\theta$, and we obtain the isospin density 
\be \label{nIrhopi}
\bar{n}_I = -\frac{\partial \bar{\Omega}}{\partial \bar{\mu}_I} =  \sin\theta \left[\kappa +\xi(\infty)\right] \, .
\ee
Stationarity with respect to the pion condensate is obtained from $x=\theta$, 
\be
\frac{\partial \bar{\Omega}}{\partial \theta} = \sin\theta\left(\frac{A}{2\lambda_0} - \bar{\mu}_I\bar{n}_I\frac{\cos\theta}{\sin^2\theta}\right) = 0 \, .
\ee
For $x=h_{3c}$ the second term in Eq.\ (\ref{dOmdx}) contributes and we must compute $\pi_{h_3}$ in the limit $u\to u_c$. To this end, we need the behavior  close to $u_c$, namely 
\begin{subequations} \label{a0h3}
\bea
{a_0^2}(u) &=& {a_{0(1)}^2}\sqrt{u-u_c}  + \ldots  \, , \\[2ex]
h_3(u) &=& h_{3c} + h_{3(1)}\sqrt{u-u_c}  + \ldots  \, ,
\eea
\end{subequations}
and 
\be\label{zetac}
\zeta = \frac{c}{\sqrt{u-u_c}} + \ldots \, ,
\ee
for a constant $c$ (whose exact expression can be computed with the help of the equations of motion, but is irrelevant here). The expansions (\ref{a0h3}) and (\ref{zetac}) imply that stationarity with respect to $h_{3c}$ yields $h_{3(1)}=0$. 

Stationarity with respect to $u_c$ must be computed more carefully  due to the fact that it sets the IR integration limit, see for instance Ref.\ \cite{Li:2015uea} for more details. One finds 
\bea
\frac{\partial \bar{\Omega}}{\partial u_c} &=& \left(\pi_{a_0^2} {a_0^2}'+\pi_{h_3} h_3'+kx_4'-{\cal L}+A\phi_Tx_4'\cos\theta \right)_{u=u_c}  \non[2ex]
&=& - \frac{u_c^{5/2}}{\zeta(u_c)} \, .
\eea
Hence, using Eq.\ (\ref{zetapirho}), imposing this derivative to vanish gives 
\be
k =  u_c^4 \sqrt{f_T(u_c)}\sqrt{1+\frac{\kappa^2}{u_c^5}} -A\phi_T(u_c)\cos\theta \, . 
\ee
Finally, by integrating the equation of motion for ${a_0^2}$ in Eq.\ (\ref{EOMpirho}) once more we get the isospin chemical potential in the form 
\be
\bar{\mu}_I = \int_{u_c}^\infty du\, \frac{\zeta\, (\kappa+\xi)}{u^{5/2}}  \, .
\ee
The equations of motion coupled with the stationarity equations can be solved numerically, and the result can be used to compute the free energy. 

To determine the onset curve of the $\pi\rho$ phase, a much simpler calculation can be performed. To this end, we approach the second-order onset from within the $\pi\rho$ phase, such that $h_3(u)$ is non-vanishing but approaches zero. Therefore, we may neglect $h_3$ in the equations of motion for $x_4$ and $a_0^2$, such that these functions (and $\zeta$) assume the form of the pure pion-condensed phase. The only nontrivial equation is thus 
\bea\label{eqh31}
\left(\frac{u^{5/2}f_T\tilde{h}_3'}{\zeta}\right)' &=& -\frac{4\lambda_0^2(a_0^2)^2\zeta  \tilde{h}_3 }{u^{1/2}f_T} \, , 
\eea
where we have introduced $\tilde{h}_3(u)\equiv h_3(u)/h_{3c}$ such that $\tilde{h}_3(0)=1$, $\tilde{h}_3'(0)=0$.  We can now find, for all $u_T$, the (smallest) $\bar{\mu}_I$ for which the pure pion solution is such that $\tilde{h}_3(\infty)=0$. This gives the critical $\bar{\mu}_I$ for the onset of a rho meson condensate on top of the pion condensate. 

We may use the same idea to calculate the vacuum mass of the rho meson. In the absence of pion condensation there is no need for the rotation (\ref{gLgR2}) and we may set $a_0^1(u)=a_0^2(u)=0$. Now the (unrotated) rho meson condensate is accounted for in the $1,2$ components in isospin space. Hence we may also set $h_3(u)=0$ and, by symmetry, denote  $h(u)\equiv h_1(u)=h_2(u)$ and consider the limit $h(u)\to 0$. This allows us to set $a_0^3(u)=\bar{\mu}_I$ and use the vacuum result for $\zeta(u)$. Then, with $\tilde{h}(u)\equiv h(u)/h(u_c)$ the only non-trivial equation is  
\bea \label{Eomh1}
\left(\frac{u^{5/2}f_T\tilde{h}'}{\zeta}\right)' &=& -\frac{4\lambda_0^2 \bar{\mu}_I^2 \zeta \tilde{h}}{u^{1/2}f_T} \, .
\eea
This equation has the same structure as Eq.\ (\ref{eqh31}). The difference sits in the nontrivial function $a_0^2(u)$ in Eq.\ (\ref{eqh31}) and in the different versions of $\zeta$, evaluated with and without pion condensation. Again, the critical chemical potential is defined as the smallest $\bar{\mu}_I$ for which the solution satisfies $\tilde{h}(\infty)=0$. The vacuum mass of the rho meson is then simply obtained by multiplying this critical chemical potential by 2. [In fact, the whole rho meson spectrum can be obtained by finding all chemical potentials for which  $\tilde{h}(\infty)=0$, but this is irrelevant for our purpose.]

\begin{figure} [t]
\begin{center}
\includegraphics[width=\columnwidth]{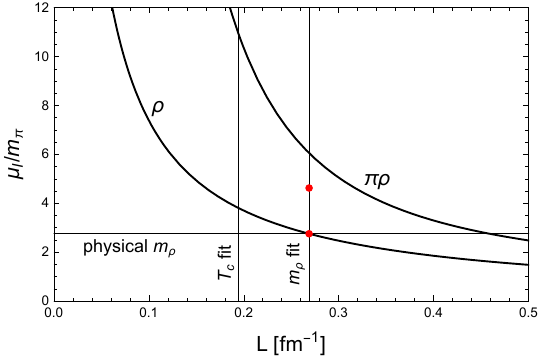}
\caption{Critical chemical potentials for the onset of rho meson condensation at $T=0$ in the vacuum (``$\rho$'') and in the pion-condensed phase (``$\pi\rho$''), as a function of the model parameter $L$, with $m_\pi$ and $f_\pi$ held fixed.  The vertical lines indicate where the physical values of $T_c$ and $m_\rho$ are reproduced. The red dots on the line indicate the results in the confined geometry  \cite{Kovensky:2023mye}, where the parameters are fitted to $m_\pi$, $f_\pi$, $m_\rho$.} 
\label{fig:Rhofit}
\end{center}
\end{figure}

Recall that the phase diagram in Fig.\ \ref{fig:PionT} was obtained with the parameter fit \eqref{Lfit}, which reproduces the physical values for $m_\pi$, $f_\pi$, $T_c$, i.e., this was a ``mixed'' fit, using both vacuum and medium properties. We find numerically that for this fit Eq.\ (\ref{Eomh1}) yields an unphysically large vacuum mass of the rho meson, namely $m_\rho \simeq 1062 \, {\rm MeV}$. Similarly, Eq.\ (\ref{eqh31}) yields an in-medium mass of about $22 m_\pi$, more than twice as large as the prediction of Ref.\ \cite{Kovensky:2023mye} (the $T=0$ plane in Fig.\ \ref{fig:phases3D}), where the model parameters were fitted only to vacuum parameters, $f_\pi$, $m_\pi$, $m_\rho$. 
Besides the different fits, the discrepancy is partially caused  by the dynamic embedding used in the decompactified limit. To isolate this geometric effect, we redo the parameter fit in the decompactified limit. This can be done in a systematic way, by continuously varying one of the model parameters, $L$, while adjusting the other two parameters $\tilde{\alpha}$ and $\tilde{\lambda}_0$ to keep $m_\pi$ and $f_\pi$ fixed. The explicit equations needed for this procedure can be found in Ref.\ \cite{Kovensky:2024oqx}.

\begin{figure} [t]
\begin{center}
\includegraphics[width=\columnwidth]{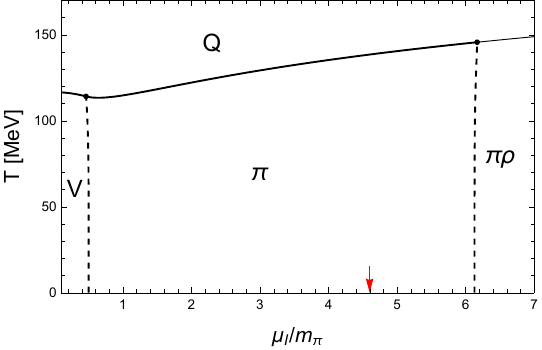}
\caption{Phase diagram in the $T$-$\mu_I$ plane using the parameter set (\ref{Lfit2}), corresponding to the ``$m_\rho$ fit'' in Fig.\ \ref{fig:Rhofit}. The red arrow indicates the zero-temperature value of the onset of the $\pi\rho$ phase in the confined geometry. The transition curve between $\pi$ and Q phases is continued beyond the tricritical point as a thin line, with the transition curve between $\pi\rho$ and Q phases to be expected slightly above that line. } 
\label{fig:RhoL}
\end{center}
\end{figure}

We present the result of this procedure in Fig.\ \ref{fig:Rhofit}, which shows the strong dependence of the $\pi\rho$ onset on the model parameters. The right vertical line indicates the ``$m_\rho$ fit'', i.e., matching the physical values of $f_\pi$, $m_\pi$, $m_\rho$. We find that this corresponds to the choice 
\be
\label{Lfit2}
\tilde{\lambda}_0 \simeq 1.28 \, , \quad \tilde{\alpha} \simeq 2.40\times 10^{-3} \, , \quad 
L \simeq  
0.267 \, {\rm fm} \, . 
\ee
A parametric form of the zero-temperature $\pi\rho$ onset was discussed within a linear sigma model \cite{Brauner:2016lkh}. For a comparison, let us use our predicted values within the ``$m_\rho$ fit'' for the confined geometry, $\mu_I \simeq 4.62\, m_\pi$, and the decompactified limit, $\mu_I \simeq 6.13\, m_\pi$. This range is reproduced by Eq.\ (41) of Ref.\ \cite{Brauner:2016lkh} (adjusting a factor 2 due to different conventions for $\mu_I$) for $g^2/\lambda\in[0.69,1.25]$, where $\lambda$ and $g$ are (unknown) self-coupling constants for $\pi$ and $\rho$, respectively, in the model used in this reference. 

\begin{figure} [t]
\begin{center}
\includegraphics[width=\columnwidth]{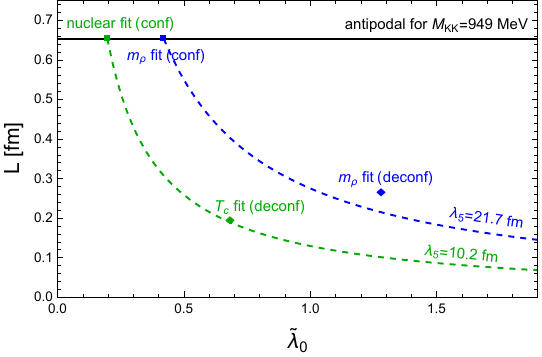}
\caption{Summary of the main parameter fits used in this review. The confined geometry requires in general three parameters, $\lambda$, $M_{\rm KK}$, $L$. We work with two parameter sets for the antipodal case $L=\pi/M_{\rm KK}$, satisfying vacuum constraints (``$m_\rho$ fit'') and nuclear matter constraints (``nuclear fit''), indicated by squares. In the decompactified limit, only the two parameters $L$ and $\tilde{\lambda}_0\propto \lambda_5/L$ are relevant, and again we employ two fits related to vacuum properties and to medium properties (diamonds). The dashed lines are obtained by varying $L$ at fixed $\lambda_5$ (given by the confined fits)  and show that the parameter sets are (roughly) related pairwise. } 
\label{fig:parameters}
\end{center}
\end{figure}

The $T$-$\mu_I$ phase structure with the parameter set (\ref{Lfit2}) is shown in Fig.\ \ref{fig:RhoL}. We observe that the critical temperature for the chiral phase transition becomes unphysically small, $T_c \simeq 117 \, {\rm MeV}$ at $\mu_I=0$. This shows a tension between fitting our model parameters to vacuum vs.\ medium properties of QCD. We also see that, just like the pion onset, the rho onset  shows a very mild temperature dependence, the critical chemical potential increases slightly from $\mu_I = 6.13 \, m_\pi$ at $T=0$ to $\mu_I = 6.17 \, m_\pi$ at the tricritical point. To complete the phase diagram we need to compute the phase transition to the quark matter phase. This requires the full formalism developed in the first part of this subsection. A simplification can be made by taking the chiral limit, which is a very good approximation at such large values of $\mu_I$. Nevertheless, we have found the numerics of this problem to be very challenging and thus simply show the continuation of the phase transition line between $\pi$ and Q phases. Since the $\rho\pi$ phase has lower free energy than the $\pi$ phase (which we {\it were} able to confirm numerically, at least at small temperatures) we expect a first-order transition curve from the $\pi\rho$ to the Q phase slightly above this continuation.

 To summarize the main conclusion of this subsection for the phase diagram in the introduction: $\mu_B=0$ and $T=0$ planes are computed with different versions of the model. The $\mu_B=0$ result shown in that diagram is computed with a parameter fit that reproduces the phase diagram from lattice QCD, see Fig.\ \ref{fig:PionT}, but which predicts a $\pi\rho$ onset much larger than given by the dashed line in the $T=0$ plane in Fig.\ \ref{fig:phases3D}. Even after re-fitting the parameters, see Fig.\ \ref{fig:RhoL}, this phase transition is, although now at smaller $\mu_I$, still slightly beyond the $\mu_I$ scale shown in Fig.\ \ref{fig:phases3D}.

It is instructive to relate the different parameter sets used in this review to each other. This is done in Fig.\ \ref{fig:parameters}. The two fits of this section, the one based on a mix of vacuum and medium properties (\ref{Lfit}), and the one based purely on vacuum properties (\ref{Lfit2}) are marked by diamonds. In Sec.\ \ref{sec: baryons with isospin}, where we review results within the confined geometry, we will also work with a vacuum fit based on $f_\pi$, $m_\pi$, $m_\rho$, and with a fit that reproduces $m_\pi$, $m_\rho$ plus basic properties of nuclear matter at saturation (but not $f_\pi$). These parameter sets are marked with squares. The dashed curves suggest that these fits are pairwise connected: We may start from the antipodal scenario and move the D8- and $\overline{\rm D8}$-branes closer together at fixed $\lambda_5$, i.e., keeping the 5-dimensional gauge theory unchanged. Then we see that we (roughly) connect the two vacuum sets and (more accurately) the two sets that involve medium properties. Since in the decompactified limit we use the deconfined geometry for all temperatures we may want to check whether indeed $\ell \ll \pi $, i.e., if the confinement and chiral scales are sufficiently separated. The figure shows that this is not satisfied for $M_{\rm KK}\sim 1 \, {\rm GeV}$, the deconfined markers are at sizable fractions of the black horizontal line for the antipodal case. This suggests that the decompactified limit should rather be considered as a separate phenomenological model, independent of the full WSS geometry.


\subsection{Towards a holographic BEC-BCS crossover}
\label{sec:BECBCS}

It has been conjectured that QCD at nonzero isospin chemical potential may undergo a BEC-BCS crossover \cite{Son:2000xc,Brandt:2019hel,Nishida:2005ds,Deng:2006ed}, first discussed in the context of low-energy condensed matter physics \cite{Eagles:1969zz,10.1007BFb0120125,Nozieres:1985zz} and experimentally observed in cold atomic gases \cite{Bartenstein:2004zza,Regal:2004zza}.
The idea is that the pion condensate, where quarks and antiquarks are bound in pionic ``molecules'', smoothly turns into a condensate of weakly coupled Cooper pairs at asymptotically large $\mu_I$ due to asymptotic freedom. In other words, there is a coexistence of up and anti-down quark degrees of freedom inside a Fermi sphere (they have a common Fermi surface for $\mu_B=0$) with a condensate of their pairs. Pairing occurs in a small vicinity of the Fermi surface at weak coupling, and involves all quarks in the limit of the strong-coupling pion-condensed phase. 

Our holographic calculation is valid at strong coupling and, as already pointed out, does not reproduce asymptotic freedom. It is thus impossible to expect our model to account for a full BEC-BCS crossover from strong to weak coupling. However, we can naturally combine quark degrees of freedom and a pion condensate, hence we may ask if at least the intermediate crossover region can be described, possibly involving a non-smooth phase transition rather than a crossover. This is exactly the region that is challenging to describe both on the lattice and with perturbative methods. The geometric setup (``$\pi$Q'') is shown in the lower left panel of Fig.\ \ref{fig:embeddings}. Since this construction has not been discussed in the previous literature, we will use this subsection to go into some details of the derivation. It is instructive to point out the analogy with quarkyonic matter, which will be discussed in Sec.\ \ref{sec: muB-T quarkyonic phase}, and which requires a similar construction, in this case a coexistence of quark and baryon degrees of freedom, see  lower right panel of Fig.\ \ref{fig:embeddings}. Our construction
may therefore also shed some light on whether in QCD there are analogies between a quarkyonic phase at large $\mu_B$ 
and a BEC-BCS crossover at large $\mu_I$, which has been discussed recently in a non-holographic context \cite{Ivanytskyi:2025cnn,McLerran:2026owy}.

For simplicity, we will ignore rho meson condensation for this discussion and thus may switch off all spatial components of the gauge fields. Our task is to combine pion condensation from Sec.\ \ref{sec: pion condensation} with string sources, which are introduced as in Eq.~\eqref{Sstring}. Here, however, our strings carry isospin charge and thus couple to the non-abelian gauge potentials. We again set $\varphi=0$ without loss of generality, such that $a_0^1(u)=0$, see for instance Eq.\ (\ref{aaaboundary}). 
It is advantageous to parameterize the remaining gauge fields in polar representation, 
\begin{equation}
    a_0^2(u) = a(u) \sin \vartheta(u) \, , \qquad
    a_0^3(u) = a(u) \cos \vartheta(u) \, ,
\end{equation}
such that the boundary conditions for the new functions $a(u)$ and $\vartheta(u)$ read 
\begin{equation}
    a(\infty) = \bar{\mu}_I \, , \qquad \vartheta(\infty) = \theta \, , \qquad \vartheta(u_c) = 0 \, .
\end{equation}
This is a convenient set of variables because $a(u)$ is precisely the field that couples to the string sources, which allows us to write the string contribution to the action as 
\begin{equation}
    S_{q} = {\cal{N}}N_f \frac{V}{T} \int_{u_c}^{\infty} du\, \bar{n}_q \left[
    u - u_T - a(u)\right]\delta(u-u_q) \, .
    \label{Sstring isospin}
\end{equation}
This formulation allows for the strings to attach at a point $u_q$ different from $u_c$, which will be determined dynamically. For symmetry reasons, this means that there are two attaching points, one on each half of the connected branes. The illustration in Fig.\ \ref{fig:embeddings} shows the simpler case $u_q=u_c$, which is included in our description and, as we shall see below, is also a valid solution of our system. We shall abbreviate the values of $a$ and $\vartheta$ at the attaching point by 
\be
a_q \equiv a(u_q) \, , \qquad \vartheta_q \equiv \vartheta(u_q) \, .
\ee 
The equations of motion can be integrated once, giving  ($a=2,3$)
\begin{subequations}
\bea
\frac{u^{5/2}{a_0^a}'}{\zeta} &=& \kappa_a + \bar{n}_q\frac{{a_0^a}(u_q)}{a_q}\Theta(u-u_q) \, , \\[2ex]
\frac{u^{11/2}f_Tx_4'}{\zeta} &=& k+A\phi_T\cos\theta \, ,   
\eea
\end{subequations}
where $\kappa_a$ and $k$ are integration constants and where 
\bea
&&\hspace{-0.3cm}\zeta \equiv \sqrt{1+u^3f_Tx_4'^2-(a_0^2)'(a_0^2)' - (a_0^3)'(a_0^3)'} \non[2ex]
&&\hspace{-0.3cm}= 
\left\{1-\frac{(k+A\phi_T\cos\theta)^2}{u^8f_T}+\frac{[\kappa_2+\bar{n}_q\sin\vartheta_q\Theta(u-u_q)]^2}{u^5} \right.\non[2ex]
&&\hspace{-0.3cm}\left.+\frac{[\kappa_3+\bar{n}_q\cos\vartheta_q\Theta(u-u_q)]^2}{u^5}\right\}^{-1/2}  \, .
\eea
Assuming that the gauge potentials are continuous -- but not necessarily smooth, since a cusp will develop at the point where the strings attach to the flavor branes -- we integrate their equations of motion  over $[u_c,\infty)$ and $[u_c,u_q]$ to get the relations 
\begin{subequations}\label{intucinf}
\bea
\bar{\mu}_I\sin\theta &=& \kappa_2\,{\cal I}_{c}^\infty  +\bar{n}_q\sin\vartheta_q \,{\cal I}_{q}^\infty \, , \label{intucinf1} \\[2ex]
\bar{\mu}_I\cos\theta -K_c &=& \kappa_3\,{\cal I}_{c}^\infty+\bar{n}_q\cos\vartheta_q \,{\cal I}_{q}^\infty \, ,
\label{intucinf2} \\[2ex]
K_q\sin\vartheta_q &=& \kappa_2\,{\cal I}_{c}^{q} \, , \\[2ex]
K_q\cos\vartheta_q -K_c &=& \kappa_3\,{\cal I}_{c}^{q} \, ,
\eea
\end{subequations}
where we have denoted
\be \label{calIdef}
{\cal I}_{c,q}^\infty \equiv \int_{u_{c,q}}^\infty du\, \frac{\zeta}{u^{5/2}} \, ,  \qquad {\cal I}_c^q \equiv \int_{u_c}^{u_q} du\, \frac{\zeta}{u^{5/2}} \, .
\ee
With the help of partial integration, the free energy density can then be written as 
\bea
&&\bar{\Omega}_{\pi{\rm Q}} = \int_{u_c}^\infty du\, u^{5/2}\zeta - \frac{A\cos\theta}{2\lambda_0}  +\bar{n}_q(u_q-u_T-a_q) \non[2ex]
&&= \int_{u_c}^\infty du\, \left(\frac{u^{5/2}}{\zeta}+A\phi_Tx_4'\cos\theta \right)-\frac{A\cos\theta}{2\lambda_0} +k\frac{\ell}{2} \non[2ex]
&&-\bar{\mu}_I[(\kappa_2+\bar{n}_q\sin\vartheta_q)\sin\theta +
(\kappa_3+\bar{n}_q\cos\vartheta_q)\cos\theta] \non[2ex]
&&+\kappa_3 a_c +\bar{n}_q(u_q-u_T) \, ,
\eea
where $a_c\equiv a(u_c)$. 
We now consider the stationarity conditions for $\bar{\Omega}$ with respect to our free parameters. We first observe that stationarity with respect to $\kappa_{2}$, $\kappa_3$ does not have to be considered separately since it is equivalent to the conditions  (\ref{intucinf1}) and (\ref{intucinf2}). Also, stationarity with respect to the remaining integration constant $k$ is equivalent to the boundary condition $x_4(\infty)=\ell/2$. Next, stationarity of $\bar{\Omega}$ with respect to $a_c$ trivially gives $\kappa_3=0$. Stationarity with respect to the location of the string sources $u_q$ can be used to eliminate $k$, 
\bea
k = u_c^4\sqrt{f_T(u_c)}\sqrt{1+\frac{\kappa_2^2}{u_c^5}}-A\phi_T(u_c)\cos\theta \, . \label{kkap1}
\eea
Combining Eqs.\ (\ref{intucinf}) with the stationarity conditions with respect to $\bar{n}_q$ and $\theta$ allows us to also eliminate $\kappa_2$,
\bea
\kappa_2 = \frac{A\sin\theta}{2\lambda_0(u_q-u_T)\cos\vartheta_q} \, .
\eea
The remaining stationarity equations with respect to $\bar{n}_q$, $u_c$, $\theta$ can be written, respectively, as 
\begin{subequations} \label{statbcs}
\bea
&&u_q-u_T = \frac{\kappa_2{\cal I}_c^q}{\sin\vartheta_q} \, , \label{statbcs1} \\[2ex]
&&0 = 1-\frac{[k+A\phi_T(u_q)\cos\theta]^2}{u_q^8f_T(u_q)}+\frac{\kappa_2^2\cos^2\vartheta_q}{u_q^5} \, , \label{statbcs2} \\[2ex]
&&\frac{A\sin\theta}{2\lambda_0}=\frac{\bar{\mu}_I(\bar{n}_I\cos\theta-\bar{n}_q\cos\vartheta_q)}{\sin\theta} \, , \label{statbcs4}
\eea
\end{subequations}
where the isospin density is 
\be \label{nIpiQ}
\bar{n}_I =  \kappa_2\sin\theta +\bar{n}_q(\sin\vartheta_q\sin\theta+\cos\vartheta_q\cos\theta) \, .
\ee
For fixed temperature and chemical potential, the 3 conditions (\ref{statbcs}), together with the self-consistency condition for $A$ (\ref{Aalpha}), the boundary condition for $x_4$ (\ref{ell2}),  and Eq.\ (\ref{intucinf1}), can be solved numerically for the 6 variables $\theta, \vartheta_q, u_q, u_c, \bar{n}_q, A$. [The variables $a_c$ and $a_q$ can be calculated afterwards from Eqs.\ (\ref{intucinf}), but are not needed for the main physical results.]

As can be seen from Eq.~\eqref{nIpiQ}, the isospin density receives contributions from the pion condensate and the quarks, as envisioned from the geometric construction. With the help of Eq.\ (\ref{intucinf1}) and recalling that in the pure pion-condensed phase the isospin density is given by Eq.\ (\ref{nIpi}), it is plausible to divide the density into the two contributions
\be \label{nIdivide}
\bar{n}_I = \bar{n}_{{\rm pion}} + \bar{n}_{{\rm quark}} \, , 
\ee
with 
\begin{subequations}
\bea
\bar{n}_{{\rm pion}} &=& \frac{\bar{\mu}_I\sin^2\theta}{{\cal I}_c^\infty} \, , \\[2ex]
\bar{n}_{{\rm quark}} &=& \bar{n}_q\left(\frac{{\cal I}_c^q}{{\cal I}_c^\infty}\sin\vartheta_q\sin\theta + \cos\vartheta_q\cos\theta\right) \, .
\eea
\end{subequations}
Our approach allows for the string sources to be attached to the tip of the flavor branes, $u_q=u_c$, $\vartheta_q=0$, which fulfills Eq.\ (\ref{statbcs2}) trivially. Rederiving Eq.\ (\ref{statbcs1}), we find that this equation becomes redundant and in this special case we solve the remaining 
equations for the variables $\theta, u_c, \bar{n}_q, A$. In this way, we actually obtain two different branches of the $\pi$Q configuration.

\begin{figure} [t]
\begin{center}
\includegraphics[width=\columnwidth]{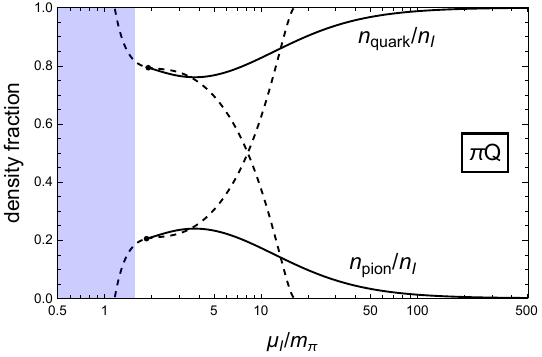}
\caption{Contributions of the pion condensate and the quarks (string sources) to the isospin density as a function of isospin chemical potential for $T=100\, {\rm MeV}$ in the $\pi$Q phase,
a possible configuration for a holographic BEC-BCS crossover.  The two variants of the $\pi$Q phase are shown as black dashed curves (string sources at the tip of the flavor branes, $u_q=u_c$) and black solid curves ($u_q>u_c$). The blue band indicates the region where the vacuum configuration is preferred over the $\pi$Q phase.} 
\label{fig:fractions}
\end{center}
\end{figure}

\begin{figure} [t]
\begin{center}
\includegraphics[width=\columnwidth]{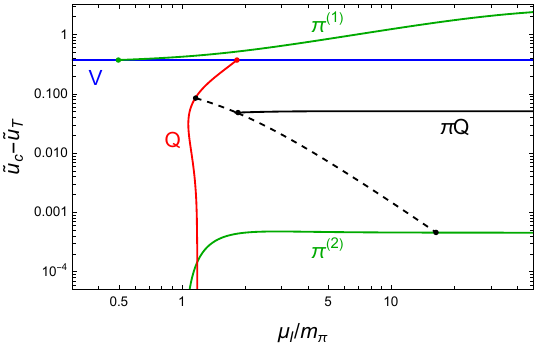}
\caption{(Dimensionless) constituent quark mass for various phases as a function of isospin chemical potential for $T=100\, {\rm MeV}$. Dashed and solid lines for the $\pi$Q phase are as in Fig.\ \ref{fig:fractions}. There are two branches of the pion-condensed phase, stable (1) and metastable (2).  } 
\label{fig:mconst}
\end{center}
\end{figure}

\begin{figure} [t]
\begin{center}
\includegraphics[width=\columnwidth]{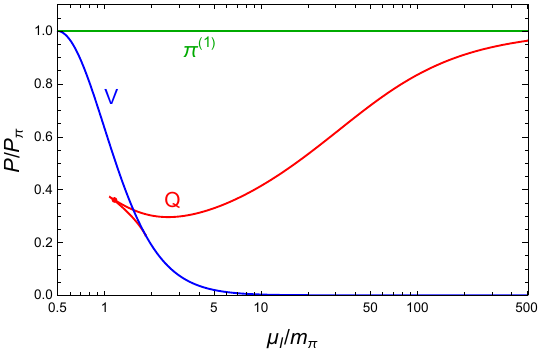}

\includegraphics[width=\columnwidth]{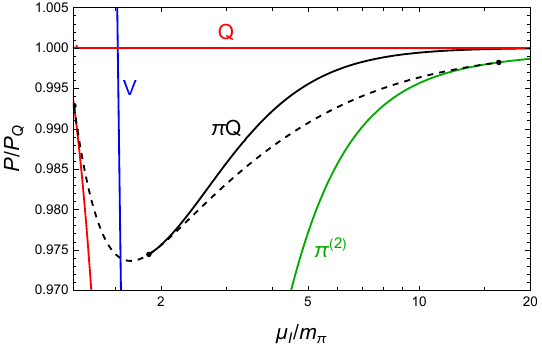}
\caption{{\it Upper panel:} Pressure of vacuum, quark phase, and (stable branch of the) pion-condensed phase, normalized to the pion-condensed phase for $T=100\, {\rm MeV}$. The dot on the red curve (on the upper branch in the two-valued domain) indicates where $u_c =u_T$.  {\it Lower panel:} Pressures at the same temperature, now normalized to the pressure of the Q configuration, including the $\pi$Q phase and the unstable $\pi$ branch. All phases shown in this panel are disfavored compared to the stable $\pi$ branch.} 
\label{fig:BCS}
\end{center}
\end{figure}

The numerical results are summarized in Figs.\ \ref{fig:fractions} -- \ref{fig:BCS}. All figures show results as functions of the isospin chemical potential at a fixed temperature $T=100\, {\rm MeV}$, using the parameter set  (\ref{Lfit}). The particle fractions according to Eq.\ (\ref{nIdivide}) are shown in Fig.\ \ref{fig:fractions}. Here and in the other figures, black dashed curves represent the solution $u_q=u_c$, while black solid curves stand for the solution $u_q>u_c$. We see that if the strings are kept at the tip, $u_q=u_c$, the pion contribution to the density increases as $\mu_I$ is increased, violating the expectation of the BEC-BCS picture. In contrast, this expectation is met if the strings are allowed to move towards the UV, $u_q>u_c$, and we see that in this case asymptotically the isospin density is formed solely by the quarks, exactly as the BEC-BCS picture suggests. 

In Figs.\ \ref{fig:mconst} and \ref{fig:BCS} the $\pi$Q configuration is put into the context of the other phases. Besides the vacuum and the pion phase from the phase diagram in Fig.\ \ref{fig:PionT}, here labeled by $\pi^{(1)}$, we have included the second (unstable) pion branch \cite{Kovensky:2019bih}, labeled by $\pi^{(2)}$, and the quark phase Q, metastable at this temperature as we know from Fig.\ \ref{fig:PionT}. The curves labeled by Q include both possible configurations, i.e., branes joining at $u=u_c$ with string sources at this point (``LTQ'') and branes reaching the horizon (``HTQ''). 

We show the distance from the horizon to the tip of the connected branes in Fig.\ \ref{fig:mconst}, abbreviating $\tilde{u}=u\ell^2$. This distance can be related to the constituent quark mass. The main point of the figure is to demonstrate that the $\pi$Q phase connects continuously -- although not smoothly -- to both the pure quark phase and the pure pion-condensed phase. This is encouraging in view of a possible holographic version of a BEC-BCS crossover. In Fig.\ \ref{fig:BCS}, we compare the pressure $P=-\Omega$ of the various phases. The upper panel sets the stage by showing vacuum, quark matter, and pion-condensed phase. The lower panel is a zoom-in (with different normalization, to make the small differences in pressure visible), now including the $\pi$Q branches. We see, firstly, that as soon as the string sources move away from the tip, this branch is favored over the one where they remain at the point $u=u_c$. Secondly, while the $\pi$Q phase is favored over the second pion branch it has lower pressure than the Q phase (the Q phase, in turn, has lower pressure than the stable pion branch, not shown in this panel). We have not found any region in the $T$-$\mu_I$ plane where this main conclusion is different: The $\pi$Q phase is never energetically preferred. This reinforces in hindsight the phase diagram of Fig.\ \ref{fig:PionT}, no new phase has to be added. We have, however, not checked the entire space of model parameters systematically. For instance, unphysically large values of the pion mass may lead to a qualitatively different phase diagram \cite{Kovensky:2020xif}. We can therefore not exclude that the $\pi$Q phase becomes stable in some regime of the phase diagram if the model parameters are chosen differently.

\section{Baryonic matter} \label{sec:baryons}

In this section we discuss the $\mu_I=0$ plane in the phase diagram of Fig.\ \ref{fig:phases3D}. The emphasis of this section is on baryonic matter. Since we are interested in the entire $T$-$\mu_B$ plane, we mainly keep working within the decompactified limit, which allows for nonzero temperature effects (we resort to the confined geometry where it is instructive). Isospin-asymmetric baryonic matter will be discussed in detail in Sec.\ \ref{sec: baryons with isospin}. Here we introduce the main approximations used to describe baryonic matter: the instanton gas, the pointlike limit, and the homogeneous approximation. In particular, we explain the configuration within the homogeneous approximation where the baryon charge distribution (``instanton layers'' \cite{Preis:2016fsp}) dynamically moves from the IR to the UV as the chemical potential is increased \cite{Ecker:2025sjb}, not unlike the ``popcorn transitions'' of Ref.\ \cite{Kaplunovsky:2012gb}. This configuration underlies the phase diagram in the introduction. Finally, we provide a holographic model of quarkyonic matter, improving the previous description that relied on the pointlike limit  \cite{Kovensky:2020xif}.

\subsection{Three ways to describe baryonic matter}

\subsubsection{Multi-instanton approximations}\label{sec:multi}

As discussed briefly in Sec.~\ref{sec: setup}, a single baryon can be understood as a topologically non-trivial instanton configuration in the spatial directions $\vec{x}$ and the holographic coordinate $z$ \cite{Douglas:1995bn,Sakai:2004cn,Hata:2007mb}.
 At large $\lambda$, it has a width of order $1/\sqrt{\lambda}$  and is centered at the point where the flavor branes connect, $z=0$. By means of the CS term \eqref{SCS}, it couples to the abelian gauge potential, whose boundary value is set by the baryon chemical potential $\mu_B$. Properties of single baryons 
in the WSS model have been studied extensively \cite{Sakai:2004cn,Hata:2007mb,Hong:2007kx,Hong:2007ay,Hashimoto:2008zw,Pomarol:2008aa,Seki:2008mu,Colangelo:2013pxk,Bolognesi:2013nja,Bigazzi:2018cpg}, mostly for $N_f=2$, but also for $N_f=3$ \cite{Hata:2007tn,Hashimoto:2009st,Lau:2016dxk,Cai:2017jjq}. Moreover, nucleon-nucleon interactions have been discussed based on the instanton picture \cite{Kim:2008iy,Hashimoto:2009ys,Kaplunovsky:2010eh,Cherman:2011ve}. Baryonic matter based on instantonic solutions is very challenging to describe and has been studied in various degrees of simplification \cite{Kim:2007vd,Rho:2009ym,Ghoroku:2012am,Ghoroku:2013gja,Bolognesi:2014dja,Kaplunovsky:2015zsa,Li:2015uea,Preis:2016fsp,BitaghsirFadafan:2018uzs,Jarvinen:2020xjh,Ghoroku:2021fos,Bartolini:2026dbr}. 

Let us start by  explaining the single-instanton solution in the decompactified geometry \cite{Preis:2016fsp,Kovensky:2021ddl}. Neglecting higher-order effects in the inverse 't Hooft coupling, we assume the instanton not to backreact on the embedding function $x_4(u)$, to be centered at $(\vec{x},z)=(0,0)$, and to have  small width such that for a power counting in $\lambda$ we can rescale spatial and holographic coordinates as $\vec{x} \to \vec{x}/\sqrt{\lambda}$, $z \to z/\sqrt{\lambda}$. The action is $S=S_{\rm DBI} + S_{\rm CS}$ with DBI and CS contributions given in Eqs.\ (\ref{DBI00}) and (\ref{SCS}), respectively. We denote the dimensionless field strengths in terms of abelian and non-abelian contributions, $\hat{F}_{\mu\nu}$ and $F_{\mu\nu}$ with $F_{\mu\nu} = F_{\mu\nu}^a\tau_a$, following the notation introduced for the gauge fields (\ref{AhatA}). The DBI action gives a purely geometric contribution plus a YM term. The leading-order contribution  is of order $\lambda$ and comes from the YM term, 
\bea \label{SYM1}
S_{\rm YM}^{(1)} &=& \frac{{\cal N}}{4\lambda_0^2M_{\rm KK}^3T}\frac{u_c\sqrt{f_T(u_c)}}{\gamma} \non[2ex]
&&\times \int d^3x \int_{-\infty}^\infty dz\, \left(\frac{1}{2} \Tr[F_{ij}^2] +\gamma^2 \Tr[F_{iz}^2]\right) \, , \;\;\;\;\;
\eea
where 
\begin{equation}
    \gamma^2 \equiv 6 u_c^3\left(
    1 - \frac{5u_T^3}{8u_c^3}
    \right) \, .
\end{equation}
The corresponding leading-order equations of motion are solved by the gauge fields 
\be \label{AzAi}
a_z^a=-\frac{1}{\gamma}\frac{x_a}{\xi^2+(\rho/\gamma)^2} \, , \quad 
a_i^a=\frac{z/\gamma\,\delta_{ia}-\epsilon_{ija}x_j}{\xi^2+(\rho/\gamma)^2} \, ,
\ee
with $\xi^2 \equiv x^2 + (z/\gamma)^2$. The resulting field strengths are
\be \label{Fsingle}
F_{zi}^a=\frac{2(\rho/\gamma)^2\delta_{ia}}{\gamma[\xi^2+(\rho/\gamma)^2]^2} \, , \quad 
F_{ij}^a=\frac{2(\rho/\gamma)^2\epsilon_{ija}}{[\xi^2+(\rho/\gamma)^2]^2} \, .
\ee
Here, $\rho$ is the (dimensionless) instanton width, to be determined with the help of the subleading contributions. The solution (\ref{AzAi}) is the BPST solution, already mentioned in Eq.\ (\ref{Amu BPST}), here with the additional factor $\gamma$. This factor indicates a deformation away from the SO(4) symmetric instanton, which was computed numerically in Refs.\ \cite{Rozali:2013fna,Bolognesi:2013jba} (``oblate instantons''). 
The next-to-leading order is given by the $\lambda^0$ terms which receive contributions from the YM and CS terms, 
\bea
&&\hspace{-0.5cm}S_{\rm YM}^{(0)}  = \frac{{\cal N}}{4\lambda_0^2M_{\rm KK}^3T}\frac{u_c\sqrt{f_T(u_c)}}{\gamma}\int d^3x \int_{-\infty}^\infty dz \non[2ex]
&&\hspace{-0.5cm}
\times    \left[\lambda_0^2\frac{\Tr[\hat{F}_{0i}^2]+\gamma^2\Tr[\hat{F}_{0z}^2]}{f_T(u_c)}+3u_cz^2\Tr[F_{iz}^2]\right.\non[2ex]
&&\hspace{-0.5cm}\left.+\frac{4u_c^6+10u_c^3u_T^3-5u_T^6}{8\gamma^2u_c^5f_T(u_c)} z^2\left(\frac{\Tr[F_{ij}^2]}{2}+\gamma^2\Tr[F_{iz}^2]\right)\right]  \, , \;\;\;\;
\eea
and
\bea
S_{\rm CS}^{(0)} = \frac{3{\cal N}}{4\lambda_0^2M_{\rm KK}^3T} \int d^3x \int_{-\infty}^\infty dz\, \hat{a}_0 \Tr[F_{iz} F_{jk}] \epsilon_{ijk}\, . 
\eea
With the help of the leading-order solutions (\ref{AzAi}) the subleading equations of motions are solved by 
\begin{subequations}
\bea
\hspace{-0.8cm}    \hat{a}_0 &=& \bar{\mu}_B-\frac{3\sqrt{f_T(u_c)}}{2\lambda_0^2u_c}\frac{\xi^2+2(\rho/\gamma)^2}{[\xi^2+(\rho/\gamma)^2]^2} \, , \\[2ex]
\hspace{-0.8cm} a_0^a &=& \frac{[(z/\gamma)^2-x^2]v_a+2x_a\vec{v}\cdot\vec{x}-2(z/\gamma)\epsilon_{abc}v_bx_c}{\xi^2+(\rho/\gamma)^2} \, , 
\eea
\end{subequations}
which leads to the field strengths 
\begin{subequations}
\bea
\hspace{-0.7cm}F_{0z}^a &=& -\frac{2\rho^2[v_a(z/\gamma)-\epsilon_{abc}v_bx_c]}{\gamma^3[\xi^2+(\rho/\gamma)^2]^2} \, , \\[2ex]
\hspace{-0.7cm}F_{0i}^a &=& \frac{2\rho^2[-\delta_{ia}\vec{v}\cdot\vec{x}+(z/\gamma)\epsilon_{iab}v_b+v_ax_i-v_ix_a]}{\gamma^2[\xi^2+(\rho/\gamma)^2]^2} \, .
\eea
\end{subequations}
Here we have imposed the boundary condition (\ref{bc potentials UV without pion}) for $\hat{a}_0$, and have generalized the boundary condition for the non-abelian gauge potential to $a_0^a (z \to \pm \infty ) = v^a$. Setting $\vec{v}=(0,0,\bar{\mu}_I)$ then accounts for a nonzero isospin chemical potential. Isospin is irrelevant for the main results of this section, but the general solution is useful for the discussion in Sec.\ \ref{sec: baryons with isospin}.  
Inserting the solutions back into the action and performing the $\vec{x}$ and $z$ integrals yields the dimensionless free energy 
\be \label{Fdeconf}
F = \frac{u_c\sqrt{f_T(u_c)}}{3}\left[1+\frac{9\gamma^2}{5\rho^2\lambda_0^2u_c^2}+\frac{u_c\beta\rho^2}{\gamma^2f_T(u_c)}\right] - \bar{\mu}_B \, , 
\ee
where the dimensionful version is obtained by multiplying with $\lambda_0N_c\MKK$, and where we have abbreviated
\be
\beta \equiv \beta_0 -\frac{\lambda_0^2\bar{\mu}_I^2}{u_c} \, , \qquad  \beta_0 \equiv 1-\frac{u_T^3}{8u_c^3}-\frac{5u_T^6}{16u_c^6} \, .
\label{beta}
\ee
We may use this result to compute the instanton width that minimizes the free energy, 
\be
\rho^2=\frac{12\pi}{\sqrt{5}\lambda}\frac{\gamma^2\sqrt{f_T(u_c)}}{u_c^{3/2}\beta^{1/2}}
\, ,
\label{rhosoldeconf}
\ee
confirming the behavior $\rho \sim 1/\sqrt{\lambda}$, which justifies the expansion a posteriori. The energy $E=F+\bar{\mu}_IN_I+\bar{\mu}_BN_B$ becomes 
 \be \label{Ebar}
 E = \frac{u_c\sqrt{f_T(u_c)}}{3}\left[1+\frac{6\beta_0}{\sqrt{5}\lambda_0u_c^{1/2}\beta^{1/2}\sqrt{f_T(u_c)}}\right] \, .
 \ee
This is the mass of a single baryon in the deconfined geometry (without quark mass correction). It does depend on temperature explicitly and implicitly through $u_c$. 

A simple approximation for a many-instanton system based on this single-instanton solution can now be obtained as follows. The main idea is to replace the non-abelian field strengths (squared) in the DBI and CS actions with a simple many-instanton version of Eq.\ (\ref{Fsingle}), where the spatial directions are averaged over, and then to solve the system self-consistently for $\hat{a}_0$ and the embedding $x_4$. In the simplest approximation of a non-interacting instanton gas, we place $N_B$ instantons at the points $(\vec{x},z)=(\vec{x}_n,0)$ and replace, for example in the CS term, 
\bea \label{gas}
&&\Tr[F_{iz}F_{jk}]\epsilon_{ijk} = -\frac{12}{\gamma} \frac{4(\rho/\gamma)^4}{[\xi^2+(\rho/\gamma)^2]^4} \non[2ex]
&&\to -\frac{1}{V}\frac{12}{\gamma}\sum_{n=1}^{N_B}\int d^3 x \frac{4(\rho/\gamma)^4}{[(\vec{x}-\vec{x}_{n})^2 + (z/\gamma)^2 + (\rho/\gamma)^2]^4} \non[2ex]
&&= -\frac{8\pi^2N_B}{V} q(z) \, , 
\eea
where
\be
q(z) \equiv \frac{3\rho^4}{4(z^2+\rho^2)^{5/2}}  \, .
\ee
This function is defined such that it integrates to $1$ over $z\in[-\infty,\infty]$. One can check that $N_B$ is indeed the baryon number according to Eq.\ (\ref{NB1}). We approximate  the squared field strengths in the DBI action analogously, and with 
\begin{equation}
     q(u)\equiv  2 \frac{\der z}{\der u} q(z)  
\end{equation}
(with a factor 2 included for convenience such that $q(u)$ integrates to 1 over  $u\in[u_c,\infty]$), we arrive at the Lagrangian 
\be \label{Linst}
{\cal L} = u^{5/2} \sqrt{(1+u^3f_T x_4'^2-\hat{a}_0'^2+g_1)(1+g_2)} - \bar{n}_B\hat{a}_0 q(u) \, ,
\ee
where 
\bea
g_1 = \frac{f_T \bar{n}_B}{3\gamma}\frac{\partial z}{\partial u}q(u) \, , \qquad 
g_2 = \frac{\gamma \bar{n}_B}{3u^3} \frac{\partial u}{\partial z} q(u)  \, ,
\eea
This Lagrangian is written in the same form as the one of Eq.\ (\ref{Lag DBI deconf}), which was  obtained by assuming homogeneous gauge fields from the beginning. In the simple approximation (\ref{gas}), used in Ref.\ \cite{Li:2015uea}, the locations of the instantons in position space are completely irrelevant and the location in the holographic direction is fixed. These assumptions can be relaxed, while retaining the form (\ref{Linst}), with more complicated forms of the function $q(u)$. In Ref.\ \cite{Preis:2016fsp} the instantons are allowed to arrange themselves into ``layers'' in the $z$ direction, while in Ref.\ \cite{BitaghsirFadafan:2018uzs} their interaction is taken into account based on the exact flat-space two-instanton solution from the Atiyah-Drinfeld-Hitchin-Manin construction \cite{Atiyah:1978ri,Kim:2008iy,Hashimoto:2009ys}. In both studies, numerical prefactors of instanton width and deformation were used as phenomenological parameters to reproduce properties of low-density nuclear matter.  While in Ref.\ \cite{BitaghsirFadafan:2018uzs} only ``defensive'' configurations were considered for simplicity -- all instantons with the same orientation in flavor space -- this assumption was relaxed in Ref.\ \cite{Bartolini:2026dbr}, in the simpler confined and antipodal geometry and using a long-distance approximation for the instanton interactions rather than the full two-instanton solution. All these approximations are expected to break down at very large baryon densities where the instantons start to overlap.

\subsubsection{Pointlike baryons}
\label{sec:pointlike}

Given that the single-instanton width goes to zero for $\lambda\to\infty$, it makes sense, as a simple approximation, to treat the baryons as pointlike objects located at $u=u_c$ \cite{Bergman:2008qv}. In this case, the action of the probe D4-branes wrapping the internal $S^4$ together with the CS term give 
\bea \label{Sb}
    &&[S_{\rm D4} + S_{\rm CS}]_{\rm pointlike}^{\rm deconf} =  {\cal{N}} N_f \frac{V}{T} \non[2ex]
    &&\times \int_{u_c}^{\infty} du  \,  \bar{n}_B \left[
    \frac{u}{3}\sqrt{f_T(u)} - \hat{a}_0(u)\right]\delta(u-u_c) \, .\;\;
\eea
In the D4 action we recover the leading-order mass of the single baryon (\ref{Ebar}). The CS term can be understood from Eq.\ (\ref{Linst}) with $q(u)\to\delta(u-u_c)$. It is also instructive to view this action in analogy to the action for string sources (\ref{Sstringdeconf}).  

As for the finite-width instantons, one may relax the condition that the pointlike baryons are placed at $u=u_c$, thus creating multiple pointlike instanton layers. This was recently studied in full generality in the confined antipodal geometry, computing number and location of the layers dynamically \cite{Ecker:2025sjb}. Even though the main results of this section will be obtained in the deconfined geometry, let us briefly review this calculation. In the confined antipodal case, the generalization of Eq.\ (\ref{Sb}), together with the YM part of the Lagrangian, can be written as 
\bea 
{\cal L} &=& -\frac{u^{5/2}\sqrt{f}\hat{a}_0'^2}{2} + \left[\bar{n}_c\delta(u-u_{\rm KK}) +\sum_{k=1}^d \bar{n}_k\delta(u-u_k)\right]\non[2ex]
&&\times \left[\frac{u}{3}-\hat{a}_0(u)\right] \, .
\eea
Here, $\bar{n}_c$ represents the baryons at $u=u_{\rm KK}$ while $\bar{n}_k$ corresponds to a baryon layer at $u=u_k$. All $\bar{n}_c$, $\bar{n}_k$ need to be determined dynamically. In total, there are $2d$ layers if $\bar{n}_c=0$ and $2d+1$ layers otherwise, and the total baryon density is  
\be \label{nBnc}
\bar{n}_B = \bar{n}_c + \sum_{k=1}^d \bar{n}_k \, .
\ee
The only dynamical field is the gauge potential $\hat{a}_0$ whose integrated equation of motion is 
\be \label{eomPoint}
\hat{a}_0' = \frac{\bar{n}_c + \sum_{k=1}^d \bar{n}_k \Theta(u-u_k)}{u^{5/2}\sqrt{f}} \, .
\ee
The baryon chemical potential can be expressed in terms of location of and gauge potential at the ``last'' layer,  closest to the UV limit,
\be \label{muBPoint}
\bar{\mu}_B = \hat{a}_0(u_d) + \frac{\bar{n}_B}{3u_{\rm KK}^{3/2}}\left(\pi-2\arctan \tilde{z}_d \right) \, ,
\ee
where $\tilde{z}\equiv z/u_{\rm KK}$. With the help of the equation of motion (\ref{eomPoint}), the free energy density becomes
\bea \label{OmPoint}
\bar{\Omega}_{\rm pointlike} &=& -\bar{\mu}_B \bar{n}_B + \bar{n}_c\frac{u_{\rm KK}}{3}+\sum_{k=1}^d \bar{n}_k\frac{u_k}{3} + \frac{\bar{n}_B^2\pi}{6u_{\rm KK}^{3/2}}\non[2ex]
&&- \sum_{k=1}^d \frac{\bar{n}_k}{3u_{\rm KK}^{3/2}}\Bigg[(2\bar{n}_c+\bar{n}_k)\arctan \tilde{z}_k\non[2ex]
&& +2\sum_{\ell=k+1}^d \bar{n}_\ell \arctan \tilde{z}_\ell\Bigg] \, .
\eea
We can minimize this expression for a given number of layers with respect to all locations $u_k$ and densities $\bar{n}_c$, $\bar{n}_k$. This yields the equations 
\begin{subequations}
\bea \label{nr3}
\bar{n}_r &=& 2(-1)^r\bar{n}_c \non[2ex]
&&+ \frac{2u_{\rm KK}^{3/2}}{3}\left[\tilde{z}_ru_r+2\sum_{k=1}^{r-1}(-1)^{r+k} \tilde{z}_k u_k\right]  \, ,  \label{nruKK} \\[2ex]
0&=&  2(-1)^r\frac{\bar{n}_c}{u_{\rm KK}^{3/2}}-\frac{u_{r+1}-u_r}{\arctan \tilde{z}_{r+1}-\arctan \tilde{z}_r} 
 \non[2ex]
&& +\frac{4}{3}\sum_{k=1}^r (-1)^{r+k} \tilde{z}_ku_k  \, ,\label{uarctan}
\eea
\end{subequations}
where $r=1,\ldots,d$ and $r=1,\ldots,d-1$, respectively, 
and where $\bar{n}_c$ is either 0 (no baryons at the tip) or given by 
\bea
\bar{n}_c  &=& \frac{u_{\rm KK}^{3/2}}{2}\frac{u_1-u_{\rm KK}}{\arctan \tilde{z}_1}  \, .
\eea
The equations look complicated but allow for a very simple numerical solution since they decouple and can be solved successively from the layers in the IR towards the layers in the UV. One can then compare the resulting free energies for different numbers of layers (for instance, at fixed baryon chemical potential). 

Remarkably, this setting has a finite limit when the number of layers is taken to infinity \cite{Ecker:2025sjb}. One finds that if the number of layers is increased the distance between the layers goes to zero while the location of the most ultraviolet layer approaches a finite location, which we denote by $u_\infty$ (and $\tilde{z}_\infty$ correspondingly). One obtains very simple expressions for the resulting baryon density 
\be
\bar{n}_B = \frac{u_{\rm KK}^{3/2}\tilde{z}_\infty u_\infty}{3} \, , 
\ee
chemical potential  
\be 
\bar{\mu}_B = \frac{u_\infty}{3}\left[1+\frac{\tilde{z}_\infty}{3}(\pi-2\arctan\tilde{z}_\infty)\right] \, , 
\ee
and free energy density 
\bea
\bar{\Omega}_\infty 
&=& -\frac{u_{\rm KK}^{7/2}}{63}\tilde{z}_\infty \left\{\frac{u_\infty^2}{u_{\rm KK}^2}+\frac{7\tilde{z}_\infty u_\infty^2}{6u_{\rm KK}^2}\left(\pi-2\arctan\tilde{z}_\infty\right)\right.\non[2ex]
&&\left.-{}_2F_1\left[\frac{1}{3},\frac{1}{2},\frac{3}{2},-\tilde{z}_\infty^2\right]\right\} \, .
\eea
The numerical evaluation shows that,  for arbitrary fixed chemical potential, the free energy decreases monotonically  with the number of layers and approaches $\bar{\Omega}_\infty$. The generalization to the deconfined geometry is straightforward, but will lead to more complicated results due to the nontrivial embedding function.

\subsubsection{Homogeneous ansatz}
\label{sec: muB-T baryonic phase}

A different way to bypass the difficulty of multi-instanton configurations is to assume a spatially homogeneous distribution of baryon charge by considering the ansatz introduced in Eq.\ \eqref{diagonal ansatz}. This was first proposed in Ref.\ \cite{Rozali:2007rx}, and subsequently employed in different variants in the WSS model \cite{Li:2015uea,Elliot-Ripley:2016uwb,Kovensky:2021ddl,Kovensky:2021kzl,Kovensky:2023mye,CruzRojas:2023ugm,Ecker:2025sjb}, the V-QCD model \cite{Ishii:2019gta,Demircik:2021zll,Bartolini:2022gdf,CruzRojas:2023ugm,Bartolini:2023wis,Bartolini:2025sag}, and a hard-wall model \cite{Bartolini:2022rkl,Bartolini:2022gdf}. We expect this approximation to be reasonably accurate at large baryon densities, where instantons start to overlap and beyond. Moreover, while crystalline many-instanton phases might be more accurate in describing the ground state at large $N_c$, a homogeneous baryon phase (a fluid as opposed to a solid) might be a more realistic description of $N_c=3$ QCD, where we know from large nuclei that baryonic matter is spatially uniform, at least at low densities.

In the isospin-symmetric case, we can simply take $h_1(u) = h_2(u) = h_3(u) \equiv h(u)$, hence the action in the deconfined geometry becomes 
\bea
\label{deconf homogeneous mui=0 action}
    &&S_{\rm DBI}^{\rm deconf} + S_{\rm CS} = {\cal{N}} N_f \frac{V}{T} \int_{u_c}^{\infty} du \non[2ex]
    &&\times \Big[u^{5/2} 
    \sqrt{(1+u^3 f_T x_4'^2 + g_1 - \hat{a}_0'^2)(1+g_2)}\non[2ex]
    && 
    -\frac{9}{4}\lambda_0 \hat{a}_0 h^2 h'\Big] \,,
\eea
with $g_1$ and $g_2$ defined in Eq.~\eqref{g123deconf}. As mentioned below Eq.\ (\ref{nBh123}), the baryon density can only be nonzero if the function $h$ is discontinuous. It is natural to take a discontinuity to be located at $z=0$. Indeed, symmetry arguments under $z\to -z$ imply that this is the only possibility if only one discontinuity is introduced. This  configuration was studied in the context of the $T$-$\mu_B$ phase diagram for the first time in Ref.\ \cite{Li:2015uea}. It was generalized to include isospin asymmetry in Ref.\ \cite{Kovensky:2021ddl} and was used to construct neutron stars in Ref.\ \cite{Kovensky:2021kzl}. At the location of the discontinuity in the bulk, the baryon charge is maximal, and the charge distribution resembles that of an instanton layer (with a cusp, generated by the discontinuity in the gauge field). Just like the instanton layers,  an arbitrary number of discontinuities at dynamically determined locations in the bulk can be introduced, and the energetically preferred solution should be identified for each chemical potential and temperature. 

\begin{figure} [t]
\begin{center}
\includegraphics[width=\columnwidth]{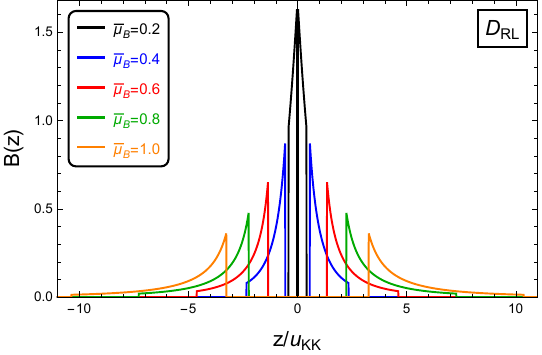}
\caption{Normalized baryon charge distribution in the bulk for different baryon chemical potentials in the $D_{RL}$ configuration. The profiles show the block-like structure that is generated by the four discontinuities in the non-abelian gauge fields and that moves towards the UV, i.e., towards large $|z|$, as the total baryon density increases. Figure taken from Ref.\ \cite{Ecker:2025sjb}. } 
\label{fig:DRLprofiles}
\end{center}
\end{figure}

\begin{figure} [t]
\begin{center}
\includegraphics[width=\columnwidth]{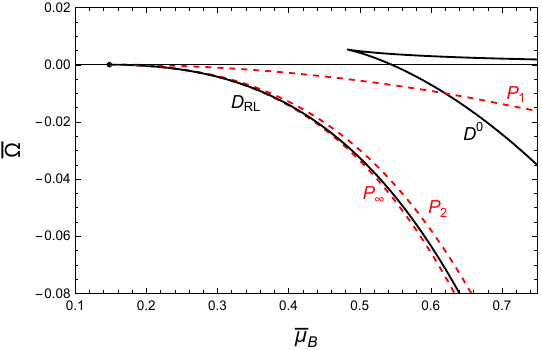}
\caption{Free energy comparison between the $D_{RL}$ phase, the single-jump solution $D^0$ and the pointlike approximation for 1,2, and infinitely many layers, all computed in the confined geometry. The vacuum has $\bar{\Omega}=0$ and thus only the $D^0$ configuration shows a first-order baryon onset (in a metastable regime). Figure taken with modifications from Ref.\ \cite{Ecker:2025sjb}.} 
\label{fig:DRLOmega}
\end{center}
\end{figure}

This was done systematically for up to 4 discontinuities (2 on each half of the flavor branes) in the confined antipodal setting with YM action in Ref.\ \cite{Ecker:2025sjb}
(where it was also argued, albeit less systematically, why phases with more discontinuities do not appear in the phase diagram). It was found that the standard single-jump configuration -- termed $D^0$ in Ref.\ \cite{Ecker:2025sjb} -- is energetically disfavored compared to one particular phase with 4 jumps. This phase was termed $D_{RL}$, referring to the behavior at the discontinuities: at the first discontinuity $h(z)$ is only nontrivial towards the UV (to the ``right'') while at the second discontinuity it is only nontrivial towards the IR (to the ``left''). This results in a block-like structure of the (normalized) baryon charge
\be
B(z) \equiv -\frac{3}{8\lambda_0^2 \bar{n}_B} \Tr[F_{ij}F_{kz}]\epsilon_{ijk} \, ,
\ee
shown in Fig.\ \ref{fig:DRLprofiles}, see also upper right panel of Fig.\ \ref{fig:embeddings}. Figure \ref{fig:DRLOmega} shows the free energy of the $D_{RL}$ phase compared to the $D^0$ phase and the pointlike configurations. The single-discontinuity configuration, routinely used in the literature and showing a first-order baryon onset as expected from QCD, is  \textit{not} the configuration with the lowest free energy.  The $D_{RL}$ phase is energetically favored but does not have a finite saturation density. The reason is that it approaches the pointlike limit for small densities (the blocks become pointlike). 

We now extend the $D_{RL}$ construction to the decompactified limit. This has not been done before in the literature and thus we will explain some of the details. First, we do so for the pure baryonic phase; in the subsequent subsection we use this configuration to generalize the holographic quarkyonic phase of Ref.\ \cite{Kovensky:2020xif}. For simplicity, we restrict ourselves to the chiral limit. This is a good approximation for the present purpose of this calculation, which is only used for $\mu_I=0$, i.e., pion condensation, which {\it does} strongly depend on the pion mass, is not expected to play a role. The generalization to a physical pion mass is  straightforward, although the numerical evaluation might be challenging. 

Starting from the action \eqref{deconf homogeneous mui=0 action}, the integrated equations of motion for $x_4'$ and $\hat{a}_0$ are 
\bea
    \hat{a}_0' = \frac{\bar{n}_B Q}{u^{5/2}}\, \zeta \, , \qquad 
x_4' = \frac{k}{u^{11/2} f_T}\zeta \, , 
\eea
with
\bea 
\zeta \equiv \frac{\sqrt{1+g_1}}{\sqrt{1+g_2-\frac{k^2}{u^8f_T}+\frac{(\bar{n}_BQ)^2}{u^5}}} \, ,
\eea
where the baryon density  $\bar{n}_B$ and the function $Q(u)$ are given by  
\bea\label{nBQ}
   \bar{n}_B &=& \frac{3\lambda_0}{4} (h_{2-}^3-h_{1+}^3) \, , \non[2ex]
   \bar{n}_B Q(u) &=& \frac{3\lambda_0}{4}\left[\Theta(u-u_1)(h^3-h_{1+}^3)\right. \non[2ex]
   &&\left.-\Theta(u-u_2)(h^3-h_{2-}^3)\right] \,.  
\eea
Here, $h_{1+} \equiv h(u_1^+)$ and $h_{2-} \equiv h(u_2^-)$, such that $Q(u<u_1)=0$ and $Q(u \to \infty) = 1$. The equation of motion for $h$ can then be expressed as 
\bea \label{EOMh}
\left(\frac{u^{5/2}f_T h'}{\zeta}\right)'  &=&  \frac{\lambda_0^2\zeta h^2}{u^{1/2}}\left(2h+\frac{3\bar{n}_BQ}{\lambda_0u^2}\right)   \, .
\eea
This equation must be solved numerically, restricting to configurations for which the free energy is stationary with respect to the free parameters of the ansatz. We assume that the abelian gauge potential $\hat{a}_0$ remains continuous and that $h(u_1<u) = h(u>u_2) = 0$. The free parameters are thus the integration constant $k$, the values $h_{1+}$ and $h_{2-}$, the locations of the discontinuities $u_1$, $u_2$, and the location of the tip of the connected branes $u_c$. As we have seen before, stationarity with respect to $k$ is equivalent to imposing the boundary condition \eqref{x4 bc}. Stationarity with respect to $h_{1+}$, $h_{2-}$ results in the conditions 
\begin{subequations}\label{dp}
\bea
    \frac{u_1^{5/2}f_T(u_1)d_{1+}}{\zeta(u_1^+)} &=& 3\lambda_0 \hat{a}_0(u_1)h_{1+}^2 \, , \\[2ex] \frac{u_2^{5/2}f_T(u_2)d_{2-}}{\zeta(u_2^-)} &=& 3\lambda_0 \hat{a}_0(u_2)h_{2-}^2 \, , 
\eea
\end{subequations}
with $d_{1+} \equiv h'(u_1^+)$ and $d_{2-} \equiv  h'(u_2^-)$. From the derivatives with respect to $u_1$, $u_2$, $u_c$,  we get 
\begin{equation}
    \zeta(u_1^+)=\zeta(u_1^-) \, , \quad 
   \zeta(u_2^+)= \zeta(u_2^-) \, , \quad 
    \frac{u_c^{5/2}}{\zeta(u_c^+)}=0\, . 
\end{equation}
Combining these conditions leads to 
\bea
   && h_{i\pm}^2 =\frac{u_i^{3/2}\sqrt{f_T(u_i)}\, d_{i\pm}}{\lambda_0\zeta(u_i)}   \, , \quad 
\hat{a}_0(u_i) = \frac{u_i\sqrt{f_T(u_i)}}{3}\, , \non[2ex] 
&&k = u_c^4\sqrt{f_T(u_c)} \, ,
\eea
where $i=1,2$, and where 
\begin{subequations}\label{zeta12}
\bea 
    \zeta(u_1) &=& \left[1-\frac{k^2}{u_1^8 f_T(u_1)}\right]^{-1/2} \, , \\[2ex]
    \zeta(u_2) &=& \left[1-\frac{k^2}{u_2^8 f_T(u_2)}+\frac{\bar{n}_B^2}{u^5}\right]^{-1/2} \, .
\eea
\end{subequations}
From the continuity of $\hat{a}_0$ we obtain an additional condition,  namely
\be \label{Acont}
\hat{a}_0(u_2)-\hat{a}_0(u_1) = \int_{u_1}^{u_2} du\, \frac{\bar{n}_BQ}{u^{5/2}}\, \zeta \, .
\ee
The baryon chemical potential can be computed from 
\begin{equation}
\bar{\mu}_B = \hat{a}_0(u_2) + \int_{u_2}^\infty du\frac{\bar{n}_BQ}{u^{5/2}}\zeta
\,, 
\end{equation}
and the free energy density can be written as 
\bea
\bar{\Omega}_{\rm B} &=& \int_{u_c}^\infty du \, u^{5/2}\left(\frac{1+g_1}{\zeta}-1\right) - \frac{2}{7}u_c^{7/2} \non[2ex]
&&+ k\frac{\ell}{2}-\bar{\mu}_B\bar{n}_B \, ,
\eea
where we have subtracted the (infinite) vacuum contribution  $\bar{\Omega}(\mu_B=T=0)$.
The results from the numerical evaluation will be discussed in Sec.\ \ref{sec:phasestr}, after the remaining phases of the $T$-$\mu_B$ phase diagram are collected.

\subsection{Quarkyonic matter}
\label{sec: muB-T quarkyonic phase}

The pressure of large-$N_c$ nuclear matter scales linearly with $N_c$, suggesting a bulk contribution of quarks, while
confinement indicates that the excitations of the system must be color singlets. This was dubbed quarkyonic matter, suggesting that the phase is partly quark-like, partly baryonic \cite{McLerran:2007qj}. Indications of quarkyonic matter were found on the lattice \cite{Philipsen:2019qqm}, but it remains unclear if quarkyonic matter plays any role in the $N_c=3$ phase diagram and which observable properties such a phase might have. In the WSS model it was speculated that quarkyonic matter is signaled by baryon charge in the bulk moving dynamically to the UV \cite{Kaplunovsky:2012gb,Kaplunovsky:2013iza}, and that it can be observed for parameterically large baryon densities \cite{deBoer:2012ij}. In an Einstein-Maxwell-Dilaton model, quarkyonic matter was identified with a chirally symmetric and yet confined phase \cite{Chen:2019rez}, based  on earlier studies in the NJL model coupled to the Polyakov loop \cite{McLerran:2008ua}. It was later found that the WSS model actually allows for a phase that is both quark-like and baryonic in a much more direct sense. This phase was constructed in Ref.\ \cite{Kovensky:2020xif}. It is given by a configuration in which quark and baryon degrees of freedom coexist, and where both contribute to the total baryon density. Instead of a construction in momentum space \cite{McLerran:2018hbz}, quarks and baryons affect different domains of the holographic direction, i.e., different energy regimes. Baryons are restricted to the UV, while quarks dominate the IR. This is reminiscent of a layered structure of the Fermi sphere, a picture often employed in phenomenological descriptions of quarkyonic matter \cite{McLerran:2018hbz,Jeong:2019lhv,Cao:2020byn,Margueron:2021dtx}. 

This holographic quarkyonic phase was originally constructed in the decompactified limit using pointlike baryons. It has 
two incarnations, chirally (approximately) restored and chirally broken. Taken together, these configurations provide a continuous interpolation between the pure baryon phase B and the high-temperature quark phase Q. It turns out that for realistic values of the pion mass, in fact for  sufficiently small $m_\pi$, only the quarkyonic configuration where the flavor branes reach the horizon is thermodynamically stable, indicating chiral restoration. While this is perhaps the most convincing realization of quarkyonic matter in holography so far, the scaling of the pressure of large-$N_c$ QCD is not correctly reproduced. Even after ignoring the dominating $N_c^2$ gluonic contribution from the background geometry, baryonic and quarkyonic phases are not distinguished by different $N_c$ scalings. 

Here we show how the quarkyonic configuration can be generalized to non-pointlike baryons. More explicitly, to the $D_{RL}$ phase, which is the favored configuration in the absence of string contributions.
For simplicity, as in the previous subsection, we will work in the chiral limit. Hence, our quarkyonic configuration is shown by the lower right panel of Fig.\ \ref{fig:embeddings}, but with straight D8- and $\overline{\rm D8}$-branes. This implies $x_4'=0$, $u_c = u_T$ and the boundary condition $\hat{a}_0(u_T)=0$. The equations of motion for $\hat{a}_0$ and $h$ are 
\begin{subequations}
\bea
    \hat{a}_0' &=& \frac{\bar{n}_q+\bar{n}_BQ}{u^{5/2}}\, \zeta \, , \\[2ex]
    \left(\frac{u^{5/2}f_T h'}{\zeta}\right)'  &=&  \frac{\lambda_0^2h^2}{u^{1/2}}\zeta \left[2h+\frac{3(\bar{n}_q+\bar{n}_BQ)}{\lambda_0u^2}\right] \, , \;\;\;\;\;\;
\eea
\end{subequations}
where 
\bea
    \zeta \equiv  \frac{\sqrt{1+g_1}}{\sqrt{1+g_2+\frac{(\bar{n}_q+\bar{n}_BQ)^2}{u^5}}} \, .
    \eea   
Here, $\bar{n}_q$ is the contribution from quarks to the baryon density, and $\bar{n}_B$ and $Q(u)$  are exactly as for the pure baryonic case (\ref{nBQ}). The behavior of $Q(u)$, which is zero in the IR, indicates that baryon number is only generated in the UV. This is in contrast to quark number, which is not suppressed in any domain of the holographic coordinate. 
The stationarity conditions with respect to the parameters $h_{1+}$, $h_{2-}$, $u_1$ and $u_2$ are derived analogously to the pure baryonic case and take exactly the same form, only distinguished by the straight brane geometry and the presence of $\bar{n}_q$, which leads to 
\begin{subequations}
\bea
    \zeta(u_1) &=& \left(1+\frac{\bar{n}_q^2}{u_1^5}\right)^{-1/2} \, ,  \\[2ex]
    \zeta(u_2) &=&  \left[1+\frac{(\bar{n}_q+\bar{n}_B)^2}{u_2^5}\right]^{-1/2}\, ,
\eea
\end{subequations}
instead of Eq.\ (\ref{zeta12}). 
The parameter $\bar{n}_q$ can be fixed by the condition  
\bea
  \hat{a}_0(u_1) &=&  \bar{n}_q\int_{u_T}^{u_1} du\, \frac{\zeta}{u^{5/2}} = u_1\, {}_2F_1\left[\frac{1}{5},\frac{1}{2},\frac{6}{5},-\frac{u_1^5}{\bar{n}_q^2}\right] \non[2ex]
&&-u_T\, {}_2F_1\left[\frac{1}{5},\frac{1}{2},\frac{6}{5},-\frac{u_T^5}{\bar{n}_q^2}\right] \, , 
\eea
and we have the continuity condition 
\be 
\hat{a}_0(u_2)- \hat{a}_0(u_1) = \int_{u_1}^{u_2} du\, \frac{\bar{n}_q+\bar{n}_BQ}{u^{5/2}}\, \zeta \, .
\ee
Finally, the dimensionless chemical potential is computed from 
\bea\label{muBQy}
\bar{\mu}_B &=& \hat{a}_0(u_2)+ (\bar{n}_q+\bar{n}_B)^{2/5}\frac{\Gamma\left(\frac{3}{10}\right)\Gamma\left(\frac{6}{5}\right)}{\sqrt{\pi}}\non[2ex]
&&-u_2\,{}_2F_1\left[\frac{1}{5},\frac{1}{2},\frac{6}{5},-\frac{u_2^5}{(\bar{n}_q+\bar{n}_B)^2}\right]  \, ,
\eea
and the renormalized free energy is 
\bea
\bar{\Omega}_{\rm Qy} &=& \int_{u_T}^\infty du \, u^{5/2}\left(\frac{1+g_1}{\zeta}-1\right) \non[2ex]
&& - \frac{2}{7}u_T^{7/2} -\bar{\mu}_B(\bar{n}_q+\bar{n}_B) \, .
\eea
For the numerical evaluation we need to fix the model parameters $L$ and $\tilde{\lambda}_0$. 
We aim at making the three-dimensional phase diagram in Fig.\ \ref{fig:phases3D} as consistent as possible. Hence we recall that the $\mu_B=0$ plane was computed with the fit (\ref{Lfit}) that reproduces $f_\pi$, $m_\pi$, $T_c$. Since in this section we work in the chiral limit, $m_\pi=0$, we readjust our parameters to hold $f_\pi=92\, {\rm MeV}$ and $T_c=160\, {\rm MeV}$ fixed. The fitting procedure is particularly simple in the chiral limit because the required numerical calculations are independent of $\tilde{\lambda}_0$ and $L$ and yield
\be
f_\pi^2 \simeq 0.01166\, \frac{\tilde{\lambda}_0}{L^2} \, , \qquad T_c \simeq \frac{0.1538}{L} \, .
\ee
Therefore, we obtain 
\be \label{fitchiral}
\tilde{\lambda}_0 \simeq 0.6711 \, , \qquad L = 0.1897 \, {\rm fm} \, .
\ee
These values deviate only very slightly from the fit with a physical pion mass (\ref{Lfit}).

\begin{figure} [t]
\begin{center}
\includegraphics[width=\columnwidth]{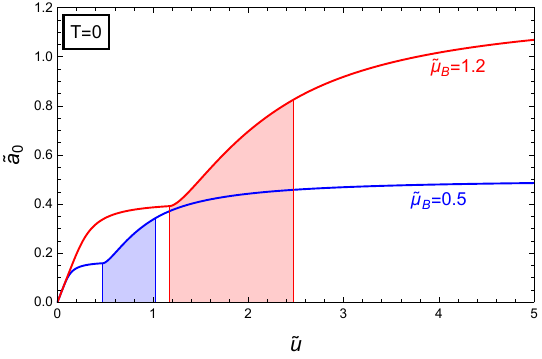}

\includegraphics[width=\columnwidth]{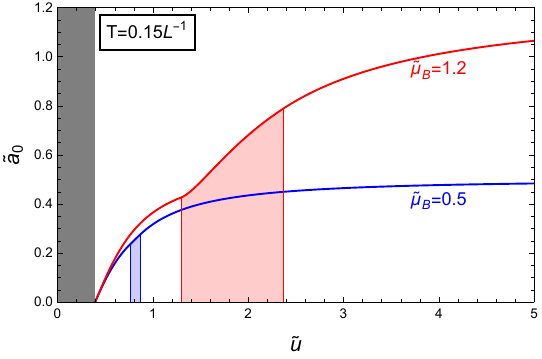}
\caption{Bulk profiles of the abelian gauge field in the quarkyonic phase for different temperatures and chemical potentials. Tilde indicates rescaled quantities, $\tilde{a}_0 = \hat{a}_0\ell^2$, $\tilde{u}=u\ell^2$, $\tilde{\mu}_B = \bar{\mu}_B\ell^2$. The black hole geometry at nonzero temperatures is represented by the gray bar in the lower panel. The shaded areas mark the region where the non-abelian gauge fields are nonzero. The layered structure of $\tilde{a}_0$ is generated dynamically by combining strings with topological baryon number. } 
\label{fig:QyProfile}
\end{center}
\end{figure}

To illustrate the characteristic behavior in the bulk, we have plotted the gauge potential $\hat{a}_0(u)$ in Fig.\ \ref{fig:QyProfile}. Most strikingly, it features a layered structure, where $\hat{a}_0$ ``wants'' to asymptote to a dynamically emerging chemical potential before the baryon charge sets in, and eventually approaches the actual chemical potential for $u\to \infty$. This structure is washed out by temperature effects, as the lower panel demonstrates. It was already observed with pointlike baryons \cite{Kovensky:2020xif}; the main difference is the smooth behavior of $\hat{a}_0$ due to the underlying $D_{RL}$ configuration, as opposed to a cusp at the location of pointlike baryons. The shaded areas in Fig.\ \ref{fig:QyProfile} indicate that for decreasing $\mu_B$ as well as for increasing $T$ the baryonic block becomes narrower. Indeed, we find that the pointlike limit is approached in these two directions. It is interesting to compute the baryon fraction $x_B = \bar{n}_B/(\bar{n}_q+\bar{n}_B)$ in the quarkyonic phase. We typically find a very large value of $x_B$. For instance, for the parameters in the upper panel of Fig.\ \ref{fig:QyProfile}, $x_B\simeq0.985,0.984$ for the small and large chemical potential, respectively, while for the lower panel $x_B\simeq0.24,0.87$, in the same order. The reason for the small value 0.24 is that the baryon fraction drops rapidly as the quarkyonic phase approaches the pure quark phase at large temperatures. This will become manifest in the phase diagram that we will discuss momentarily.  

\subsection{Vacuum and quark matter}

For completeness we also mention the remaining phases needed for the phase diagram in the $\mu_I=0$ plane. The vacuum V and quark matter Q are essentially the same as discussed in Sec.\ \ref{sec: muB=0 plane}, with $\mu_I$ replaced by $\mu_B$ and with the additional simplification of working in the chiral limit, see for instance Refs.\ \cite{Horigome:2006xu,Li:2015uea}.

The baryonic vacuum  is obtained for instance by setting $\bar{n}_B=h(u)=0$ in the formalism of Sec.~\ref{sec: muB-T baryonic phase}. This yields  
\begin{equation}
    \hat{a}_0(u) = \bar{\mu}_B \, , \quad x_4' = \frac{k}{u^{11/2} f_T} \zeta \, , \quad 
\zeta = \left(1-\frac{k^2}{u^8f_T}\right)^{-1/2} \, , 
\end{equation}
with $k = u_c^4\sqrt{f_T(u_c)}$. The resulting free energy is independent of $\bar{\mu}_B$ and takes the form 
\bea
\bar{\Omega}_{\rm V} = \int_{u_c}^\infty du \, u^{5/2}\left(\zeta^{-1}-1\right) - \frac{2}{7}u_c^{7/2} + k\frac{\ell}{2}\, .
\eea
One needs to compute $u_c$ from the boundary condition (\ref{ell2}) and insert the result into $\bar{\Omega}_{\rm V}$. At zero temperature, this can be done analytically. 

The quark phase can be obtained by setting $\bar{n}_B=0$ in Sec.~\ref{sec: muB-T quarkyonic phase}, i.e.~by removing the baryons from the quarkyonic phase. This gives
\begin{equation}
    \frac{u^{5/2}\hat{a}_0'}{\sqrt{1-\hat{a}_0'^2}} = \bar{n}_q \, , \quad x_4(u) = \frac{\ell}{2}  \, .  
\end{equation}
We can solve this case analytically for all temperatures and chemical potentials. With $\hat{a}_0(\infty)=\bar{\mu}_B$ we obtain 
\be
\hat{a}_0(u) = \bar{\mu}_B-\frac{\bar{n}_q^{2/5}\Gamma\left(\frac{3}{10}\right)\Gamma\left(\frac{6}{5}\right)}{\sqrt{\pi}}
+ u\,{}_2 F_1\left[\frac{1}{5},\frac{1}{2},\frac{6}{5},-\frac{u^5}{\bar{n}_q^2}\right] \, . 
\ee
From $\hat{a}_0(u_T) = 0$ we get the condition 
\be \label{ceq}
0 = \bar{\mu}_B-\frac{\bar{n}_q^{2/5}\Gamma\left(\frac{3}{10}\right)\Gamma\left(\frac{6}{5}\right)}{\sqrt{\pi}}
+ u_T\,{}_2 F_1\left[\frac{1}{5},\frac{1}{2},\frac{6}{5},-\frac{u_T^5}{\bar{n}_q^2}\right] \, .
\ee
Finally, the renormalized free energy reads 
\bea
\bar{\Omega}_{\rm Q}  = -\frac{2}{7}u_T^{7/2} {}_2F_1 \off{
-\frac{7}{10},-\frac{1}{2},\frac{3}{10},-\frac{\bar{n}_q^2}{u_T^5}
} - \bar{\mu}_B \bar{n}_q \, .
\eea

\subsection{Phase structure} \label{sec:phasestr}

\begin{figure} [t]
\begin{center}
\includegraphics[width=\columnwidth]{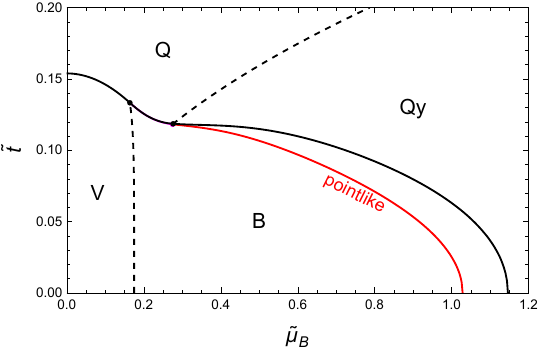}
\caption{Phase diagram in the plane of dimensionless temperature $\tilde{t}=t\ell=TL$ and baryon chemical potential $\tilde{\mu}_B=\bar{\mu}_B\ell^2 = \mu_B L/(N_c\tilde{\lambda}_0)$. Solid (dashed) curves are first- (second-) order phase transitions. Baryonic matter B and quarkyonic matter Qy are obtained from the $D_{RL}$ configuration. If instead pointlike baryons are used, the phase transition between B and Qy phases is given by the red curve. } 
\label{fig:Qy}
\end{center}
\end{figure}

We are now prepared to compute the phase structure in the $T$-$\mu_B$ plane, using the model parameters  (\ref{fitchiral}). The result is shown in Fig.\ \ref{fig:Qy} in terms of chemical potential $\tilde{\mu}_B = \bar{\mu}_B\ell^2$ and temperature $\tilde{t} = t\ell$. In these units, the phase diagram is valid for all $L$ via simple rescalings (but depends nontrivially on $\tilde{\lambda}_0$).
 The only difference to the phase diagram with pointlike baryons \cite{Kovensky:2020xif} lies in the transition from baryonic to quarkyonic matter (besides a small difference, hardly visible in the plot, in the transition from baryonic to quark matter). Both baryonic and quarkyonic phases approach the pointlike limit.  Baryonic matter becomes pointlike for small $\mu_B$, which results in a second-order baryon onset. Baryons in the quarkyonic phase become pointlike for large temperatures, resulting in a second-order transition to quark matter. The generalization from the pointlike limit (one layer on each of the flavor branes) to the $D_{RL}$ configuration has made quarkyonic matter less favorable, as can be seen by comparing the red with the black phase transition curve. 

The phase diagram has some obvious unphysical features. Besides the first-order chiral transition at small $\mu_B$, already seen the $\mu_B=0$ plane, the liquid-gas phase transition expected from QCD is replaced by a second-order transition from the vacuum to baryonic matter. The critical chemical potential at this transition corresponds to the (medium-dependent) baryon mass $u_c/3\,\sqrt{f_T(u_c)}$ (\ref{Sb}). At zero temperature, $f_T(u)=1$ and $u_c/3 \simeq 0.17495/\ell^2$, independent of the choice for $\tilde{\lambda}_0$ and $L$. For the parameters used here (\ref{fitchiral}) this gives a baryon mass of $0.17495\, N_c\tilde{\lambda}_0/L\simeq 366\, {\rm MeV}$, far below the physical nucleon mass $m_N=939\, {\rm MeV}$. If we were to reproduce the physical nucleon mass  together with $T_c=160\, {\rm MeV}$ we would need to work with $\tilde{\lambda}_0\simeq 1.72$ and the same $L$ as in Eq.\ (\ref{fitchiral}). 

\begin{figure} [t]
\begin{center}
\includegraphics[width=\columnwidth]{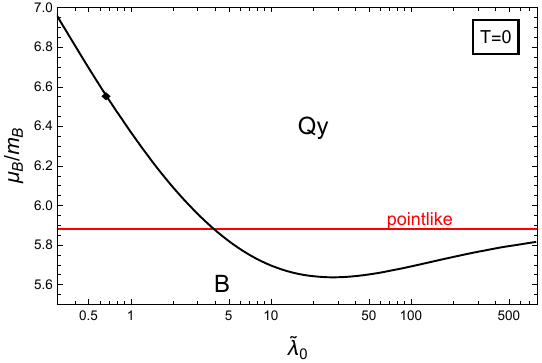}
\caption{Zero-temperature phase transition between quarkyonic and baryonic phases as a function of the coupling $\tilde{\lambda}_0$. The diamond marks the coupling used in Fig.\ \ref{fig:Qy}. The critical chemical potential is given in terms of the vacuum mass of the baryon $m_B$. } 
\label{fig:BQyT0}
\end{center}
\end{figure}

The transition from the B to the Qy phase is of first order. The brane embedding changes abruptly from connected to straight and disconnected, and thus also the baryon charge in the bulk rearranges discontinuously. For applications to neutron stars, it is interesting to ask whether the this transition occurs at realistic, i.e., sufficiently low densities. In Fig.\ \ref{fig:BQyT0} we show the critical chemical potential for $T=0$ in units of the baryon mass as a function of the coupling $\tilde{\lambda}_0$. In these units, the pointlike approximation predicts the same transition point for all $\tilde{\lambda}_0$. The improved finite-width approximation shows a dependence on the coupling with a minimal critical chemical potential at $\tilde{\lambda}_0 \simeq 27.8$. At this point the transition occurs at $\mu_B \simeq 5.63 \, m_B$, which is, translated to a realistic baryon mass, far too large to be relevant for neutron stars, where $\mu_B \lesssim 1.5 \, {\rm GeV}$. To put this into context, however, we point out that in the absence of the quarkyonic phase, our calculation would predict no chiral phase transition at all for $T=0$, indicating that baryonic matter is energetically very favorable in the present approximation. Introducing quarkyonic matter induces a low-temperature chiral transition, albeit at ultra-large densities.

In the three-dimensional phase diagram of the introduction, which reflects the result of Fig.\ \ref{fig:Qy}, we have rescaled $\mu_B$ by the zero-temperature baryon onset, in order to match this onset to the one of the $T=0$ plane. The latter is computed within the confined geometry and using the $D^0$ version of baryonic matter, as will be explained in detail in the next section. Therefore, this rescaling of the $\mu_B$ axis should be viewed as an ad-hoc fix, useful for a discussion based on currently available results, but to be removed after a more consistent calculation of the three-dimensional phase structure in the future.

\section{Baryonic matter with isospin and application to neutron stars}
\label{sec: baryons with isospin}

In this section, the $T=0$ plane of the phase diagram in Fig.~\ref{fig:phases3D} is discussed. The main difficulty  is to generalize baryonic matter to isospin-asymmetric configurations sourced by a nonzero $\mu_I$. To this end, we work in  the confined geometry of the WSS model, with antipodal flavor branes, and within the YM approximation. We employ the homogeneous approximation for baryonic matter, which needs to be generalized compared to Sec.\ \ref{sec: muB-T baryonic phase} due to the isospin asymmetry. In large parts, this section is a review of Refs.\ \cite{Kovensky:2023mye,Kovensky:2021kzl}. Like for the $\mu_B=0$ plane, we include a physical pion mass, which is important to describe pion condensation realistically. Pion and rho meson condensation are, in the setting of this section, simpler versions of what we have already discussed in Sec.\ \ref{sec: muB=0 plane}. However, here we  couple them with baryons to allow for phases where meson condensates and baryonic matter coexist. In the last part of this section, we apply the results to neutron stars. 

\subsection{Mesonic phases}
\label{sec:mesconf}

We start with purely mesonic phases, $n_B=0$. It is expected that for nonzero isospin chemical potentials and sufficiently small values of $\mu_B$, baryons are absent and the isospin number density $n_I$ is fully generated by the pion and rho meson condensates. A priori, it is not excluded that the system finds its energetically favored state by populating baryonic states with nonzero isospin number, even at small $\mu_B$, thus creating a nonzero baryon density $n_B$. We shall see, however, that, at least for realistic parameter sets, this is not the case. 

There are three distinct phases with $n_B=0$: the vacuum, the pion-condensed phase, and the phase where pion and rho meson condensates coexist. We expect a similar situation as in the low-temperature region of the $\mu_B = 0$ plane in Sec.~\ref{sec: muB=0 plane}, i.e., an onset of pion condensation followed by an onset of rho meson condensation in the background of the pion condensate.  In the vacuum, all gauge fields are either constant or zero. Together with the fixed embedding in the case of antipodal separation, this implies that the vacuum does not depend on $\mu_B$ and $\mu_I$ and its free energy density is zero. 
The pion-condensed phase $\pi$ was already discussed in passing in Sec.\ \ref{sec: pion condensation}, see Eqs.~\eqref{piconfined} and \eqref{thOmnI}. The only nonzero fields are the gauge potentials, and the thermodynamics is exactly like in leading-order chiral perturbation theory. In particular, there is a second-order transition from the vacuum to the pion-condensed phase at $\mu_I=m_\pi/2$, independent of $\mu_B$ (this phase transition will be modified when $\mu_B$ is large enough for baryons to appear).   

Including rho meson condensation requires us to take into account the  spatial components of the non-abelian gauge fields. In analogy to Sec.\ \ref{sec: rho meson deconfined}, in the presence of a pion condensate, and taking into account our rotation (\ref{gLgR2}), the rho meson sits in the $\tau_3$ direction, hence the only nonzero component is $h_3$. The CS term  plays no role since $h_1=h_2=0$, and the equations of motion are  derived from Eq.\ (\ref{Lag DBI conf}). After setting $\varphi=0$ without loss of generality, the only nontrivial equations of motion are 
\begin{subequations}
\bea
    \left(u^{5/2}\sqrt{f}{a_0^2}'\right)'&=& \frac{\lambda_0^2 a_0^2h_3^2}{u^{1/2}\sqrt{f}}   \, , \\[2ex]
\left(u^{5/2}\sqrt{f} h_3'\right)' &=& -\frac{4\lambda_0^2\of{a_0^2}^2h_3}{u^{1/2}\sqrt{f}} \, .
\eea
\end{subequations}
The critical value of $\bar{\mu}_I$ associated with the rho meson onset in the pionic medium is obtained by inserting the pure pionic solution for $a_0^2$ (\ref{piconfined}) into the equation of motion for $h_3$. The resulting equation takes a particularly simple form in the variable $\tilde{z}\equiv z/u_{\rm KK}$,  
\bea
    \tilde{h}_3''+\frac{2\tilde{z}}{1+\tilde{z}^2}\tilde{h}_3' = -\frac{16\Lambda\tilde{h}_3 \arctan^2 \tilde{z}}{9\pi^2 u_{\rm KK}(1+\tilde{z}^2)^{4/3}} \, , 
\eea
with $\tilde{h}_3 = h_3/h_{3c}$, and where 
\bea
\Lambda \equiv 4\lambda_0^2\bar{\mu}_I^2\left(1-\frac{\bar{m}_\pi^4}{16\bar{\mu}_I^4}\right) \, .
\eea
The relevant (smallest) eigenvalue for the boundary conditions $\tilde{h}_3(0)=1$, $\tilde{h}_3(\infty)=0$ is $\Lambda \simeq 1.90176$. Similarly, and in analogy to Eq.\ (\ref{Eomh1}), the rho meson onset from the vacuum is found from 
 \be\label{tildeh}
\tilde{h}''+\frac{2\tilde{z}}{1+\tilde{z}^2}\tilde{h}' =-\frac{4\Lambda^{(0)}\tilde{h}}{9u_{\rm KK}(1+\tilde{z}^2)^{4/3}}  \, , \quad \Lambda^{(0)} = 4\lambda_0^2\bar{\mu}_I^2\, .
\ee
Here the numerical evaluation for $\tilde{h}(0)=1$, $\tilde{h}(\infty)=0$ gives the smallest eigenvalue $\Lambda^{(0)} \simeq  0.669314$. This translates into the vacuum mass of the rho meson, 
\be \label{mrhoconf}
m_\rho^2 = M_{\rm KK}^2\Lambda^{(0)} \, .
\ee
 The critical chemical potential for the onset of the $\pi\rho$ phase can be written as
\be
\mu_I =  \frac{m_\rho}{2} \sqrt{\frac{\Lambda}{2\Lambda^{(0)}} + \sqrt{\left(\frac{\Lambda}{2\Lambda^{(0)}}\right)^2+\frac{m_\pi^4}{m_\rho^4}}} \simeq 4.67 \, m_\pi \, .
\ee
Hence, if expressed in terms of $m_\pi$, this critical chemical potential does not depend on the model parameters. It defines the second-order line between $\pi$ and $\pi\rho$ phases in Fig.\ \ref{fig:phases3D}, and deviates from the result obtained in the decompactified limit, where it {\it does} depend on the model parameters, as discussed in detail in Sec.\ \ref{sec: rho meson deconfined}.

\subsection{Isospin-asymmetric holographic baryons}

Single baryons with nonzero isospin number were addressed  in the WSS model not long after the model was established \cite{Hata:2007mb}, employing a quantization in the moduli space of the instanton known from the Skyrme model \cite{Adkins:1983ya}. This quantization induces separate states for neutrons and protons, which has for instance been used to compute their mass difference \cite{Bigazzi:2018cpg}. A full-fledged many-baryon description including this quantization is difficult and has not yet been developed in the WSS model. Hence, existing studies essentially build on the classical isospin spectrum. In the deconfined geometry, this spectrum is given by Eq.\ (\ref{Ebar}) and can be written in terms of isospin number $N_I$ as 
\be
E = e_0 + \frac{b}{\lambda}\sqrt{\frac{N_I^2}{6}+\frac{2}{15}} \, .
\ee
Written in this general form, with constants $e_0$ and $b$ that do not depend on $\lambda$ and $N_I$, it is also valid in the confined geometry relevant for this section, where $e_0 = u_{\rm KK}/3$ and $b=6\pi u_{\rm KK}^{1/2}$ [upon reinstating the dimensionful factors and using $u_{\rm KK}=4/9$ one recovers the baryon mass (\ref{mB})]. In this classical limit, the spectrum $E(N_I)$ is continuous, and thus isospin-symmetric matter can be made of $N_I=0$ baryons, as opposed to $N_I=1/2$ protons and $N_I=-1/2$ neutrons. Isospin asymmetry then requires populating heavier states, rather then adjusting the relative number of neutrons and protons. This artifact is important to keep in mind for instance for the symmetry energy of our holographic baryonic matter. 

Early attempts to account for isospin-asymmetric baryonic matter can be found in Refs.\ \cite{Seo:2012sd,Seo:2013nka}. A more systematic study was performed in Ref.\ \cite{Kovensky:2021ddl}, in the chiral limit and assuming isotropy in the non-abelian gauge fields, $h_1=h_2=h_3$. 
This approach was later refined in Ref.\ \cite{Kovensky:2023mye}, where a nonzero pion mass was included and a less rigid ansatz for baryonic matter was adopted by allowing for the functions $h_i$ in Eq.\ \eqref{diagonal ansatz} to be different. This is necessary since these functions describe different components in isospin space. With $\mu_I$  pointing in the $\tau^3$ direction, it is sufficient to work with two different functions, $h_1=h_2\equiv h$ and $h_3$. As a consequence, with this ansatz, isospin-asymmetric matter is spatially anisotropic, since the functions $h_i$ also constitute different components in position space. This allows for a continuous crossover to a phase with rho meson condensation. 

The relevant action is given by Eqs.~\eqref{SCS homogeneous}, \eqref{Lag DBI conf}, and \eqref{S_mpi conf}. The equation of motion for the abelian gauge potential is easily integrated,  
\begin{equation}
\label{Asol}
    \hat{a}_0'(u) = \frac{\bar{n}_B Q(u)}{u^{5/2}\sqrt{f(u)}} \, , \qquad Q(u)\equiv 1-\frac{h^2(u)h_3(u)}{h_{c}^2h_{3c}} \, ,
\end{equation}
where $h_c=h(u_c)$ and $h_{3c}=h_3(u_c)$. The product $h^2h_3$ must be discontinuous in order to create baryon number. In this section, we assume that there is only one discontinuity, located at $u=u_{\rm KK}$, such that  
\be
\label{defnB}
\bar{n}_B  =-\frac{3\lambda_0}{4} h_{c}^2h_{3c} \, .
\ee
In other words, the baryonic matter discussed here is the $D^0$ configuration from Sec.\ \ref{sec: muB-T baryonic phase} and Fig.\ \ref{fig:DRLprofiles}, generalized to nonzero isospin. The generalization of the $D_{RL}$ configuration with 4 discontinuities has not been done yet. 
The equations of motion for the abelian spatial components can also be integrated to obtain ($i=1,2$) 
\bea \label{A123prime}
\hat{a}_i' &=& -\frac{3\lambda_0^2hh_3a_0^i}{2u^{5/2}\sqrt{f}} \, , \qquad  
\hat{a}_3' = -\frac{3\lambda_0^2h^2a_0^3}{2u^{5/2}\sqrt{f}} \, .
\eea
We set $\hat{a}_1(\infty)=\hat{a}_2(\infty)=0$, and determine $\hat{a}_3(\infty)$ dynamically such that the current associated with this component vanishes, as it should be in the absence of external fields. 
For the non-abelian potentials we have ($i=1,2$)
\begin{subequations}\label{eomK}
\bea
    \left(u^{5/2}\sqrt{f}{a_0^i}'\right)'&=& \frac{\lambda_0^2 {a_0^i}(h^2+h_3^2)}{u^{1/2}\sqrt{f}} +\frac{9\lambda_0^2a_0^i h^2h_3^2}{4u^{5/2}\sqrt{f}}    \, , \;\;\;\;\;\;\\[2ex]
\left(u^{5/2}\sqrt{f}{a_0^3}'\right)'&=& \frac{2\lambda_0^2 {a_0^3} h^2}{u^{1/2}\sqrt{f}}  +\frac{9\lambda_0^2a_0^3 h^4}{4u^{5/2}\sqrt{f}}  \, , 
\eea
\end{subequations}
with boundary conditions given by Eq.\ (\ref{aaaboundary}). Finally, abbreviating $v_0^2\equiv (a_0^1)^2+(a_0^2)^2$, the equations of motion for $h$ and $h_3$ are 
\begin{subequations} \label{eomh}
\bea
 (u^{5/2}\sqrt{f}h')' &=& \frac{\lambda_0^2h}{u^{1/2}\sqrt{f}}\Big[ h^2+h_3^2-2v_0^2-4(a_0^3)^2 \non[2ex]
&& \hspace{-1.7cm}-\frac{9}{2u^{2}}\left(h_3^2 v_0^2  + 2h^2 (a_0^3)^2 - \frac{2h_3 \bar{n}_B Q}{3 \lambda_0}\right) \Big]\, , \label{eomh1}  \\[2ex]
(u^{5/2}\sqrt{f}h_3')'  &=& \frac{\lambda_0^2}{u^{1/2}\sqrt{f}}\Big[h_3(2h^2-4v_0^2)\non[2ex]
&&-\frac{9h^2}{u^{2}}\left(h_3v_0^2-\frac{\bar{n}_B Q}{3 \lambda_0} \right) \Big]   \, . 
\label{eomh3}
\eea
\end{subequations}
Equations (\ref{eomK}) and (\ref{eomh}) must be solved numerically, while simultaneously minimizing the free energy with respect to the free parameters.  This can be done for the pure baryonic phase B, where $\theta=0$, as well as the coexistence phase $\pi$B, where isospin-asymmetric baryons coexist with a pion condensate given by $\theta$. Both phases can also accommodate rho mesons at large $\mu_I$. 
The total isospin density is 
\begin{equation} \label{nIk}
\bar{n}_I  = -(\kappa_1\sin\varphi-\kappa_2\cos\varphi)\sin\theta+\kappa_3\cos\theta \, ,   
\end{equation}
where $\kappa_i \equiv (u^{5/2}\sqrt{f}{a_0^i}')_{u=\infty}$ for $i=1,2,3$, and the stationarity condition with respect to $\theta$ reads
\begin{equation}
    0 = (\kappa_1\sin\varphi-\kappa_2\cos\varphi)\cos\theta + \kappa_3\sin\theta+\frac{2\bar{m}_\pi^2}{9\pi\bar{\mu}_I}\sin\theta  \, .
\end{equation}
The baryon chemical potential can be expressed as  
\begin{equation}
\label{muBdef}
    \bar{\mu}_B  =  \frac{u_{\rm KK}^2h_{(1)}}{2\sqrt{3}\lambda_0h_{c}h_{3c}} +\int_{u_{\rm KK}}^\infty du\,  \frac{\bar{n}_B Q}{u^{5/2}\sqrt{f}} \, ,
\end{equation}
where $h_{(1)}$ is defined in the IR expansion 
\be
h(u) =   h_c+h_{(1)}\sqrt{u-u_{\rm KK}} +\ldots \, , 
\ee
and the free energy density is  
\bea
\bar{\Omega} &=& \int_{u_{\rm KK}}^\infty du \frac{u^{5/2}}{2\sqrt{f}}\left[g_1+g_2+\frac{(\bar{n}_BQ)^2}{u^5}\right]\non[2ex]
&&-\bar{\mu}_B\bar{n}_B-\frac{\bar{\mu}_I\bar{n}_I}{2}  - \frac{2\bar{m}_\pi^2}{9\pi}(\cos\theta-1)\, .
\label{OMfinal}
\eea
This expression is used to compute the free energies of the B and $\pi$B phases, which can then be compared to the mesonic free energies from Sec.\ \ref{sec:mesconf} for all values of $\bar{\mu}_B$ and $\bar{\mu}_I$.

\subsection{Parameter fit and phase structure}\label{sec:Pfit}

The parameter fit in this section needs to be discussed independently from the other sections because we work in the confined geometry, where the parameters are $\MKK$ and $\lambda$ (plus the pion mass $m_\pi$ which is set to its physical value). The fit is complicated by the fact that we are interested in the entire $\mu_B$-$\mu_I$ plane. We do not know much about this plane from first-principle QCD, but we do know that isospin-symmetric nuclear matter has a saturation density $n_0\simeq 0.15\, {\rm fm}^{-3}$ and a binding energy $E_B\simeq -16\, {\rm MeV}$, which leads to an onset baryon chemical potential of $\mu_0 \simeq 923\, {\rm MeV}$. Since we work with the $D^0$ configuration in this section, our baryonic matter does have a first-order onset, see Fig.\ \ref{fig:DRLOmega}, and we may seek to reproduce these physical values.

Fitting our model parameters to the pion decay constant (\ref{fpiconf}) and the rho meson mass (\ref{mrhoconf}) yields $M_{\rm KK} = 949\, {\rm MeV}$ and $\lambda=16.6$. As in the decompactified limit, this vacuum fit is in tension with medium properties. In fact, it gives the unphysical values $n_0 \simeq 0.43 \, {\rm fm}^{-3}$ and $\mu_0\simeq 1430\, {\rm MeV}$.  It was pointed out in Ref.\ \cite{Kovensky:2023mye} that, while keeping $m_\rho$ physical, i.e., keeping $M_{\rm KK} = 949\, {\rm MeV}$, one can find a value of the 't Hooft coupling, $\lambda=7.8$, for which $n_0$ and  $\mu_0$ are both in excellent agreement with real-world nuclear matter. With this choice, however, the pion decay constant is unphysically low, $f_\pi \simeq 63.6\, {\rm MeV}$. 

\begin{figure} [t]
\begin{center}
\includegraphics[width=\columnwidth]{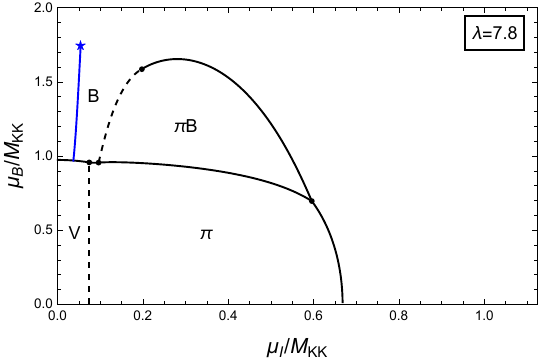}

\includegraphics[width=\columnwidth]{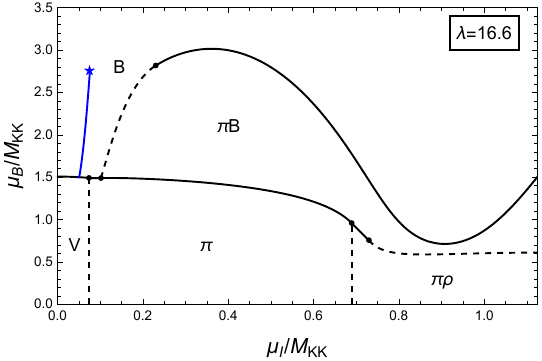}
\caption{Phase diagram in the $\mu_B$-$\mu_I$ plane for the confined geometry and two different values for $\lambda$. Black lines are phase transitions (solid and dashed corresponding to transitions of first and second order), while the blue line shows electrically neutral, $\beta$-equilibrated matter, with an endpoint (asterisk) corresponding to the interior of the heaviest neutron star, based on the construction of Sec.\ \ref{sec:neutron}. With $\lambda=7.8$, basic properties of isospin-symmetric nuclear matter at saturation are reproduced, but $f_\pi$ is unphysical, while $\lambda=16.6$ implies a physical $f_\pi$ but unphysical saturation properties. Plots taken from Ref.\ \cite{Kovensky:2023mye}, with $\mu_I$ adjusted by a factor of 2 to the conventions of this review.} 
\label{fig:lambda}
\end{center}
\end{figure} 

We show the $\mu_B$-$\mu_I$ phase diagram for both choices of $\lambda$ in Fig.\ \ref{fig:lambda}. Since $M_{\rm KK}$ does not enter the numerical calculation explicitly, both axes scale trivially with $M_{\rm KK}$. We see that the topology of the phase structure is the same for both parameter choices at small $\mu_I$, but differs significantly at large $\mu_I$. For $\lambda=7.8$, the $\pi\rho$ phase is metastable and entirely covered by the B phase. This is not per se unphysical, but in this region of the phase diagram, where the baryon density is low, the vacuum fit can be expected to be more trustworthy. Therefore, the $\lambda=16.6$ result, i.e., the lower panel of Fig.\ \ref{fig:lambda}, has been used for the three-dimensional phase diagram in the introduction. We know that for this coupling the region of small $\mu_I$ is quantitatively incorrect. In Fig.\ \ref{fig:phases3D}, this has been taken into account by rescaling the  $\mu_B$-axis  with the unphysical $\mu_0$. As a consequence, the first-order baryon onset of the $D^0$ configuration aligns, somewhat arbitrarily, with the second-order onset from the $D_{RL}$ configuration, where the rescaling has been done with a different $\mu_0$. The resulting small-$\mu_I$ regime is, qualitatively, the same as the one for $\lambda=7.8$, where the baryon onset is correct also {\it quantitatively}.

Figure \ref{fig:lambda} predicts that the effective pion mass in the baryonic medium increases. This can be seen from the second-order transition from the B to the $\pi$B phase. Anticipating the results from the next subsection, we have indicated the location of neutron stars in this phase diagram. We see that they live in the purely baryonic region, i.e., the current approach does not predict pion condensation in neutron stars. More precisely, our ansatz corresponds to isotropic $s$-wave pion condensation, which is usually assumed to be suppressed due to the repulsive pion-neutron interaction, although it cannot be ruled out \cite{Voskresensky:2022gts}. In contrast, $p$-wave condensation is considered a more viable option \cite{1972JETP...34.1184M,Sawyer:1972cq,Scalapino:1972fu}, but has not yet been studied holographically. 

For the interpretation of the phase diagram it is important to keep in mind that the calculation in the confined geometry does not allow for chiral restoration. Hence, for very large $\mu_B$, Fig.\ \ref{fig:lambda} differs from our deconfined calculation for the $\mu_I=0$ plane. In particular, there is no analogue of the (chirally restored) quarkyonic phase in the confined antipodal background. This explains why the phase transition curve between B and Qy phases in Fig.\ \ref{fig:phases3D} does not extend into the $\mu_B$-$\mu_I$ plane. In any case, very large chemical potentials need to be interpreted with care because they induce large gauge fields on the flavor branes and thus the probe brane approximation, used for all our results,  becomes questionable.

\subsection{Neutron stars}\label{sec:neutron}

Isolated neutron stars have temperatures much smaller than the baryon chemical potential and thus their thermodynamics is well approximated by zero temperature. In the simplest scenario, they contain a crystalline crust of nuclei and homogeneous matter made of neutrons, protons, and electrons in the core. This scenario can be described using the ingredients developed in this section. Neutron star matter has been studied within the WSS model and other holographic approaches \cite{Hoyos:2016zke,BitaghsirFadafan:2019ofb,BitaghsirFadafan:2020otb,Jokela:2021vwy,Hoyos:2021uff,Tootle:2022pvd,Bartolini:2022rkl,Zhang:2022uin,Bartolini:2023wis,Bartolini:2025sag,BitaghsirFadafan:2025rpm}, for reviews see Refs.\  \cite{Jarvinen:2021jbd,Hoyos:2021uff}. Here we review the WSS construction of Refs.\ \cite{Kovensky:2021kzl,Kovensky:2021ddl}, which is unique in the sense that the entire star -- including the crust -- is built from the same model. It is based on the $D^0$ configuration of baryonic matter in the confined antipodal setup of the previous subsections. Isospin is included as explained, with the additional simplification that the functions $h_1(u), h_2(u), h_3(u)$ are assumed to be identical. This is a good approximation for the relatively small values of $\mu_I$ involved here \cite{Kovensky:2023mye}.

Since leptons interact weakly with each other and with the baryons, we may simply add a gas of free electrons and muons to the thermodynamic potential obtained from holography. With $\ell=e,\mu$ denoting electrons and muons, pressure and number density are 
\begin{subequations}
\bea
P_\ell  &=& \frac{\Theta(\mu_\ell-m_\ell)}{24\pi^2}\left[(2\mu_\ell^2-5m_\ell^2)\mu_\ell\sqrt{\mu_\ell^2-m_\ell^2}\right.\non[2ex]
&&\left.+3m_\ell^4\ln\frac{\sqrt{\mu_\ell^2-m_\ell^2}+\mu_\ell}{m_\ell}\right]\,, \\[2ex] 
n_\ell &=& \Theta(\mu_\ell-m_\ell) \frac{(\mu_\ell^2-m_\ell^2)^{3/2}}{3\pi^2} \, ,
\eea
\end{subequations}
where electron and muon masses are $m_e=511\, {\rm keV}$ and  $m_\mu=106\, {\rm MeV}$, and their chemical potentials are denoted by $\mu_e$ and $\mu_\mu$.

Electric charge neutrality 
and equilibrium with respect to the electroweak interaction are imposed as follows. We recall that neutron and proton states are not explicitly present in our calculation since isospin degrees of freedom are treated classically in the bulk. Nevertheless we can assign electric charges $0$ and $+1$ to the two isospin components. To this end, we define neutron and proton chemical potentials by $\mu_n = \mu_B + \mu_I$, $\mu_p = \mu_B-\mu_I$, which yields the corresponding number densities $n_n = (n_B+ n_I)/2$, $n_p = (n_B-n_I)/2$. 
Electroweak equilibrium is imposed with respect to the processes $p+\ell\to n+\nu_\ell$, $n\to p+\ell+\bar{\nu}_\ell$. Neglecting the neutrino chemical potentials, this yields $\mu_e=\mu_\mu$ and $\mu_n=\mu_p+\mu_e$. Electric charge neutrality is $n_p=n_e+n_\mu$. These constraints are implemented in terms of our holographic dimensionless quantities as 
\be\label{neutral}
\mu_e  = -2N_c\lambda_0M_{\rm KK} \bar{\mu}_I \, , \quad 
\frac{\bar{n}_B-\bar{n}_I}{2} = \frac{3\pi^2(n_e   + n_\mu)}{\lambda_0^2 M_{\rm KK}^3} \, . 
\ee
We need to solve the holographic equations of motion coupled to these constraints. The scaling with $N_c$ requires a comment. The above definitions for neutron and proton chemical potentials together with Eq.\ (\ref{neutral}) follow our conventions of Table \ref{table: dimensionless defs}, except for the relation between $\mu_I$ and $\bar{\mu}_I$. Following the convention of Ref.\ \cite{Kovensky:2021kzl}, this relation is taken to be $\mu_I=N_c\lambda_0 M_{\rm KK} \bar{\mu}_I$. This implies that neutrons are composed of $N_c-1$ down quarks and 1 up quark, and vice versa for protons. Hence, at large $N_c$, $\mu_n \simeq N_c\mu_d$, $\mu_p\simeq N_c\mu_u$. It is  possible to define large-$N_c$ nucleons differently, with the neutron having $(N_c+ 1)/2$ down quarks and $(N_c- 1)/2$ up quarks. (In both cases, assigning electric charges +1 and 0 has to be understood in the $N_c=3$ case.) This would lead to  different factors $N_c$ in Eq.\ (\ref{neutral}): The factor $N_c$ in front of $\bar{\mu}_I$ would be absent and a factor $N_c$ would appear in front of $\bar{n}_I$. Since eventually in the calculation we set $N_c=3$, this is not a fundamental difference, but will  change the results quantitatively. Since $\bar{\mu}_I$ and $\bar{n}_I$ scale inversely with respect to $N_c$, this different convention leads for instance to a symmetry energy smaller by a factor $N_c^2$ \cite{Bartolini:2022gdf}. The physical reason is easy to understand: Within the scalings employed here, moving away from the isospin-symmetric case is done in units of the baryons, much more costly energetically as if it is done in smaller units of the quarks. 

Imposing the conditions (\ref{neutral}) allows us to identify neutron star matter in the phase diagram. The blue line in Fig.\ \ref{fig:phases3D} (and in Fig.\ \ref{fig:lambda}) has been computed in this way (note that the endpoint cannot be obtained from this calculation alone). To actually construct a neutron star, we need the equation of state, which requires the calculation of the energy density $\epsilon$. In the presence of $\beta$-equilibrium and charge neutrality, we have 
\begin{equation}
    \epsilon = - P +\mu_n n_B \, , 
\end{equation}
where $P = -\Omega + P_e + P_\mu$ includes the holographic contribution from the baryons $\Omega$ and the lepton contributions. Coupling with gravity is done by inserting the equation of state $P(\epsilon)$ as well as its derivative, the squared speed of sound $c_s^2$, into the Tolman-Oppenheimer-Volkoff (TOV) equations \cite{tolman,PhysRev.55.364,PhysRev.55.374} together with the equation needed for the tidal deformability \cite{Hinderer:2007mb,Postnikov:2010yn,Zacchi:2020dxl,Margueron:2021dtx},
\begin{subequations} \label{TOVs}
\bea
&&\frac{\partial P}{\partial r} = -\frac{G}{r^2} \frac{(M+4\pi Pr^3)(\epsilon+P)}{1-\frac{2GM}{r}} \, ,  \label{TOV1}\\[2ex]
&& \frac{\partial M}{\partial r} = 4\pi r^2 \epsilon \, ,  \label{TOV2}\\[2ex]
&&0= r\frac{\partial y}{\partial r}+y^2+\frac{4\pi G r^2\left(5\epsilon+9P+\frac{\epsilon+P}{c_s^2}\right)-6}{1-\frac{2GM}{r}}\non[2ex]
&&+\frac{y\left[1-4\pi G r^2(\epsilon-P)\right]}{1-\frac{2GM}{r}}-\frac{4G^2(M+4\pi Pr^3)^2}{r^2\left(1-\frac{2GM}{r}\right)^{2}} \, , \hspace{0.8cm}\label{TOV3}
\eea
\end{subequations}
where $G=6.709 \times 10^{-39}\, {\rm GeV}^{-2}$ is the gravitational constant. This system of coupled differential equations must be solved for mass $M(r)$, pressure $P(r)$, and the dimensionless function $y(r)$, where the radial variable $r$ measures the distance from the center of the star. The boundary conditions are  $M(0)=0$, $P(0)=P_c$, $y(0)=2$, where $P_c$ is the central pressure of the star. The dimensionless tidal deformability 
\be
\Lambda = \frac{2k_2}{3c^5} 
\ee
can then be computed from the so-called Love number
\bea
&&k_2 = \frac{8c^5}{5}(1-2c)^2[2-y_R+2c(y_R-1)]\non[2ex]
&&\times \Big\{2c[6-3y_R+3c(5y_R-8)]\non[2ex]
&&+4c^3[13-11y_R+c(3y_R-2)+2c^2(1+y_R)]\non[2ex]
&& +3(1-2c)^2[2-y_R+2c(y_R-1)]\ln(1-2c)\Big\}^{-1} \, , \hspace{1cm}
\eea
with $y_R=y(r=R)$, $R$ being the radius of the star, and where $
c= GM/R$ is the compactness of the star.

\begin{figure} [t]
\begin{center}
\includegraphics[width=\columnwidth]{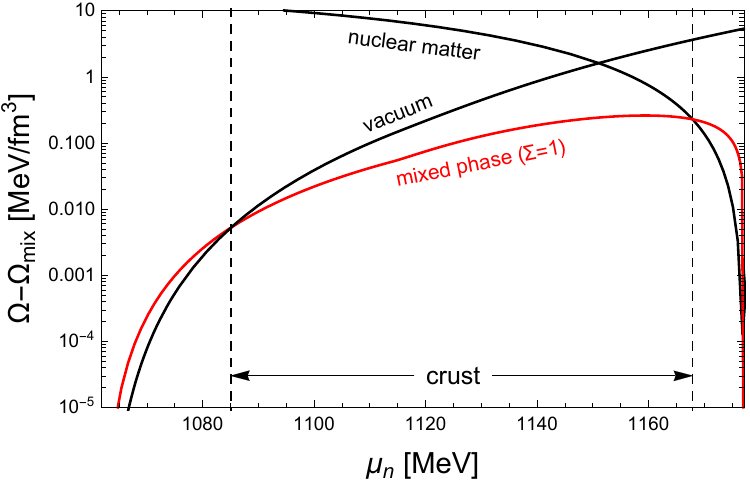}
\caption{Free energy comparison between the pure phases (holographic nuclear matter and the baryonic vacuum) and the mixed phase with surface tension $\Sigma=1 \, {\rm MeV}/{\rm fm}^2$, all relative to the mixed phase where Coulomb and surface effects are ignored. The first-order phase transitions marked with dashed vertical lines define the boundaries of the neutron star crust in terms of the neutron chemical potential $\mu_n$.} 
\label{fig:crust}
\end{center}
\end{figure}

Rather than patching together holographic dense nuclear matter with  low-density nuclear matter from ``traditional'' methods, it is in fact possible to obtain realistic stars from holography alone. To this end, we construct the crust in an approximation originally used for the quark-hadron mixed phase \cite{Glendenning:1992vb,PhysRevLett.70.1355,Schmitt:2020tac}.  The idea is to find solutions where holographic nuclear matter (plus leptons) is positively charged and occupies a volume fraction $1-\chi$, while a pure lepton gas, negatively charged,  occupies a volume fraction $\chi$, such that the system is globally, not locally, neutral, 
\be
(1-\chi)\left[\frac{\bar{n}_B-\bar{n}_I}{2} - \frac{3\pi^2(n_e   + n_\mu)}{\lambda_0^2 M_{\rm KK}^3}\right] = \chi \frac{3\pi^2(n_e   + n_\mu)}{\lambda_0^2 M_{\rm KK}^3}  
\, .
\ee
One finds mixed phase solutions for $\chi\in [0,1]$ that are energetically favored over the pure phases in a vicinity of the first-order baryon onset. Here it is important that  we work with the $D^0$ configuration; the $D_{RL}$ phase does not allow for this construction because its baryon onset is of second order, see Fig.\ \ref{fig:DRLOmega}. The resulting free energy of the mixed phase is denoted by $\Omega_{\rm mix}$ in Fig.\ \ref{fig:crust}, where it is used as reference value. So far, this construction ignores the geometric structure of the mixture and the energy costs from separating the two phases spatially. These effects can be approximated by assuming bubbles with infinitely thin interfaces.  Then, the competition, say at fixed $\chi$,  between Coulomb energy (favoring many small bubbles) and surface energy (favoring fewer large bubbles) yields an energetically favorable bubble size that depends on $\mu_n$. This calculation requires an additional input, namely the surface tension, which we take to be $\Sigma = 1\, {\rm MeV}/{\rm fm}^2$, a typical value for ordinary nuclei \cite{Hua:2000gd,Drews:2013hha}, although in reality the density variations in the crust are expected to affect this value (see Ref.\ \cite{Kovensky:2021kzl} for how the results change with different values of $\Sigma$). 

\begin{figure} [t]
\begin{center}
\includegraphics[width=\columnwidth]{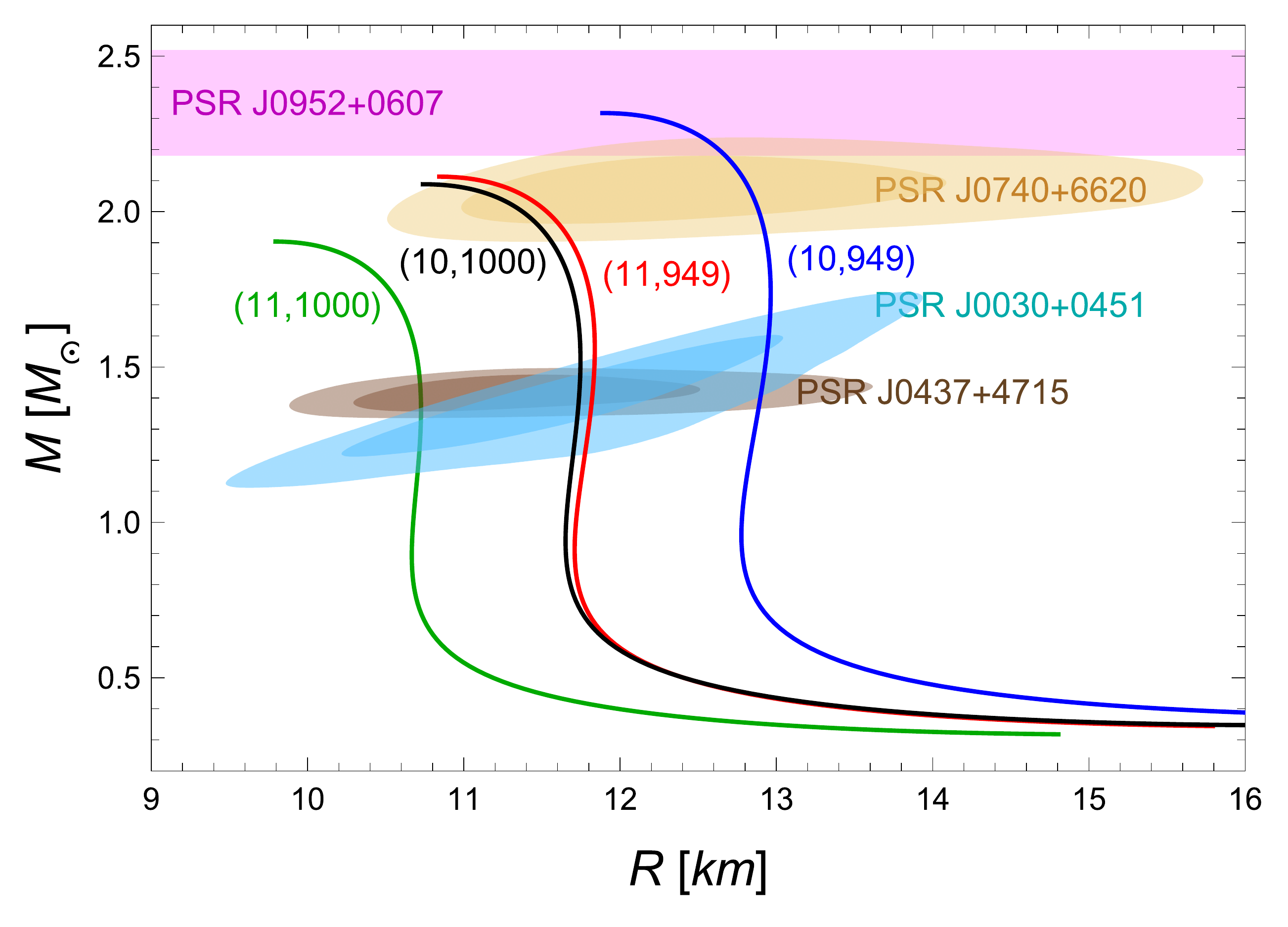}

\includegraphics[width=\columnwidth]{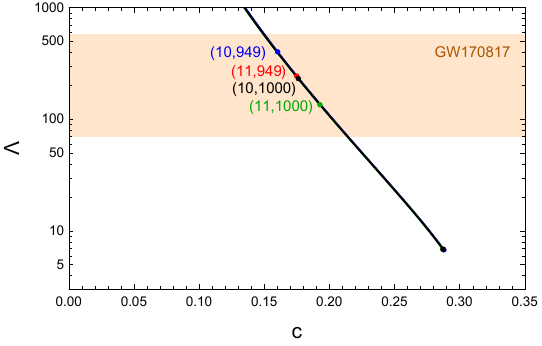}
\caption{{\it Upper panel:} Mass-radius curves of holographic neutron stars, for different parameter sets, with the first (second) number referring to $\lambda$ ($M_{\rm KK}$ in units of MeV). Astrophysical observations are included for the mass of PSR J0952+0607 \cite{Romani:2022jhd} and NICER data (68\% and 95\% confidence regions) for PSR J0030+0451 \cite{Miller:2019cac}, PSR J0740+6620 \cite{Miller:2021qha}, PSR J0437+4715 \cite{Choudhury:2024xbk}. {\it Lower panel:} Tidal deformability as a function of compactness for the same parameter sets (the four lines are almost exactly on top of each other). The markers are for stars with $M=1.4\, M_\odot$ and all are within the range inferred from the neutron star merger GW170817 \cite{LIGOScientific:2018hze}.} 
\label{fig:NS}
\end{center}
\end{figure}

These effects increase the free energy of the mixed phase, but there is still a regime in $\mu_n$ where it is favored, as shown in Fig.\ \ref{fig:crust}. This regime refines the equation of state for low densities. Coupling the full equation of state (mixed phase at low densities, homogeneous nuclear matter at large densities) with the TOV equations yields a dynamical calculation of the crust-core transition and thus a holographic prediction for the size of the crust. For the four parameter sets used in Fig.\ \ref{fig:NS}, the thickness of the crust for the most massive star is $\Delta R \simeq (0.77 - 0.88)\, {\rm km}$, while for a star with $M=1.4\, M_\odot$, where $M_\odot$ is the mass of the sun, we find $\Delta R \simeq (1.69 - 2.48)\, {\rm km}$, somewhat larger than predicted from traditional approaches \cite{Steiner:2014pda,Fortin:2016hny,Grams:2021lzx}. A possible reason for this discrepancy is the suppression of the so-called inner crust, where the nuclei are immersed in a neutron (super)fluid. The holographic formalism allows for the definition of pure neutron matter and thus it is possible to improve the holographic neutron star in this regard in the future.

The global properties of the holographic neutron stars are shown in Fig.\ \ref{fig:NS}, together with astrophysical constraints for mass, radius, and tidal deformability. Four different curves are shown to illustrate the dependence on the model parameters. We see that there are parameter choices that satisfy all astrophysical constraints. As $\lambda$ is decreased at fixed $M_{\rm KK}$, the radii and the tidal deformability become larger and start to violate the constraints as we approach the parameter set compatible  with saturation properties of nuclear matter, which, as we have seen in Sec.\ \ref{sec:Pfit}, are well reproduced with $(\lambda,M_{\rm KK})=(7.8,949\, {\rm MeV})$. A more systematic discussion of the results in view of astrophysical data can be found in Ref.\ \cite{Kovensky:2021wzu}.

As pointed out in Ref.\ \cite{Ecker:2025sjb}, the incompressibility at saturation density of the holographic nuclear matter used here is much larger than in real-world nuclear matter, see also Ref.\  \cite{Kim:2008bv}. It is thus somewhat surprising that realistic stars can nevertheless be obtained. Presumably the main reason is that the large stiffness at low densities is compensated by a softer equation of state at high densities compared to non-holographic models \cite{Ecker:2025sjb}.

\section{Future directions}

The results reviewed here are a snapshot of the current understanding of QCD-like phases from ``holographic QCD'' based on the WSS model. Numerous open questions remain, some of them in fact raised by the novelties presented here. These include questions concerning the calculations within the WSS model, as well as questions  in the wider context of other holographic approaches to QCD or in the context of QCD itself. Let us list some future directions, from specific refinements to more difficult broader extensions.

\begin{itemize}

\item The three-dimensional $T$-$\mu_B$-$\mu_I$ phase diagram shown here is incomplete and in parts inconsistent. A straightforward extension -- perhaps numerically challenging -- is to use the decompactified limit throughout. In this way the $T=0$ plane can be made consistent with the nonzero-temperature results and the phase transition surfaces in the full three-dimensional space can in principle be calculated.

\item We have presented the first attempt at a holographic version of the BEC-BCS crossover at nonzero isospin chemical potential. There seems to be no fundamental reason why the crossover region (without the weak-coupling BCS limit) cannot be described within holography. However, our $\pi$Q construction with string sources,  although a valid solution of the model, turns out to be metastable, and hence does not appear to alter or replace the purely mesonic pion-condensed phase. Improved constructions or a deeper understanding of this observation are clearly desirable.

\item The recent improvement of the homogeneous approximation of baryonic matter encourages further studies. It is relatively straightforward to include isospin into the favored  $D_{RL}$ configuration. This would be a step towards an isospin-asymmetric version of quarkyonic matter, which has not been considered within holography so far, and is potentially relevant for the physics of neutron stars.

\item The generalization of the quarkyonic phase presented here can be further exploited. From a conceptual point of view, a deeper understanding of the layered structure in the bulk and its relation to the weak-coupling Fermi surface picture is required. Observables such as the speed of sound can also be re-calculated with our improved construction. Since our phase diagram predicts quarkyonic matter at relatively large temperatures, its effect for heavy-ion collisions and neutron star mergers  should be investigated. 

\item The neutron star crust within the WSS model contains various simplifications which suggest interesting extensions. Most directly, an inner crust can be constructed, adding a phase that combines nuclear clusters immersed in pure neutron matter, in addition to the existing crust where nuclear clusters are immersed in a lepton gas. Conceptually more challenging, it would be interesting to compute the surface tension of the nuclear clusters within the model itself rather than taking it as an external input. A related study in the WSS model was done recently for the interfaces between chirally restored and chirally symmetric matter \cite{Bigazzi:2026ppu}.

\item So far, neutron stars from holographic nuclear matter have only been constructed in the zero-temperature limit. For applications to neutron star mergers, it would be important to include nonzero-temperature effects, which can be done by employing the decompactified limit for isospin-asymmetric matter. No further approximations would be necessary for this calculation, which is interesting in comparison to non-holographic models, where nonzero-temperature effects in neutron star mergers are often treated within simple approximations.

\item Several of the recent advances discussed here can be combined with earlier studies including a magnetic field. For instance, inverse magnetic catalysis at nonzero baryon chemical potential was discussed with pointlike baryons, and it would be interesting to redo the analysis with more realistic baryonic approximations or in the presence of quarkyonic matter. This can be combined with the chiral soliton lattice, predicted to appear in QCD at large magnetic fields and baryon chemical potentials and recently studied in the WSS model \cite{Amano:2025iwi}. 

\item The results presented here were all obtained with two flavors. Obviously, including strangeness  would open up several possibilities for predictions and comparisons with QCD. For most applications it would be crucial to include a strange quark mass different from the light quark masses. This can be done in a purely mesonic setting and, as a more difficult extension, in dense hyperonic matter, along the lines of Refs.~\cite{Hata:2007tn,Hashimoto:2009st}. A concrete question with  astrophysical relevance would be whether the WSS model -- or holography in general -- predicts kaon condensation in nuclear matter.

\item We have solely discussed phases in thermodynamic equilibrium. Non-equilibrium or near-equilibrium properties are crucial to further refine the holographic description and to make predictions that can be tested against data. It is one of the strengths of the holographic approach that transport properties can be calculated without major conceptual problems. This has been done extensively in the context of fundamental questions in hydrodynamics and, e.g., applications to heavy-ion collisions. Applications of holographic transport to neutron stars have been explored more recently as well \cite{Hoyos:2020hmq,Hoyos:2021njg,CruzRojas:2024etx,Hoyos:2024pkl}. It is mainly this regime where the WSS results in equilibrium provide a basis for innovative studies, for instance to understand dissipative phenomena in strongly-coupled nuclear matter and potentially in quarkyonic matter.

\item Several candidate phases suggested by field-theoretic models or by perturbative QCD are not part of the WSS phase diagram and have not yet been constructed in this model. Most prominently, at high baryon densities, QCD is expected to be a color superconductor due to Cooper pairing of quarks. Holographic approaches, more or less far away from QCD, have been suggested \cite{Chen:2009kx,Faedo:2018fjw,BitaghsirFadafan:2020otb,Ghoroku:2023uxr,Preau:2025ubr,CruzRojas:2025fzs} but a realistic holographic version is difficult to construct. Perhaps our $\pi$Q phase, where quarks in the IR coexist with a pion condensate, i.e., a condensate of quark-antiquark pairs, can help conceptually to construct a phase with quark-quark pairing. 
A similar question arises in nuclear matter, where Cooper pairing is suggested by general principles for low temperatures as well. It would be important to develop a convincing strong-coupling picture of nuclear superfluidity or superconductivity, possibly based on the models for baryonic matter discussed in this review. An early attempt in the WSS model was made in Ref.\ \cite{KalyanaRama:2011ny}.

\item In cold and dense QCD, spatially anisotropic or  inhomogeneous phases are expected, for example through a spatially dependent chiral condensate and/or Cooper pair condensate. A candidate phase is the chiral density wave, brought up in the holographic context in Refs.\ \cite{Chuang:2010ku,Ooguri:2010xs,Bu:2018trt}, and one may ask if such a modulation affects baryonic matter in  the WSS model \cite{progress}. Pion condensation in the $p$-wave channel is a related phenomenon and can be modeled straightforwardly, due to the fundamental connection of the WSS model with chiral perturbation theory.

\item The WSS phase diagram discussed here has obvious unphysical features that are very difficult to fix, such as a first-order chiral phase transition. It is therefore important to understand which predictions to take seriously and which ones to discard as artifacts, and to view the results as part of a joint effort with other holographic studies. For instance, bottom-up models can reproduce the QCD chiral crossover (by construction) and have also been used for a global view on the QCD phase diagram. Mutually benefiting from the improvements in various models is thus important; for instance, one can ask whether the WSS constructions for meson condensation, the quarkyonic phase, or the BEC-BCS crossover can also be implemented in V-QCD or a D3/D7 model.

\end{itemize}

From a more general point of view, one should keep in mind the strengths and limitations of the gauge-gravity duality. While refinements have been made over the years and new methods have been developed, a gravity dual of real-world QCD is currently out of reach. Nevertheless, holography remains a valuable approach to improve our understanding of strongly coupled matter. One general question is whether matter inside a neutron star is strongly coupled or if it can be well described with weak-coupling methods, or whether both descriptions are needed depending on the observable under consideration. For questions like these a strong-coupling approach, even if it is limited to the infinite-coupling limit and even if it is only valid in some distorted version of QCD, is useful and should be further exploited. 

Models like the WSS model are not controlled approximations to QCD in a strict sense and thus should not be taken too literally in their predictions for QCD. Several simplifications are usually employed, on top of the already existing limitations of the gauge-gravity duality. 
Nevertheless, as emphasized throughout this review and manifest in the phase diagram in the introduction, the WSS model is  able to describe a variety of physical QCD-like phases and non-perturbative phenomena within the same framework. This is helpful for a global understanding of the phase diagram and is often difficult to achieve in phenomenological field-theoretic models.  Furthermore, holographic models can be crucial, at least on the same level as other phenomenological models, to point out novel candidate phases or previously unknown exotic effects and provide a dynamical and consistent description of them. In a joint effort with full QCD calculations and experimental data it then remains to be seen whether these effects play a role in $N_c=3$ QCD away from the strong-coupling limit.

\section*{Acknowledgments}
We thank Orestis Papadopoulos for valuable comments and discussions. The work of N.K.~is supported by CONICET-Argentina.

\bibliography{references}

\end{document}